\documentclass[]{aa}
\pdfoutput=1
\usepackage{natbib}
\usepackage{graphicx}
\usepackage{caption}
\usepackage{subcaption}
\usepackage{color}
\usepackage{amsmath}
\usepackage{url}
\usepackage{enumitem}

\usepackage{txfonts}

\bibpunct{(}{)}{;}{a}{}{,} 

\newcommand\bb[1] {   \mbox{\boldmath{$#1$}}  }
\newcommand\del{\bb{\nabla}}
\newcommand\bcdot{\bb{\cdot}}
\newcommand\btimes{\bb{\times}}

\definecolor{brown}{rgb}{0.42,0.24,0.07}
\definecolor{darkgreen}{rgb}{0.0,0.6,0.00}

\DeclareMathOperator{\sech}{sech}

\begin{document}


\title{Shear--driven instabilities and shocks in the atmospheres of
  hot Jupiters}
\author{S\'ebastien Fromang \inst{1,2}, Jeremy Leconte \inst{3,4} and
  Kevin Heng \inst{5}} 

\institute{Laboratoire AIM, CEA/DSM-CNRS-Universit\'e Paris 7,
  Irfu/Service d'Astrophysique, CEA-Saclay, 91191 Gif-sur-Yvette,
  France \and Institut d’Astrophysique de Paris and UPMC, CNRS (UMR 7095)
, 98 bis Boulevard Arago, 75014, Paris, France \and Univ. Bordeaux, LAB, UMR 5804, F-33270, Floirac, France \and
  CNRS, LAB, UMR 5804, F-33270, Floirac, France \and
  University of Bern, Center for Space and Habitability, Sidlerstrasse
  5, CH-3012, Bern, Switzerland \\ \email{sebastien.fromang@cea.fr}
} 

\offprints{s.fromang}

\date{Accepted; Received; in original form;}

\label{firstpage}

\abstract
{General circulation models of the atmosphere of hot
  Jupiter have shown the existence of a supersonic eastward equatorial
  jet. Yet these results have been obtained using
  numerical schemes that filter out vertically propagating sound waves
  and assume vertical hydrostatic equilibrium, or with fully
  compressive codes that use large dissipative coefficients.}
{In this paper, we remove both limitations and investigate the effects
  of compressibility on the atmospheric dynamics by solving the
  standard Euler equations.}
{This is done by means of a series of simulations performed
  in the framework of the equatorial $\beta$--plane approximation
  using the finite volume shock--capturing code RAMSES.} 
{At low resolution, we recover the classical results
  described in the literature: we find a strong
  and steady supersonic equatorial jet of a few km.s$^{-1}$ that
  displays no signature of shocks. We next show that the jet zonal velocity depends
  significantly on the grid meridional resolution. When 
  that resolution is fine enough to properly resolve the jet,
  the latter is subject to a Kelvin-Helmholtz instability. The jet
  zonal mean velocity displays regular oscillations with a typical
  timescale of few days and a significant amplitude of about $15 \%$
  of the jet velocity. We
  also find compelling evidence for the development of a vertical
  shear instability at pressure levels of a few bars. It seems to be
  responsible for an increased downward kinetic energy flux,
  significantly affecting the temperature of the deep atmosphere,
  and appears to act as a form of drag on the equatorial jet. This instability
 also creates velocity fluctuations that propagate upward and steepen into
  weak shocks at pressure levels of a few mbars.} 
{We conclude that hot Jupiter equatorial jets are potentially unstable
  to both a barotropic Kelvin-Helmholtz instability and a
  vertical shear instability. Upon confirmation using more realistic
  models, both instabilities could result in significant time
  variability of the atmospheric winds, may provide a small scale
  dissipation mechanism in the flow, and might have consequences for
  the internal evolution of hot Jupiters.} 

\authorrunning{S.Fromang et al.}
\titlerunning{Hot jupiter atmosphere variability}
\maketitle

\section{Introduction}
\label{introduction}

In the last decade, hot Jupiters observations have evolved from
characterizing only the orbital (semi major axis, period,
eccentricity) and structural (mass and radius) properties of the newly
discovered planets to constraining also the physical properties of the
atmosphere such as the composition and the temperature--pressure
profile \citep{madhusudhanetal14,heng&showman15}, as well as
observables that are directly connected to the dynamics such as 2D
brightness maps \citep{knutsonetal07,dewitetal12} and even (although
maybe still tentatively) atmospheric winds \citep{snellenetal10}. With
new instruments such as the James Webb Space 
Telescope (JWST) soon available to the community, future 
observations will become more and more constraining. Their
interpretation and understanding will require elaborate and well
understood models of the dynamical structure of their upper
atmosphere. Such a need explains the recent development by several
groups of numerical models describing the atmospheric dynamics of hot
Jupiters
\citep[among which ][]{showman&guillot02,menouetal03,choetal08,showmanetal08,showmanetal09,dobbsdixonetal08,rauscher&menou10,hengetal11}, a comprehensive review of which has recently been
published by \citet{heng&showman15}. Taken as
a whole, these global circulation models (GCM) have produced a few
robust results that characterize the dynamics of hot Jupiter
atmospheres. Probably the 
most notable is the existence of an equatorial jet with typical
eastward gas velocity of  the order of a few kilometers per second as
first shown by \citet{showman&guillot02}. This is faster than the
speed of sound by a factor of a few (i.e. the Mach number of the jet,
defined as the ratio between the wind and the sound speed, is larger
than unity). This remarkable result comes from the fact that hot
Jupiters are both tidally locked and strongly irradiated by their
central star as a result of the small semi major axis of their
orbit. It has been given theoretical grounds by the semi--analytical
calculations of \citet{showman&polvani10,showman&polvani11} who could
explain the origin of the equatorial jet by the excitation of a
stationary planetary scale Rossby wave by stellar
irradiation, a result recently confirmed and extended to include three
dimensional effects by \citet{tsaietal14}. One reason of the success
of these models is that they appear to be consistent with some of the
observations of hot Jupiter atmospheres, such as the eastward shift of
the hot spot at the photosphere \citep{knutsonetal07} or the
day--night temperature contrast measurements
\citep{perezbecker&showman13}. 

Despite these significant findings, several questions remain. First,
the supersonic nature of the equatorial jet have raised the question
of the existence of shocks in hot Jupiters atmospheres. The presence
of such structures could have important consequences for the
atmosphere time variability and act as an efficient 
drag mechanism on the flow \citep{rauscher&menou12}.
This question is made even more accute because many of the published 
results have been obtained using approaches borrowed from the
community studying the earth atmosphere climate and use approximations
that are well adapted to study the earth atmospheric dynamics. The most
widely used is based on solving a reduced set of equations, the
so--called primitive formulation of the hydrodynamics equations, that
filters out vertically propagating sound waves and assumes that the 
gas is in hydrostatic equilibrium in the vertical direction. Such
codes are only able to capture hydraulic jumps as opposed to real shocks
\citep[see][ for a discussion of this issue]{rauscher&menou12}. Using
such approaches, several groups have reported the existence of regions
where significant compression occurs and take the form of standing
discontinuities \citep{showmanetal09,rauscher&menou10}, suggesting
that shocks might indeed develop, in agreement with theoretical
arguments such as given by \citet{heng12}. In addition to these
models, a few papers have been published and present simulations that
do not rely of these standard GCM approximations and thus do not
suffer from these limitations
\citep{dobbsdixonetal08,dobbsdixonetal10,mayneetal14,mayneetal14b}.
These groups solved the Navier-Stokes equations on the
sphere and explicitly include some form of dissipation to stabilize
the scheme and/or to account for the effect of unresolved features of
the flow. Even though the issue of shocks is not the main focus of either of these
papers, \citet{dobbsdixonetal10} report shock--like
features in some of their models. However, the use of high dissipation
coefficients in these particular simulations significantly changes the
mean flow so that the issue of shocks formation in hot Jupiter
atmospheres remains open.

In addition, it is possible that the equatorial jet is destabilized by
various hydrodynamic instabilities. For example, \citet{li&goodman10}
have recently shown using analytical arguments coupled with idealized
numerical simulations that fast supersonic equatorial jets are
vulnerable to a vertical shear instability, a possibility first mentionned by
\citet{showman&guillot02}\footnote{It is worth
reminding that such a vertical shear instability is by definition absent in any
simulation that rely on the standard primitive formulation of 
hydrodynamics.}. As shown by \citet{li&goodman10}, shocks
might develop during its nonlinear evolution and have significant
consequences for the jet mean velocity. As explained by
\citet{li&goodman10}, properly resolving this instability in a GCM is
a tremendous task because it requires that the numerical grid is
sufficiently fine to resolve the vertical pressure scale height of the
atmosphere {\it in the horizontal direction}. Given these difficulties
and because of the very idealized nature of the numerical setup used by
\citet{li&goodman10}, it is perhaps not surprising that the question
of the possible development and potential consequences of a vertical
shear instability in hot Jupiter atmospheres has never been addressed
in any GCM so far.

In addition to their importance as possible drag mechanisms, the questions of
the existence of shocks and/or of the growth of hydrodynamic
instabilities such as described above may have some potential observational
consequences in terms of the flow variability, but also to explain the
origin of the inflated hot Jupiters \citep{baraffeetal10}. Indeed, it
has been suggested that downward transport of kinetic energy possibly
associated with these instabilities \citep{showman&guillot02,GS02} along
with turbulent mixing of heat \citep{youdin&mitchell10} might
transport and deposit sufficient amount of thermal energy in the deep
layers of the atmosphere to account for such highly inflated radii
\citep[for a detailed discussion, see][]{ginzburg&sari15}.

The purpose of this paper is to develop an idealized model in order to
investigate the issues described above. For that purpose, we will
solve the compressible Euler equations. Since we 
are interested in shocks, we will use the finite volume shock capturing
scheme RAMSES \citep{teyssier02} that is based on the Godunov method
\citep[][see also section~\ref{sec:num_scheme}]{toro97}. In
addition to being well adapted to resolving shocks in supersonic flows
such as encountered here, finite volume code like 
RAMSES conserve the gas total energy, the sum of its kinetic and
thermal energy, even in the absence of explicit dissipation. This
means in particular that all the kinetic energy 
that might be numerically dissipated (as would be the case, for
example, in a turbulent flow) goes back into heat. As noted by
\citet{goodman09}, this is of particular importance in the context of
hot Jupiter atmospheres. Our strategy in this first paper is to keep
the numerical setup as simple 
as possible while retaining the basics ingredients that drive the flow
dynamics. We will thus make two important
simplifications regarding the thermodynamics and the geometry. While
most published results now use a complex but realistic treatment
of radiative transfer effects, we will model these effects with a
simple parametrized cooling function. For practical purpose, we will
use a form of the cooling function that is linear in the temperature,
also known as Newtonian cooling. This has proved very useful in
modelling atmospheric flows in 
general and hot Jupiter general circulation in particular
\citep{showmanetal08,rauscher&menou10,hengetal11}. Published
simulations using that approximation compares well with more realistic
models that include a detailed treatment of radiative transfer effects
\citep{dobbsdixonetal08,showmanetal09,hengetal11b,rauscher&menou12}. Our second 
approximation will be to solve the equations in a Cartesian geometry
using the equatorial beta plane model. This is a well--known
approximation in atmospheric dynamics and one that been successfully
used recently in the context of hot Jupiter atmospheric flow
\citep{showman&polvani10,showman&polvani11,tsaietal14,heng&workman14}. The
idea is to 
focus on the equatorial region of the planet and to expand the
vertical projection of its angular velocity linearly to the first
order in the distance $y$ to the equator:
\begin{equation}
\bb{\Omega}=\frac{1}{2} \beta y \bb{e_z} \, ,
\label{eq:beta_plane}
\end{equation}
where $\beta$ is a constant and a free parameter of the model, while
$\bb{e_z}$ stands for the unit vector in the vertical direction. For hot
Jupiter atmospheres, this is a particularly useful approximation
because of their slow rotation and since the dynamics develops mainly
at the vicinity of the equator. Using these simplifications but
including the effect of compressibility, the goal of this paper is to
investigate such questions as the occurrence of shocks, the stability
of the equatorial jet and the origin of the flow variability, if any. 

The plan of the paper is as follows. In section~\ref{sec:setup}, we
detail our numerical setup and the parameters we use. The idea is to
use a set of numerical parameters as close as possible to the
benchmark models presented by \citet{hengetal11} and to investigate
the potential effects induced by compressibility in this model. In
section~\ref{sec:LR}, we present a low resolution model and show that
we recover many of the features described in the literature despite
our simplistic geometry. These results validate our approach and our
approximations. In section~\ref{sec:res_study}, we make a quick
resolution study before presenting a high resolution simulation in
section~\ref{sec:HR}. As we shall see, the flow features variability at
different spatial and temporal scales that we relate to well known
hydrodynamics instabilities. We then conclude and discuss the
limitations of our work and the perspectives it opens in
section~\ref{sec:conclusion}.

\section{Physical model and numerical implementation}
\label{sec:setup}

\subsection{Equations and notations}

As explained in the introduction, we solve the
hydrodynamic equations in a Cartesian coordinate system denoted by
$(\bb{e_x},\bb{e_y},\bb{e_z})$, in a frame rotating with angular
velocity $\bb{\Omega}$. We consider a grid that extends over the ranges
$[-L_x/2,L_x/2]$, $[-L_y/2,L_y/2]$ and $[0,L_z]$ in the $x$, $y$ and
$z$ directions, respectively. The $x$--direction should be thought
of as representing longitudes in traditional GCM models, $y$
as a proxy for latitudes and $z$ would stand for the radial direction 
(aligned with the gravitational acceleration). Since hydrostatic
equilibrium is not built into the equations, we do not 
use pressure coordinates in the vertical direction as is common in
atmospheric sciences when solving the primitive formulation of
hydrodynamics. The time evolution for the gas density $\rho$, the
velocity $\bb{v}$ and the total energy $E$ writes: 
\begin{eqnarray}
\frac{\partial \rho}{\partial t} + \del \bcdot (\rho \bb{v})  &=&  0 \, , \\
\frac{\partial \rho \bb{v}}{\partial t} + \del \bcdot
(\rho\bb{v}\bb{v} + P )  &=& - \rho g \bb{e_z} - 2\rho  \bb{\Omega} \btimes \bb{v} \, , \\
\frac{\partial E}{\partial t} + \del \bcdot \left[
  (E+P)\bb{v}) \right] &=& {\cal L} \, .
\label{eq:hyd_eq}
\end{eqnarray}
where $g$ is the constant vertical acceleration due to the planet
gravitational field, ${\cal L}$ is the cooling function (see
section~\ref{sec:cooling_func}) and $P$ denotes thermal
pressure. It is related to the total energy with the relation: 
\begin{equation}
E=\frac{1}{2}\rho \bb{v}^2+\rho e = \frac{1}{2}\rho \bb{v}^2+
\frac{P}{\gamma-1} \, .
\end{equation}
In the above equation, we have introduced the adiabatic exponent of
the gas $\gamma$ and have assumed an ideal equation of state in
writing the internal energy $e$ as a function of thermal pressure. In
that case, $P$ also relates directly to the gas temperature $T$ 
through the following equation:
\begin{equation}
P=\frac{\rho k T}{\mu m_H}=\rho {\cal R} T \, .
\end{equation} 
Here $k$ is the Boltzmann constant, $\mu$ is the mean molecular weight,
$m_H$ the mass of the hydrogen atom and ${\cal R}$ is the specific gas
constant. Throughout this paper, we will assume the atmosphere is
composed of a mixture of hydrogen and helium and we will take for
${\cal R}$ the same numerical value as \citet{hengetal11}: ${\cal
  R}=3779$ J.kg$^{-1}$.m$^{-1}$. We will also set 
$\gamma=1.4$. Finally, as explained in the introduction, we work in
the framework of the equatorial $\beta$--plane model, so that
$\bb{\Omega}$ is given by Eq.~(\ref{eq:beta_plane}). In the remainder
of this paper, we will thus refer to the $y=0$ plane as ``the equator''.

\subsection{The cooling function ${\cal L}$}
\label{sec:cooling_func}

${\cal L}$ varies linearly with the departure from an
equilibrium temperature, a prescription that is also known as
Newtonian cooling:
\begin{equation}
{\cal L}=\frac{\rho {\cal R}}{\gamma-1}\frac{T-T_{eq}}{\tau_{rad}} \, ,
\end{equation}
where $T_{eq}$ and $\tau_{rad}$ are the radiative equilibrium
temperature and radiative cooling timescale, respectively. They both
depend a priori on the location in the atmosphere. ${\cal L}$ is meant to
provide a rough but computationally straightforward description of the
balance between heating due to stellar irradiation and radiative
cooling. Our choice for the spatial dependence of $\tau_{rad}$ is
discussed in section~\ref{sec:num_params} and
appendix~\ref{app:radiative_eq}. To calculate $T_{eq}$, we followed  
a procedure that is largely inspired by (but not identical to,
essentially because we use a different coordinate system) the
benchmark calculation presented 
by \citet{hengetal11}. We describe it here briefly for completeness. We
started from a reference pressure--dependent temperature profile
$T_{P}^0$ that is usually computed from 1D radiative transfer
calculations. We next defined: 
\begin{eqnarray}
T_{\textrm{day/night}} &=& T_P^0 \pm \Delta T \, ,
\end{eqnarray}
where $\Delta T$ is a constant and a free parameter of the
simulation. For simplicity, we did not prescribe any
pressure variation for $\Delta T$, which is different from papers
published in the literature that use Newtonian cooling to model
radiative effects. Because the planet is tidally locked, the
substellar point is fixed in time. We chose its horizontal
coordinates to be $(x,y)=(0,0)$. Using these definitions for $T_{day}$
and $T_{night}$, we calculated $T_{eq}$ as follows: on the night side 
of the planet ($|x|>L_x/4$), we followed \citet{hengetal11} and set
$T_{eq}=T_{\textrm{night}}$. On the planet day side ($|x|<L_x/4$), we
used the relation: 
\begin{eqnarray}
T_{eq}^4 &=& T_{night}^4 \nonumber \\
 &+& (T_{day}^4-T_{night}^4) \cos \left(\frac{2 \pi x}{L_x} \right) \exp
  \left(-\frac{y^2}{2 L_{th}^2} \right) \, .
\end{eqnarray}
Note that this profile is different from the functional
form of $T_{eq}$ introduced by \citet{hengetal11}. While the
sinusoidal x--dependence is reminiscent of their longitudinal profile,
a similar dependence would have little meaning in the y--direction. It
is more natural to use an exponential variation in that direction, as
done recently by \citet{showman&polvani11}. The drawback of this
formulation is that it introduces an additional length scale $L_{th}$
in the problem that is another free parameter of the model. The
specific value we chose in this paper is discussed below in
section~\ref{sec:num_params}. 

\subsection{Numerical implementation}
\label{sec:num_scheme}

We solve the above equations using a uniform grid version of the
community code RAMSES \citep{teyssier02}. RAMSES uses a finite volume
scheme based on the MUSCL-Hancock Godunov method \citep{toro97}. By
systematically upwinding all the waves that enter in the problem,
finite volume codes are intrinsically stable (provided they satisfy the
Courant--Friedrich--Lax, or CFL, condition) and do not require using
any explicit dissipation. In fact, the algorithm is constructed to add
the minimum amount of dissipation that is necessary to stabilize the
numerical scheme. This is important for one of the problems we are interested
in, namely the issue of shock formation. However, a well--known
difficulty of Godunov codes is their inability to properly handle vertical
hydrostatic equilibrium \citep[see, e.g. ][]{leveque98}. This is
because the latter results from a balance between two terms that are
treated differently by the numerical algorithm, namely the pressure
gradient (treated as a flux of momentum computed as part of the
solution of a Riemann problem) and gravity (treated as a source term
by a simple 
split Crank-Nicholson algorithm). Starting from a solution initially
in exact equilibrium balance, this situation immediately creates a
mass flux at cell interfaces that rapidly leads to spurious vertical
velocities (even in 1D) and compromises the simulation. Recently,
\citet{kappeli14} developed an algorithm for the 
MUSCL--Hancock scheme that is especially designed to fix that
problem. The idea is to suppose that the density and pressure profiles 
within each cells are isentropic, and to use the extrapolated values at cell
faces as inputs for the Riemann problem. We have implemented their
solution in RAMSES and found that it gives very satisfactory results,
in the sense that the code is now able to keep a stratified atmosphere
in hydrostatic equilibrium with almost vanishing deviations.

The boundary conditions (BC) we used are periodic in the x--direction. Because
the Coriolis force increases in amplitude with $y$, the dynamics is 
confined to the vicinity of the equator. For this reason, the 
boundary conditions in the y--direction are not critical provided the
domain is chosen wide enough. In the simulations presented in this
paper, we set the horizontal velocity to zero there and assumed
zero--gradient for all other variables. The vertical boundary
conditions are more subtle to implement and somewhat arbitrary. We
empirically found that the following produces good results, in the
sense that artifacts introduced by the boundaries could not be
detected: we extrapolated both density and pressure assuming an
isentropic vertical profile,
imposed zero gradient horizontal velocity gradient and reflective BC
on the vertical momentum. In addition, we forced the mass, momentum
and energy fluxes to vanishes at the vertical boundaries. We note that  
the two conditions (on the variables themselves and on the fluxes) are
not necessarily consistent with each other, but did not lead to any
problem in the simulations.

The numerical implementation described above was tested by reproducing
standard results of the literature. We present two of
them in the appendix of the paper, namely the growth of a baroclinic
wave in an adiabatic atmosphere as described by \citet{polichtchouketal14}
in appendix~\ref{app:baroclinic_wave} and a shallow hot Jupiter model
such as presented by \citet{menou&rauscher09,hengetal11,mayneetal14} in
appendix~\ref{app:shallow_hj}. Although these tests do not have an
analytical solution, the similarity between our results and the
published calculations, along with the analysis of a low resolution
deep model that we present in section~\ref{sec:LR} give confidence in
our setup.

\subsection{Model parameters}
\label{sec:num_params}

The physical model presented above contains several free
parameters: $g$ and $\beta$ characterize the planet, while $T_P^0$,
$\tau_{rad}$, $\Delta T$ and $L_{th}$ describes the heating and
cooling processes. Rather than presenting a complete survey of the
associated parameter space, the idea of the present paper is to choose
a unique set of parameters that matches as closely as possible those of
\citet{hengetal11} and to show that the flow properties are similar to 
the results presented in that paper. We defer a detailed study of the
influence of these free parameters on the results to a future
publication. 

\paragraph{Planet parameters: }we follow the shallow 
hot Jupiter model of \citet{hengetal11} and consider a
planet with a radius $a_p=10^8$ meters (m) and for which $g=8$
m.s$^{-2}$. We assume the planet rotates with a frequency
$\Omega_p=2.1 \times 10^{-5}$ s$^{-1}$. Expanding $\Omega=\Omega_p
\sin (\phi)$, where $\phi$ is the latitude, close to the equator, and
using Eq.~(\ref{eq:beta_plane}), we thus have:
\begin{equation}
\beta=\frac{2 \Omega_p}{a_p}= 4.2 \times 10^{-13}
\textrm{ m}^{-1}.\textrm{s}^{-1} \, .
\end{equation}

\paragraph{Newtonian cooling parameters: }we next need to prescribe
the parameters entering in the definition of $T_{eq}$ and
$\tau_{rad}$. To do so, we use an analytic relationship between
$T_{P}^0$ and $P$, as well as between $\tau_{rad}$ and pressure. Both 
functions are given in appendix~\ref{app:radiative_eq} and have
numerical parameters chosen to give an approximate match to the
profiles of \citet{hengetal11}. We also set $\Delta T=300$ K in
all our simulations. The curves so obtained for $T_{night}$ and
$T_{day}$ are plotted on figure~\ref{fig:Teq_tau_prof} and are close
to the profiles used by \citet[][see their figure 7]{hengetal11}. It
is worth noting that $\tau_{rad}$ becomes infinite below $10$
bars. This choice creates an ``inert layer'' at location where $P>10$
bars, in the sense that gas cannot cool radiatively in that region. As
we shall see, this point turned out to have rather 
important consequences in the simulations. Finally, the cooling
function ${\cal L}$ depends on the parameters $L_{th}$ that governs
its $y$--dependence. Apart from being of order $a_p$, there is no
obvious way of choosing $L_{th}$ and its influence on the flow
properties deserves further investigations. After a few tests at low
resolution, we found a good agreement with published results for 
$L_{th}=0.7 a_p$. We will use this value in the remainder of the
paper. 

\paragraph{Computational domain and initial conditions: }we
need to specify the extend of the computational domain and the 
initial conditions. We used $L_x=2 \pi a_p$
in the $x$ direction and found that $L_y=2.5 a_p$ was wide enough for
the lateral boundaries to have no effect on the flow topology. As
noted by \citet{mayneetal14}, the vertical extent of the box has to be
chosen carefully, because hot gas of the planet day side is inflated
and expands upward. We thus followed these authors and used $L_z=9
\times 10^6$ meters. Except for one model (see section~\ref{sec:HR}),
our calculations were all initiated from an atmosphere at rest
($\bb{v}=\bb{0}$ everywhere), with a uniform temperature $T=1800$ K
and in vertical hydrostatic equilibrium. We set the density in the
bottom layer (i.e. at $z=0$) so that $P=220$ bars. With these
parameters, we typically found that the pressure at the top of the
domain ranges between $6 \times 10^{-3}$ and $0.1$ mbars, i.e. well
below the typical pressures we will be interested in in the remainder
of the paper. We note that there is a current debate in the literature
regarding the sensitivity of the flow to the initial conditions
\citep{thrastarson&cho10,liu&showman13}, particularly when explicit
dissipation is small or vanishing as we consider here
\citep{choetal15}. Here, we leave these questions aside and consider
the same initial conditions as used by \citet{hengetal11}.

\paragraph{Simulation post treatment:} Unless otherwise stated, we
typically produce snapshots of all the physical variables every planet day
during the simulation. To ease comparison with published papers, we
present all of our results interpolated in the vertical direction on a
pressure grid that extends from $200$ bars at the bottom to $1$ mbars
at the top using $48$ levels uniformly spaced in the logarithm of the
pressure.\\

Finally, we measure time in unit of one planet day, which corresponds
to $299200$ seconds with our choice of parameters, or about $3.5$
Earth days. In the simulations we present below, the timestep is
always limited by sound waves propagating in the vertical
direction. It is typically of the order of $25$ seconds in model
LowRes (section~\ref{sec:LR}) and $15$ seconds in model HighRes
(section~\ref{sec:HR}). This should be contrasted with the typical
timesteps of $120$ seconds used by, for example,
\citet{hengetal11}. Although markedly larger when using the primitive
set of hydrodynamics equations, the difference with a fully
compressible set of equations is not large enough to be prohibitive
when performing hot Jupiter atmosphere simulations. This is because
the typical pressure scale height $H$ is large in this case (typically
$10^3$ km) so that the ratio between $H$ and the horizontal scales
involved in the problem is not as small as it would be for an
Earth--like simulation.

\section{Low resolution model}
\label{sec:LR}

In order to validate our numerical implementation and our choice for
the numerical parameters entering in the problem, we start by
presenting a low resolution model (labelled LowRes in the
following). The grid spatial resolution is set to
$(N_x,N_y,N_z)=(64,33,48)$ in this section. This is smaller by about a
factor of two than the resolution used by the finite difference core
of \citet{hengetal11}. However, we did not used any explicit
dissipation here, so that a one to one correspondence is difficult to
establish a priori. The model was integrated for $500$ planet days, which
corresponds to about $1700$ earth days for our choice of
$\Omega_p$. 


\begin{figure*}[!t]
\begin{center}
\includegraphics[scale=0.45]{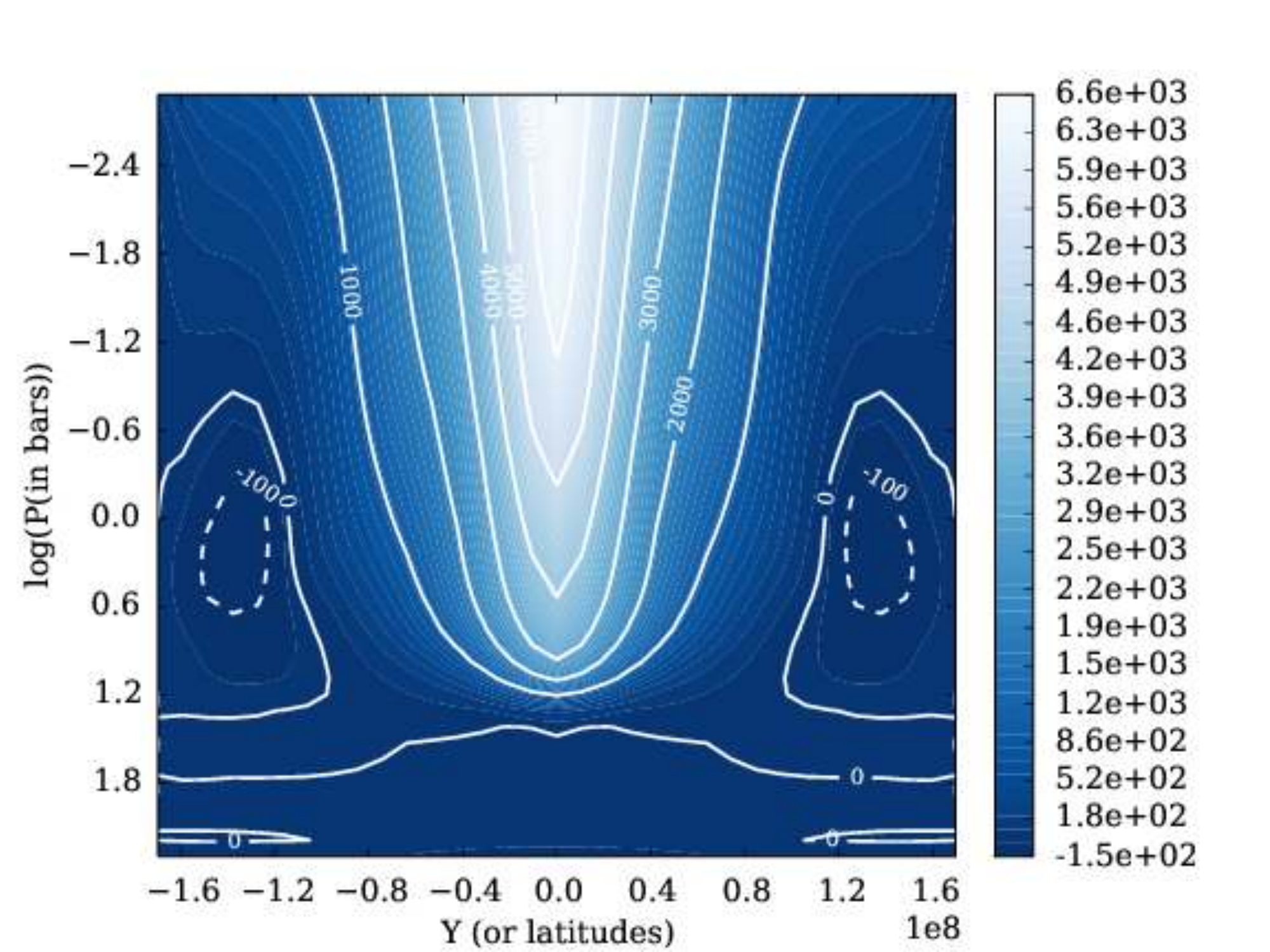}
\includegraphics[scale=0.45]{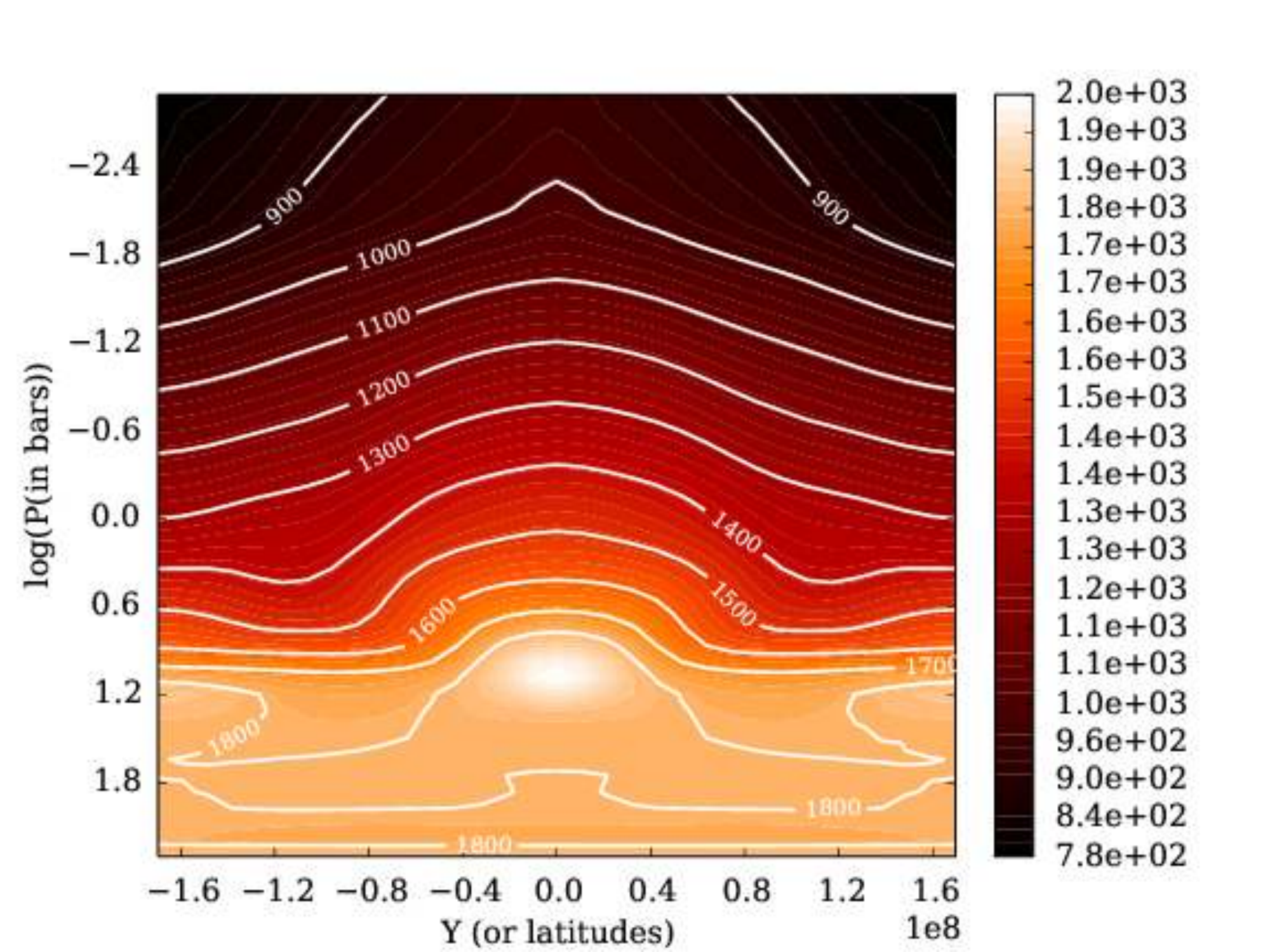}
\caption{Color contours of the time averaged zonal mean zonal velocity
  ({\it left panel}) and temperature ({\it right panel}) in the
  $(y,P)$ plane for the low resolution model. The time average is performed
  from $t=200$ days to $t=500$ days using $300$ snapshots.} 
\label{lowres_zonalMean_YP_fig}
\end{center}
\end{figure*}

\begin{figure}[!t]
\begin{center}
\includegraphics[scale=0.44]{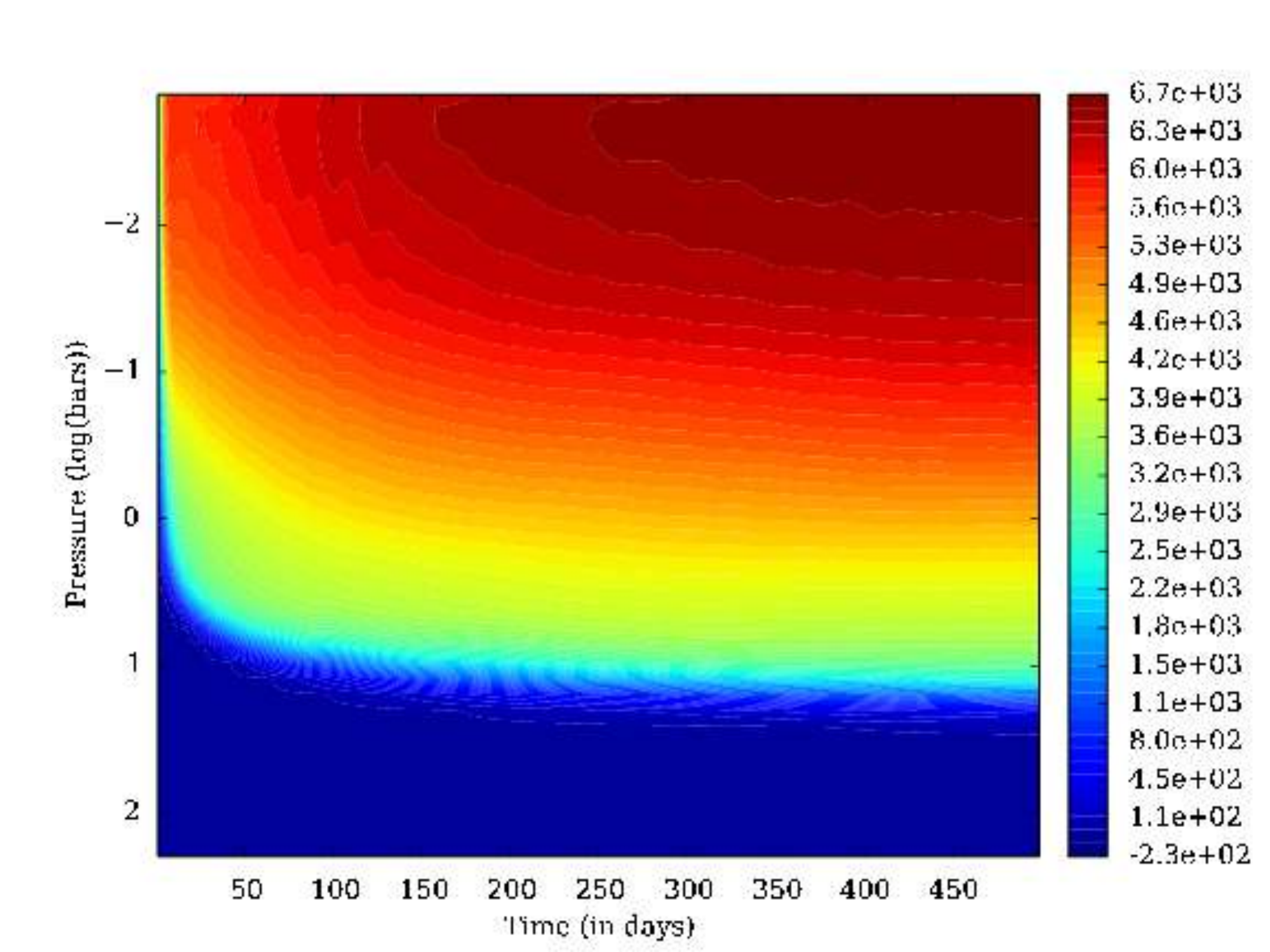}
\caption{Spacetime diagram (i.e time evolution of the pressure
  profile) of the zonally averaged zonal wind at the equator in the
  low resolution model.} 
\label{lowres_spacetimeP_vx_fig}
\end{center}
\end{figure}

\begin{figure}[!t]
\begin{center}
\includegraphics[scale=0.33]{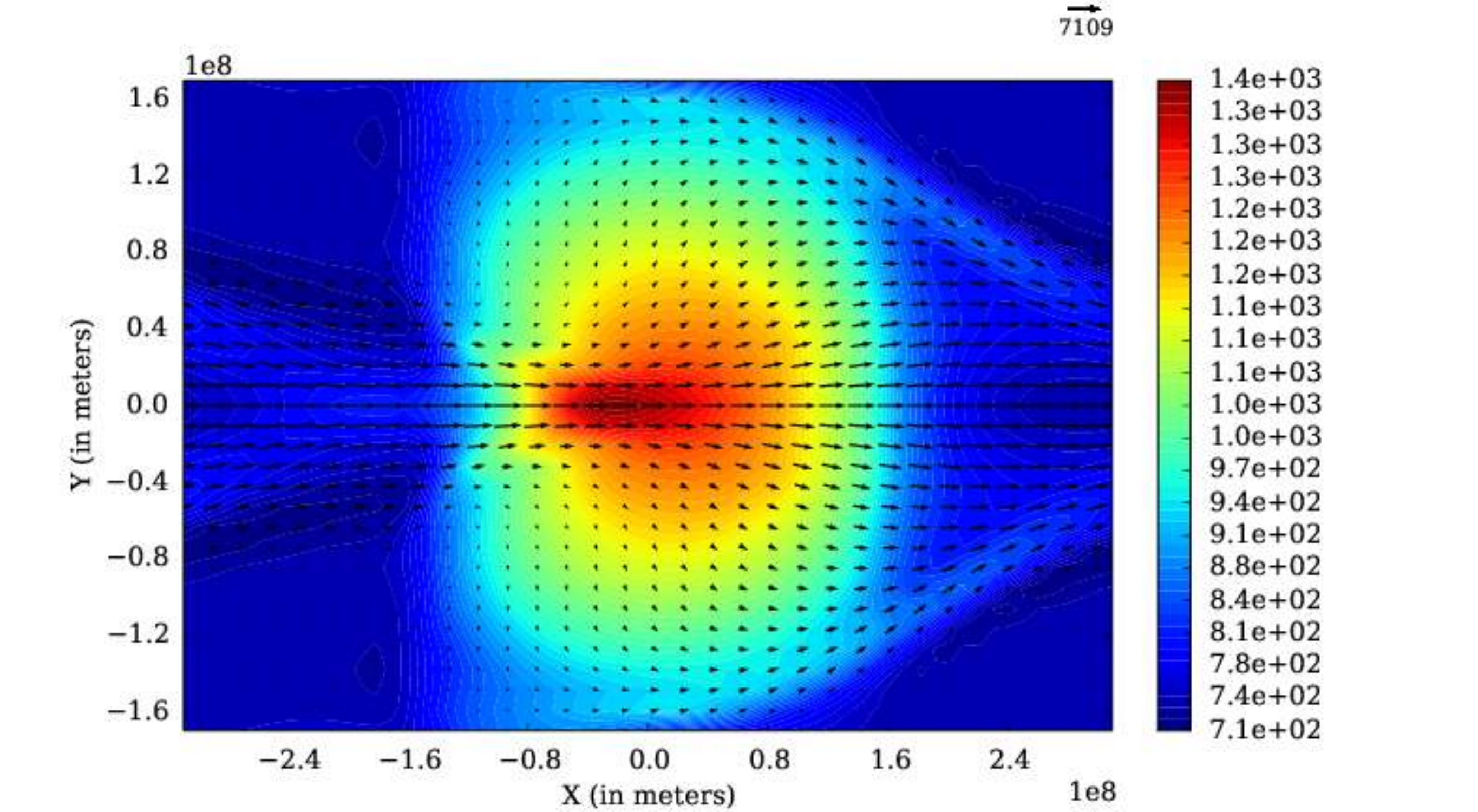}
\includegraphics[scale=0.33]{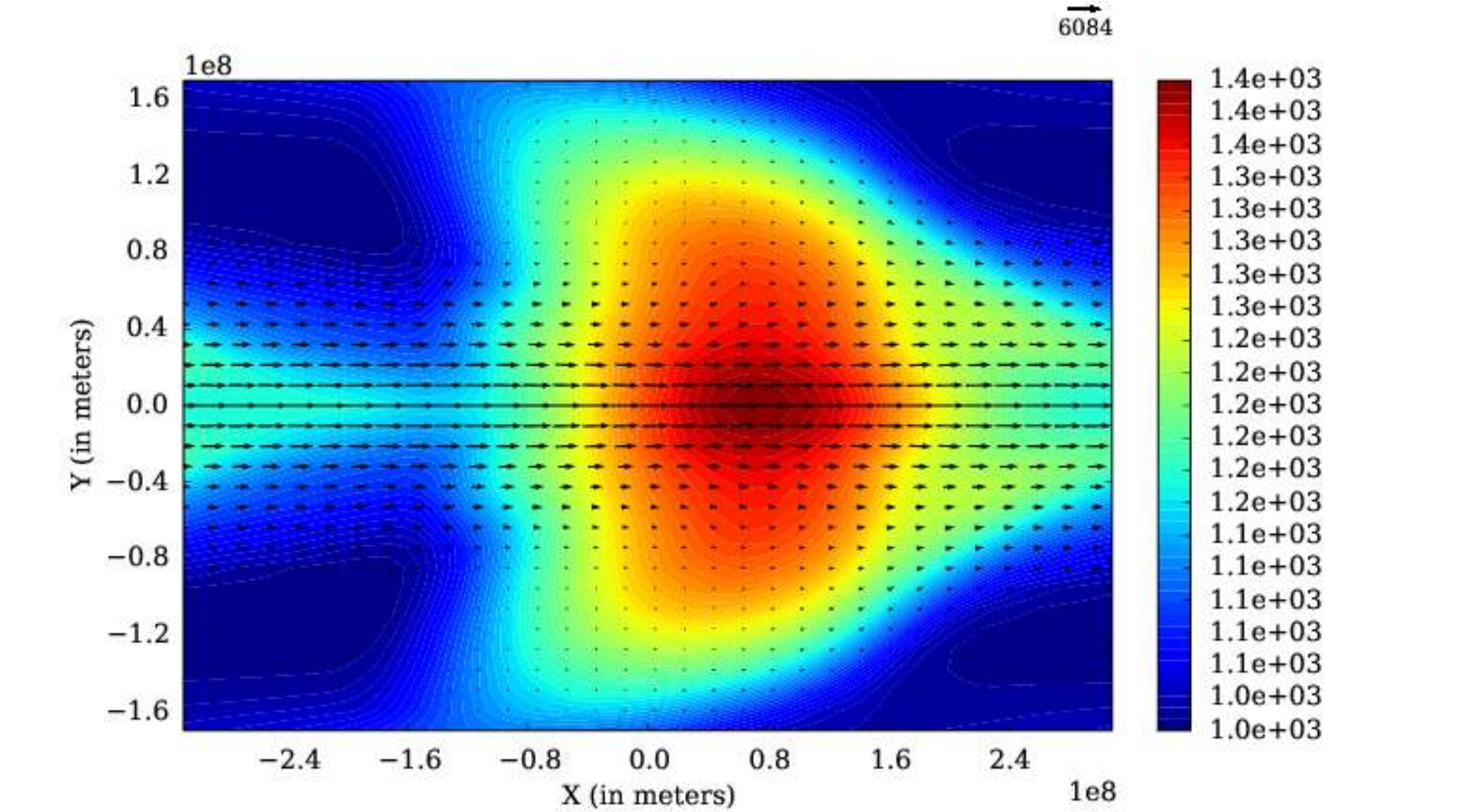}
\includegraphics[scale=0.33]{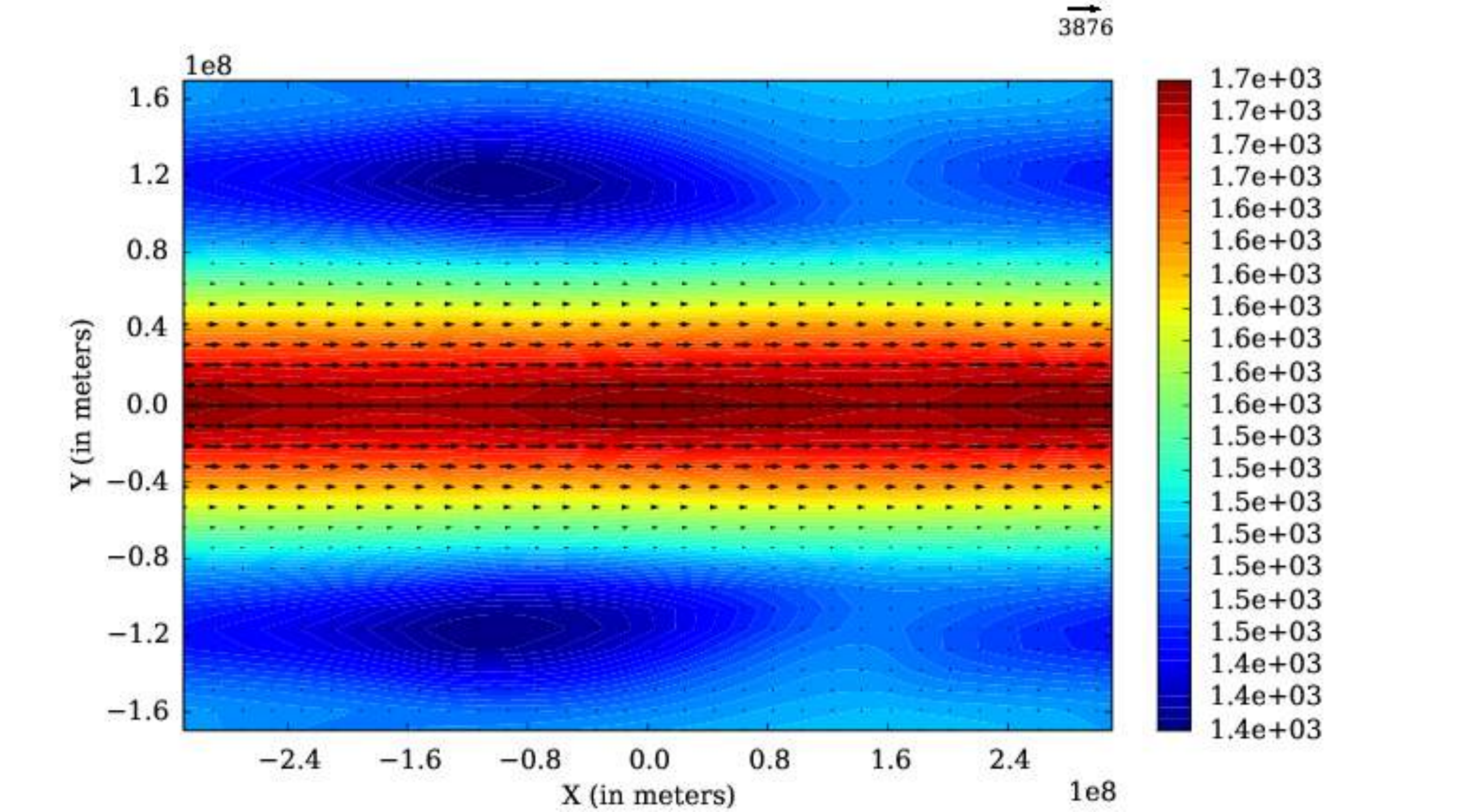}
\caption{Time averaged temperature (color contour) and horizontal wind
  (arrows) for the low resolution model at pressure level $P=1.66$
  mbars ({\it top panel}), $P=97$ mbars ({\it middle panel}) and
  $P=4.4$ bars ({\it bottom panel}). The time average is performed
  from $t=200$ days to $t=500$ days using $300$ snapshots.} 
\label{lowres_XY_fig}
\end{center}
\end{figure}

We now describe the climatology of the atmosphere. As previously
documented in the literature, the flow develops a 
strong eastward jet around the planet equator
(figure~\ref{lowres_zonalMean_YP_fig}, left panel). In about $200$ 
days, it reaches a steady state at pressure level above $1$ bar, while 
larger pressure levels gradually accelerate until the end of the
simulation (figure~\ref{lowres_spacetimeP_vx_fig}). Based on this
result, we computed the mean properties of the flow by calculating
time averages from $t=200$ until the end of the simulation. The
zonally averaged jet velocity at the equator reaches about $6.6$
km.s$^{-1}$ at the top of the atmosphere (i.e. at a pressure level of
about $1$ mbar), which corresponds to a mean Mach number of
about $2$. As a result of the short radiative timescale in the atmosphere
(compared to the dynamical timescale), the temperature structure
displays a structure that is close to the zonally averaged radiative
equilibrium temperature, except in the deep layers of the atmosphere (at
$P=10$ bars and below) where a hot spot is seen with $T$ up to $2000$
K. We find very little velocity fluctuations in the flow above $1$
bar after $200$ days which is consistent with, for example, 
\citet{showmanetal08}. Even if we tend to find slightly faster
equatorial jets, the zonally averaged structure described above
compares favorably with results published in the literature that use a
similar cooling function
\citep{showmanetal08,rauscher&menou10,hengetal11,mayneetal14}. One
difference, though, is that we find weaker westward jets at high
latitudes (with maximum westward winds of $~100$ m.s$^{-1}$ compared to
typical published values that are about one order of magnitude
larger). This difference is most likely due to the fact that we are
using a Cartesian coordinates system, as opposed to the spherical
geometry that is commonly used. We have checked that the total
(i.e. volume integrated) angular momentum is conserved in our
calculations. In the equatorial $\beta$--plane approximation, the
latter takes the form 
\begin{equation}
M=\iiint \rho \left( u - \frac{1}{2} \beta y^2 \right) d\tau \, ,
\end{equation}
where the integral should be taken over the computational domain. We
have found that $M$ is conserved to within $2 \%$ in model 128x33 over 
the $500$ days of integrations (with the small change being associated with
a small leakage through the meridional boundaries of the domain). The
weakness of westward jets in our simulation is something to keep in
mind when comparing our results with previously published models.

The day/night heating contrast results in strong zonal asymmetries in
both the dynamical fields and the temperature. These asymmetries are
stronger in the atmosphere upper layers and gradually decrease
downward. This is illustrated in figure~\ref{lowres_XY_fig}, which
shows the temperature and horizontal winds in the $(x,y)$ plane at the
pressure levels $1.66$ mbars, $100$ mbars and $4.4$ bars. The first
two panels zonal asymmetries are large because of the short
radiative timescale at those locations. For example, at $1.66$ mbar,
the maximum (resp. minimum) velocity amounts to $7.1$ km.s$^{-1}$
(resp. $5.5$ km.s$^{-1}$) and temperatures range from about $700$ K to
$1400$ K. By contrast, the equatorial jet is essentially
zonally symmetric at $P=4.4$ bars, in agreement with previous
results. At $100$ mbars, we recover the chevron shape structure in the 
temperature reminiscent of previously published calculations. The
structure of the flow at $1.66$ mbars is interesting, as 
it illustrates one of the consequence of using the equatorial $\beta$
plane model: while simulations performed on the sphere typically
display a planetary scale Hadley cell, with upward motions in the
substellar point and downward motions at the antistellar point, such a
flow is impossible to obtain by definition in the framework of the
equatorial $\beta$--plane model. Instead, meridional velocity are
always deflected back to the equator since the Coriolis force becomes
gradually stronger as $y$ increases.

To conclude on section~\ref{sec:LR}, the good agreement between the
flow structure in our simulation, as shown on
figures~\ref{lowres_zonalMean_YP_fig} and \ref{lowres_XY_fig}, 
with equivalent figures reported by \citet{hengetal11}, along with the
fact that these results broadly agree with several studies of the same
kind published in the literature, both validate our numerical
implementation and the choice of our model parameters.

\section{Resolution study}
\label{sec:res_study}

\begin{table}[t]\begin{center}\begin{tabular}{@{}ccccc}\hline\hline
Model & $N_x$ & $N_y$ & $U^{mean}_{50}$ & $U^{\prime}_{50}$  \\
 &  & & m.s$^{-1}$ & m.s$^{-1}$  \\
\hline\hline
LowRes  & 64 & 33 & 6158 & - \\
\hline
128x33 & 128 & 33 & 6117 & 1.1 \\
128x65 & 128 & 65 & 7390 & 18.6 \\
128x109 & 128 & 109 & 7703 & 66.8 \\
128x195 & 128 & 195 & 7957 & 124.3 \\
\hline
256x65  & 256 & 65 & 7371 &  - \\
512x65  & 512 & 65 & 7226 &  - \\
1024x65  & 1024 & 65 & 7511 &  - \\
\hline
HighRes & 1024 & 195 & 7731 & 207.3 \\ 
\hline\hline
\end{tabular}
\caption{Model properties. Col. 1 gives the model label. Col. 2 and 3
  features the horizontal resolution. Col. 4 and col. 5 respectively
  gives the zonally and time averaged zonal wind and zonal wind
  fluctuations at (see text for details on its calculation) $50$
  mbars. All the other parameters of the models are identical and
  described in section~\ref{sec:setup}. Data are averaged between
  $t=200$ and $t=400$, except for the models having a $N_y=195$ for
  which the averaging is performed between $t=170$ and $t=280$.} 
\label{tab:runs}
\end{center}
\end{table}

\begin{figure}[!t]
\begin{center}
\includegraphics[scale=0.4]{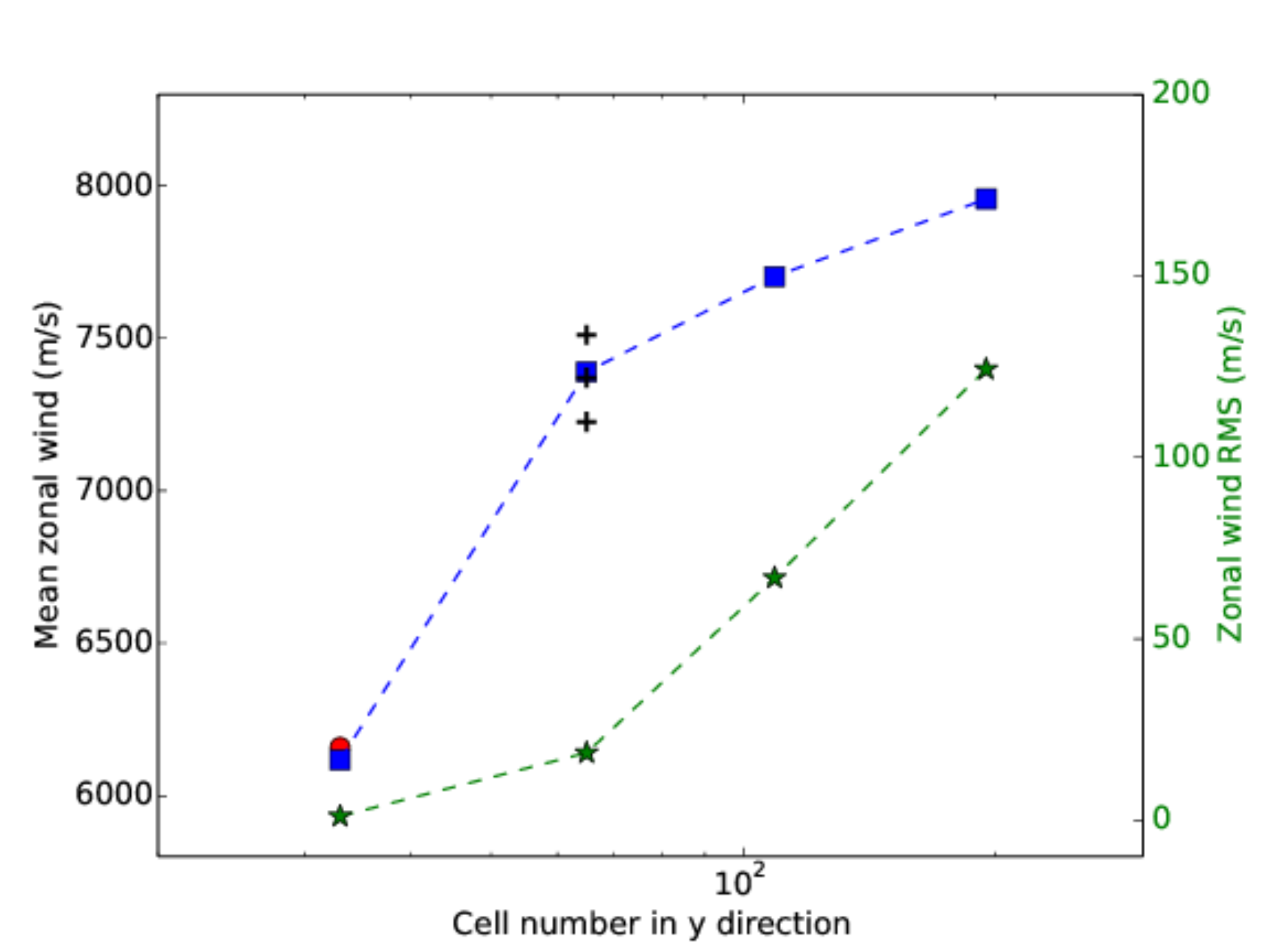}
\caption{Left axis: variations of the $50$ mbars zonally and time
  averaged zonal mean wind at the equator with $N_y$. The blue squares
  corresponds to the models 
  having a zonal resolution $N_x=128$, the red circle shows model
  LowRes and the black pluses shows the results of the models having
  a meridional resolution $N_y=65$ and varying number of cells in the
  zonal direction. Right axis (green stars): variations of the
  amplitude of the high frequency fluctuations of the zonally averaged
  zonal wind fluctuations (see text for details on its calculation)
  with $N_y$ for the series of models having $N_x=128$.}
\label{fig:meanwind_res}
\end{center}
\end{figure}

We next investigate the sensitivity of the properties described above
to variations in the horizontal spatial resolution. All of the other
parameters of the simulations are kept fixed. We vary $N_x$
from $64$ to $1024$ and $N_y$ from $33$ to $195$. The simulations
parameters and their outcomes are summarized in Table~\ref{tab:runs}.
All runs but one are integrated for $400$ days. Model HighRes, with 
a resolution of $(N_x,N_y)=(1024,195)$, is computationally
demanding. To reduce part of the burden, we restarted
model 1024x65 at $t=100$, multiplying the number of cells in the $y$
direction by a factor of three. Integrations of the equations was then
performed for another $350$ days.

\subsection{Effect on the climatology}

All models qualitatively display the same climatology as described
above. However, quantitative measures vary. As an example, we report
on Table~\ref{tab:runs} (forth column, see also
fig.~\ref{fig:meanwind_res}) the zonally and time averaged (between
$t=200$ and $t=400$) zonal wind at the equator at the pressure level
$P=50$ mbars. We note that the equatorial jet is still being slightly
accelerated over that period and has not yet  reached a perfect
equilibrium. For example, in model 128x65, the zonal wind at $50$
mbars increases from $7.1$ to $7.4$ km.s$^{-1}$ over that period
(which represents less than $5 \%$, a rather small acceleration that
justify the claim made above that the flow is close to a steady
state). Taken as a whole, this resolution study shows a strong
sensitivity to the meridional resolution $N_y$ (see the blue 
squares) but a very weak sensitivity to the zonal resolution $N_x$ (see
the symbols that correspond to $N_y=65$ and varying $N_x$, all
clustered at a mean zonal wind of about $7.2$--$7.3$
km.s$^{-1}$). This difference comes  
from the different spatial scales of the equatorial jet in the $x$ and
$y$ directions, namely a sharp meridional velocity gradient but a
large scale longitudinal structure. In model LowRes, grid cells have a
meridional extend of $dy=10^{7}$ meters which means that the jet is only 
resolved with a handful of grid points (see the jet meridional size in
figure~\ref{lowres_zonalMean_YP_fig}) and is severely affected by
numerical dissipation. Quite differently, the zonal variations of the
jet have a typical scale comparable to the zonal computational domain
itself (simply because thermal forcing is modulated on that scale) and
is easily resolved with a few tens of cells in all of our models.  

\subsection{Shocks}
\label{sec:shocks}

\begin{figure}[!t]
\begin{center}
\includegraphics[scale=0.32]{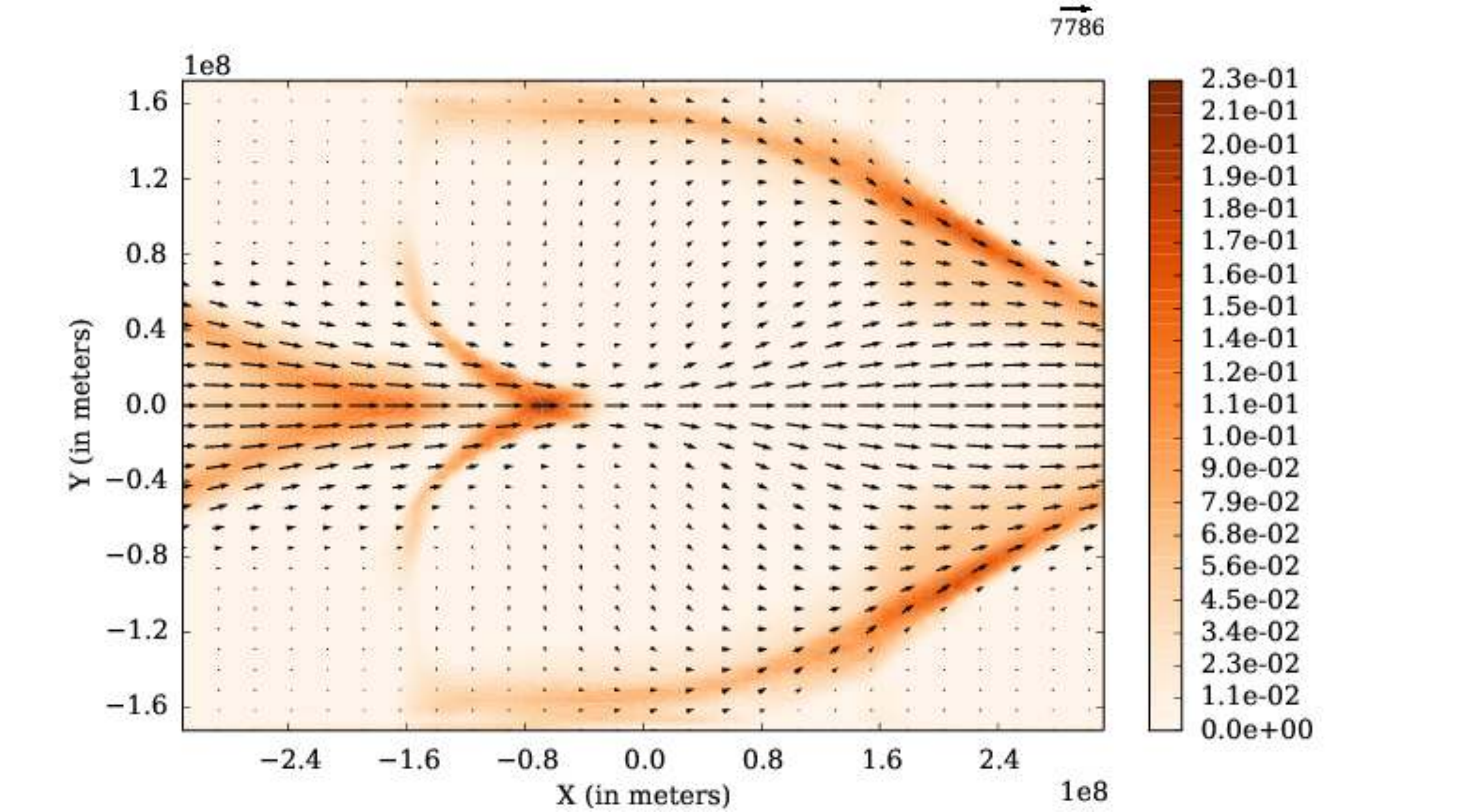}
\caption{Spatial distribution of $\eta$ at the pressure level $P=1.66$
  mbars in model 128x65 (color contours). The arrows show the
  amplitude and direction of the hozizontal wind.}
\label{fig:eta_128x65}
\end{center}
\end{figure}

As discussed in the introduction, it has been a long standing issue to
determine whether or not shocks exist in hot 
Jupiter atmospheres. In our simulations, we have not found any 
signatures of shocks that would be traced by a sudden decrease of the
local Mach number or a rapid increase of the density. As a more
quantitative diagnostic, we have computed a dimensionless measure of
the flow divergence, as done recently by \citet{zhuetal13}:
\begin{equation}
\eta=\textrm{max} \left( 0,-\frac{dx}{c_s}\del \bcdot \bb{v} \right) \, ,
\label{eq:eta_shocks}
\end{equation}
where $dx$ denotes the cell size in the radial direction. In shock
capturing scheme such as RAMSES, the numerical 
algorithm is designed to spread shocks over only a handful of cells,
so we would expect $\eta$ to reach at least a few tens of percent in
shocks and be {\it independent of spatial resolution}. In fact, 
\citet{zhuetal13} argue that $\eta>0.2$ is a good criterion for
detecting strong shocks in their simulation. Since shocks are
  spread over a few cells in finite volume codes such as RAMSES, such
  a threshold indeed corresponds to a velocity jump larger than the
  sound speed across a discontinuity (or, using the Rankine-Hugoniot
  conditions, to an upstream Mach number larger than $\sim 1.5$).

In general, we have found that $\eta$ only amounts to a few percents
at pressure larger than a few tens of mbars, and decreases
downward. For example, in model 128x65, its maximum value 
reaches $0.045$, $0.034$ and $0.022$ at pressure levels $P=50$ mbars,
$100$ mbars and $1$ bars, respectively. $\eta$ only takes significant
values at pressure of a few mbars, as illustrated on
figure~\ref{fig:eta_128x65}. At such a low pressure, three zones
feature values of $\eta$ in excess of $10 \%$. The first two are
symmetric around the equator and located on the jet wings, for
positive values of $x$. We note that they also manifest themselves by
an increased temperature (see figure~\ref{lowres_XY_fig}). It is
likely that these two compressive regions are an artefact of the
equatorial $\beta$--plane model that prevents an hemispheric Hadley cell
from developing around the planet. The last zone with a large $\eta$ is
located at $x \sim -0.8$ around the equator and corresponds to the region
where the zonal wind is strongly decelerated when entering the day
side of the planet. $\eta$ reaches $23 \%$ at that location in model
128x65. As emphasized by \citet{heng12}, this is the location where
shocks would most easily form. However, here, it appears that this
zone is only a region of large compression, but not a shock. Indeed,
we found that $\eta$ gradually decreases when the zonal resolution is
increased, reaching $12 \%$ in model 256x65 and 
only $6 \%$ in model 512x65. This gradual decrease shows that the
zonal wind decline when entering the planet day side is smooth and
converged upon increasing the resolution, as opposed to a shock that
would steepen and be caracterized by a constant, large $\eta$
value. We note that the two symetric high $\eta$ regions on
figure~\ref{fig:eta_128x65} show the same trend with resolution, which
means that these two zones are also regions of adiabatic compression. We
conclude from this analysis that the mean flow in the atmosphere of
hot Jupiters displays no shocks.

\subsection{Time variability}

We also computed an estimate of the high frequency rms fluctuations
$U^{\prime}_{50}$ of the zonally averaged equatorial zonal wind at
$50$ mbars (see last column on table~\ref{tab:runs} and green symbols on
figure~\ref{fig:meanwind_res} for the models with $N_x=128$). To do so, we
first subtracted the low frequency component of the wind by smoothing
the raw data with a Hamming function having a $20$ days
width. $U^{\prime}_{50}$ is then simply the root mean square of that
signal. While it is very small when $N_y=33$, we find that
$U^{\prime}_{50}$ increases steadily and amounts to more than $100$
m.s$^{-1}$ for model 128x195 and about twice that value for our
highest resolution model HighRes (not shown on
figure~\ref{fig:meanwind_res}). We now focus on this last model in
order to investigate the physical origin of that variability.

\section{High resolution model}
\label{sec:HR}

\begin{figure}[!t]
\begin{center}
\includegraphics[scale=0.42]{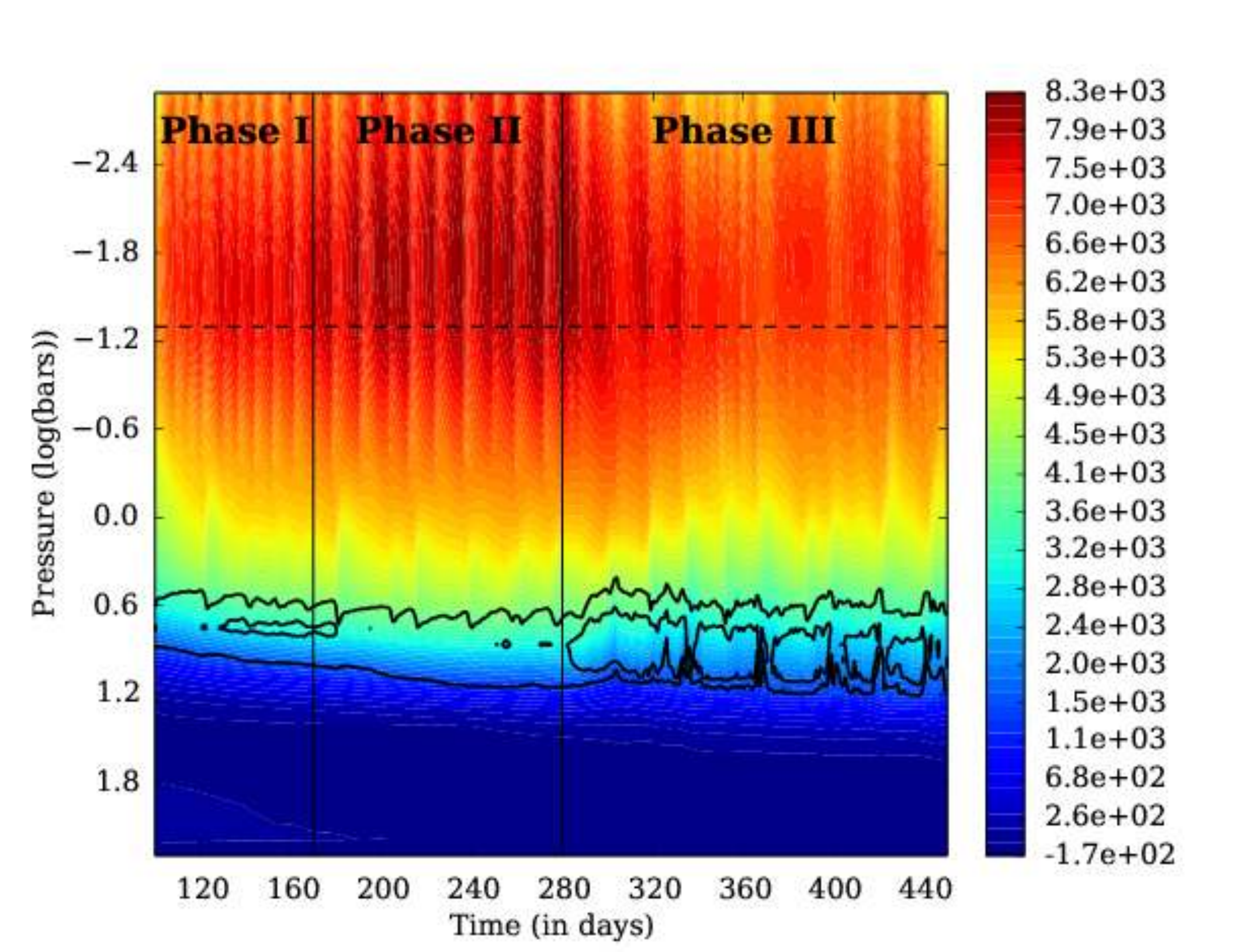}
\includegraphics[scale=0.43]{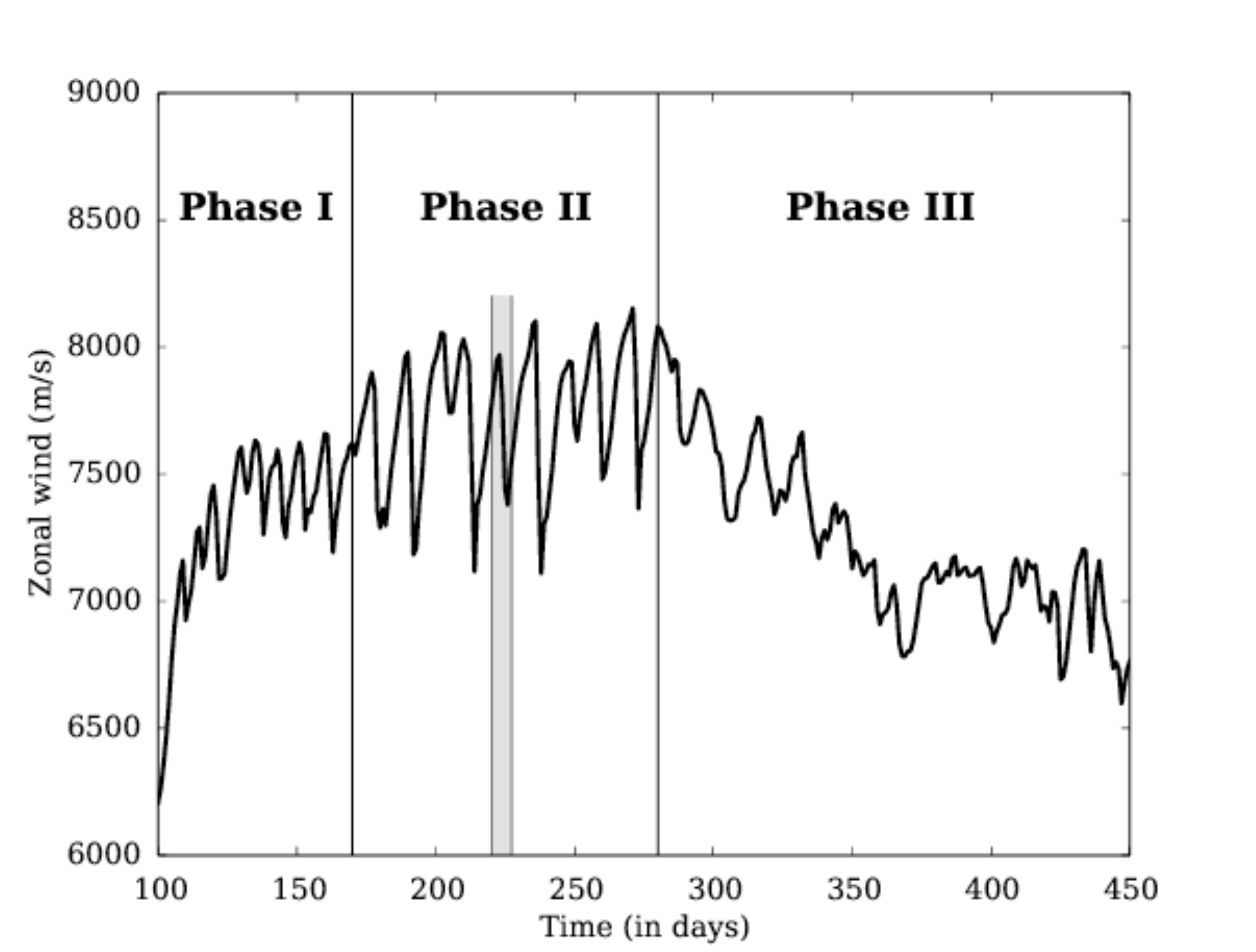}
\caption{Top: evolution of the vertical profile of the zonally averaged
  zonal wind at the equator in a time-pressure plane for model
  HighRes. The horizontal dashed line marks the $50$ mbars level. The
  solid lines are contours of the zonally averaged Richardson number
  (see text). The contours are for $0.1$ (interior contour) and
  $0.25$ (exterior contour). Bottom: time variations of the zonally
  averaged zonal wind at the equator at the pressure level $P=50$
  mbars for model HighRes (see text for a discussion of the three
  phases displayed on both panels).} 
\label{fig:timeEvolHIGHRES}
\end{center}
\end{figure}

The nature of the zonally averaged zonal wind variability in model
HighRes is best illustrated by showing the time variation of the
vertical profile of the zonal wind at the equator $\overline{U_{eq}}$ 
(figure~\ref{fig:timeEvolHIGHRES}, top panel). The difference with
figure~\ref{lowres_spacetimeP_vx_fig} is striking: $\overline{U_{eq}}$
shows significant variations during the entire duration of the
simulation. These variations takes the form of quasi--periodic
oscillations of the zonal wind, with a period of about $10$ days, 
during the first half of the simulation ($t<300$) and are less
regular, but still significant, at later times. The zonal wind
variations are in phase from the top of the planet atmosphere down to
the inert layer below $10$ bars, which is most likely a result of the
fast communication timescale across the atmosphere provided by sound
waves. The time history of $\overline{U_{eq}}$ at $P=50$ mbars is
shown on figure~\ref{fig:timeEvolHIGHRES} (bottom panel). In fact, we
can distinguish three phases of the zonal wind evolution (indicated
with labels on the upper part of both panels): ``Phase I'' ($t=100$ to
$t=170$) is a 
spin--up phase during which the flow adjusts to the sudden change in
meridional resolution. In agreement with the results of
section~\ref{sec:res_study}, the zonal wind rapidly increases by about
$1$ km.s$^{-1}$ over a period of just a few days. We call ``Phase II''
the period that extends between $t=170$ and $t=270$. The flow appears
to have reach a ``quasi--steady state'' over that period (see also the
top panel of the same figure that suggests that this is the case at
all levels above a few bars) and the zonal wind
displays quasi--periodic oscillations around a well--defined mean
value. At $50$ mbars, the zonally averaged zonal wind
fluctuations are significant and amount to almost $1$ km.s$^{-1}$
between its largest and smallest values. We investigate in detail the
flow properties during ``Phase II'' in section~\ref{sec:phaseII}. At
time $t \sim 270$, this quasi--periodic evolution of the wind seems to come
to an end and the mean zonal wind decreases. This is what we call
``Phase III'' of the flow evolution. As shown by the solid contours on
figure~\ref{fig:timeEvolHIGHRES} (top panel), the beginning of that
phase coincides with a significant decrease of the Richardson number
$Ri$. The latter is defined by the relation
\begin{equation}
Ri=\frac{N^2}{(\partial U/\partial z)^2} \, ,
\label{eq:ri_number}
\end{equation}
where $N^2$ stands for the Brunt--Vaisala frequency and is calculated
according to
\begin{equation}
N^2=\frac{\rho g^2}{P} \left[ \left( \frac{d\ln
    T}{d\ln P}\right)_{S}- \left(\frac{d\ln
    T}{d\ln P}\right) \right] \, ,
\end{equation}
is which the first term inside the square bracket denotes the
isentropic gradient. As discussed by \citet{showman&guillot02} in the context
of hot Jupiter atmospheres and later investigated in more details by
\citet{li&goodman10}, sheared flow in stratified atmospheres are 
prone to a vertical shear instability when $Ri$ is smaller than
a threshold value that depends on the gas thermodynamics but is
of order $1/4$. In our simulation, the coincidence between the end of
Phase II and the appearance of zonally averaged Richardson numbers
smaller than $0.1$ suggests that such an instability develops at that
time and starts perturbing the flow. In section~\ref{sec:longterm}, we
will investigate ``Phase III'' in more details and give compelling
evidences that this is indeed the case.

\subsection{A barotropic Kelvin-Helmholtz instability (phase II)}
\label{sec:phaseII}

\begin{figure}[!t]
\begin{center}
\includegraphics[scale=0.33]{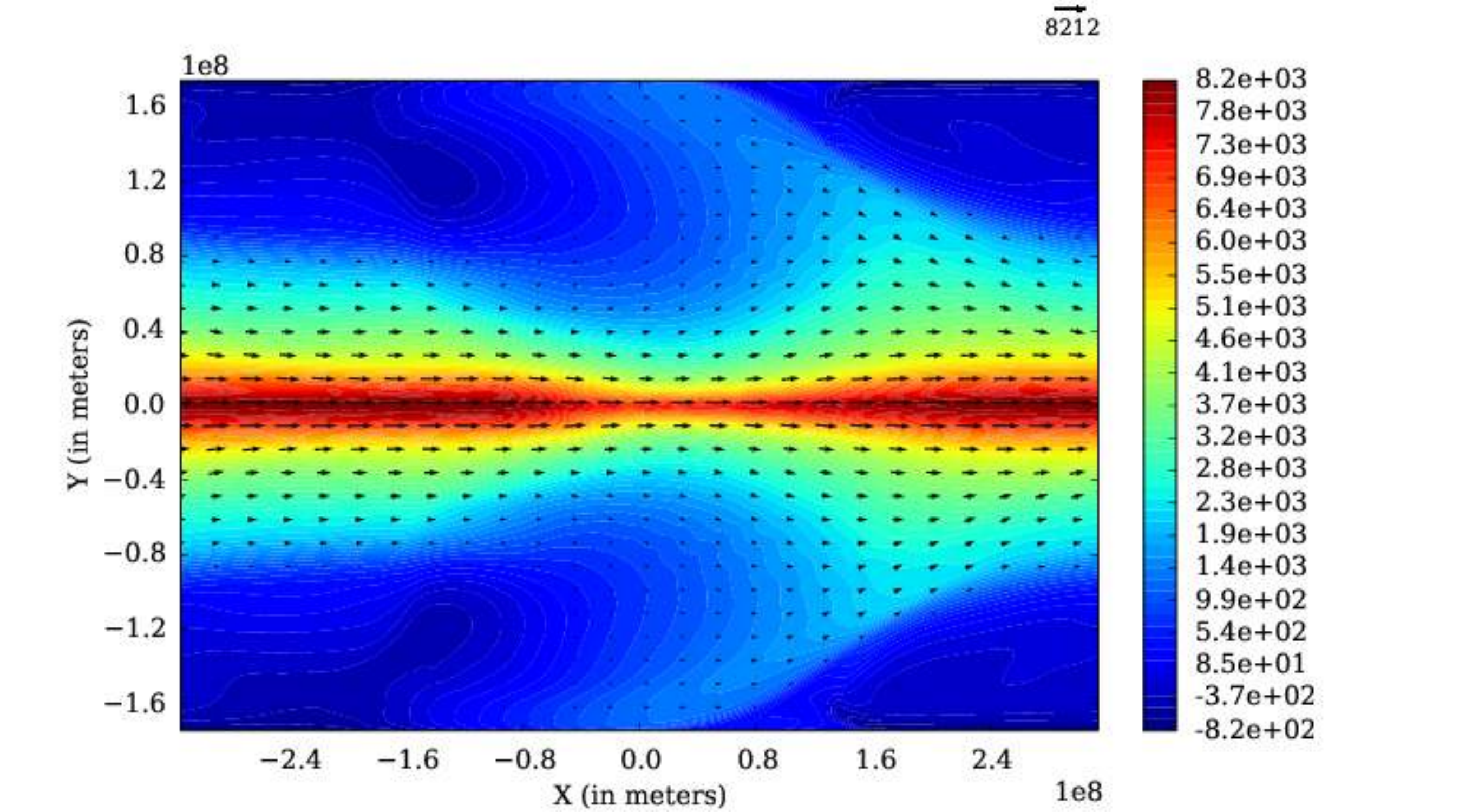}
\includegraphics[scale=0.33]{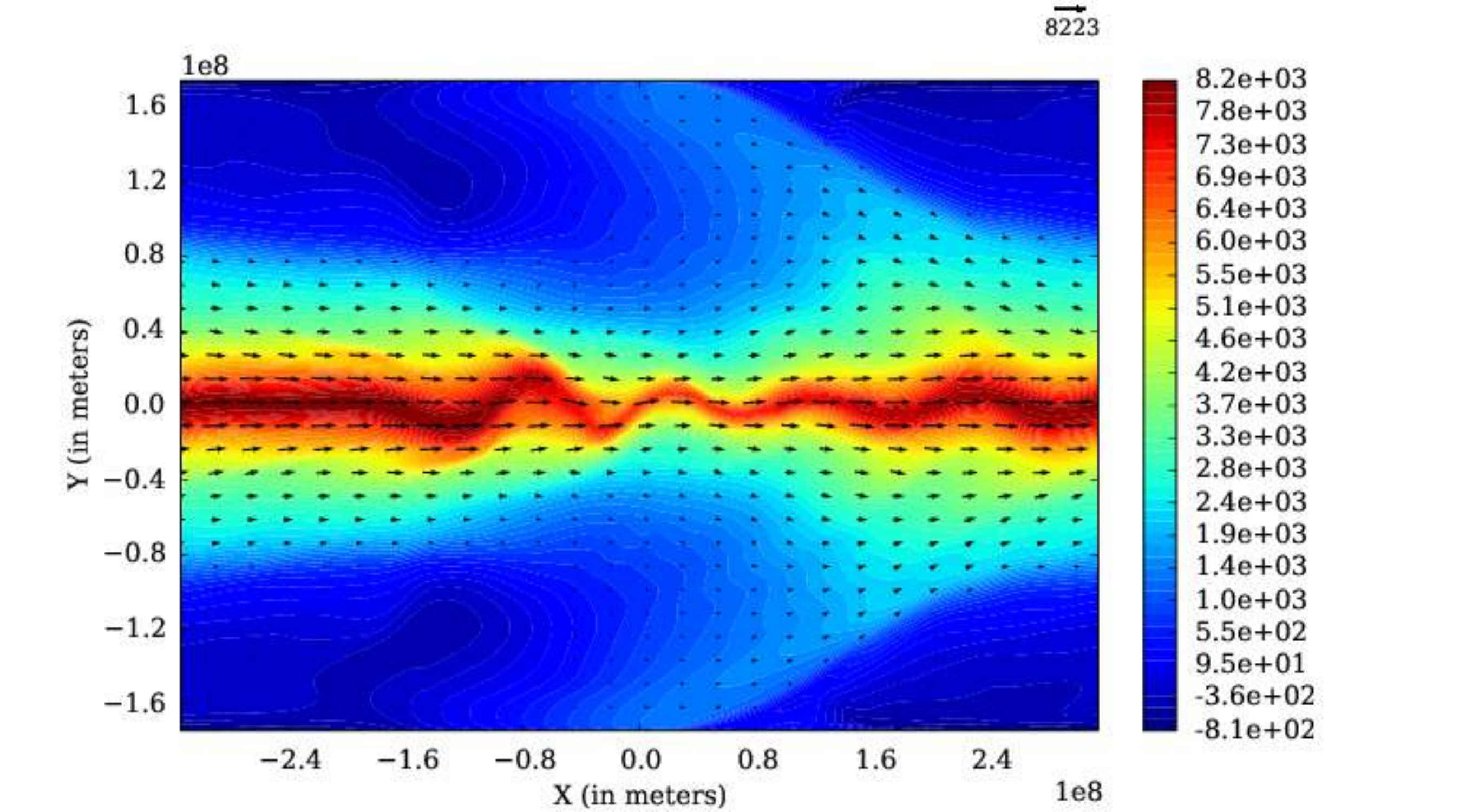}
\caption{Color contours showing the horizontal distribution of
  the zonal wind at time $t=208$ ({\it top panel}) and $t=213$ ({\it
    bottom panel}) in model HighRes at $P=10$ mbars. The 
  arrows display the horizontal wind vectors. Note the clear meanders
  of the equatorial jet in the bottom panel as opposed to the more
  zonal structure of the equatorial jet in the top panel.}
\label{fig:jet_phaseII}
\end{center}
\end{figure}

\begin{figure}[!t]
\begin{center}
\includegraphics[scale=0.45]{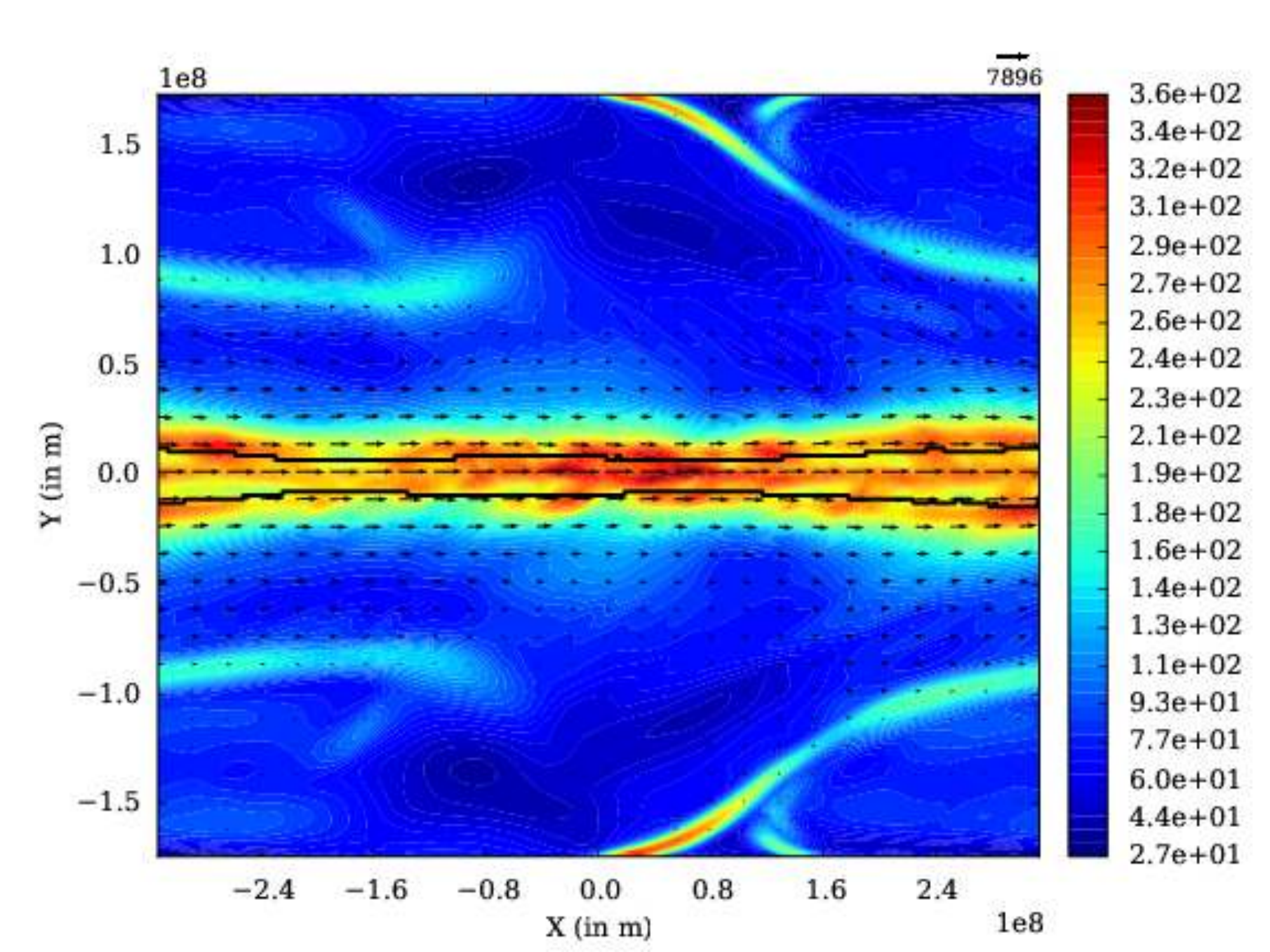}
\includegraphics[scale=0.26]{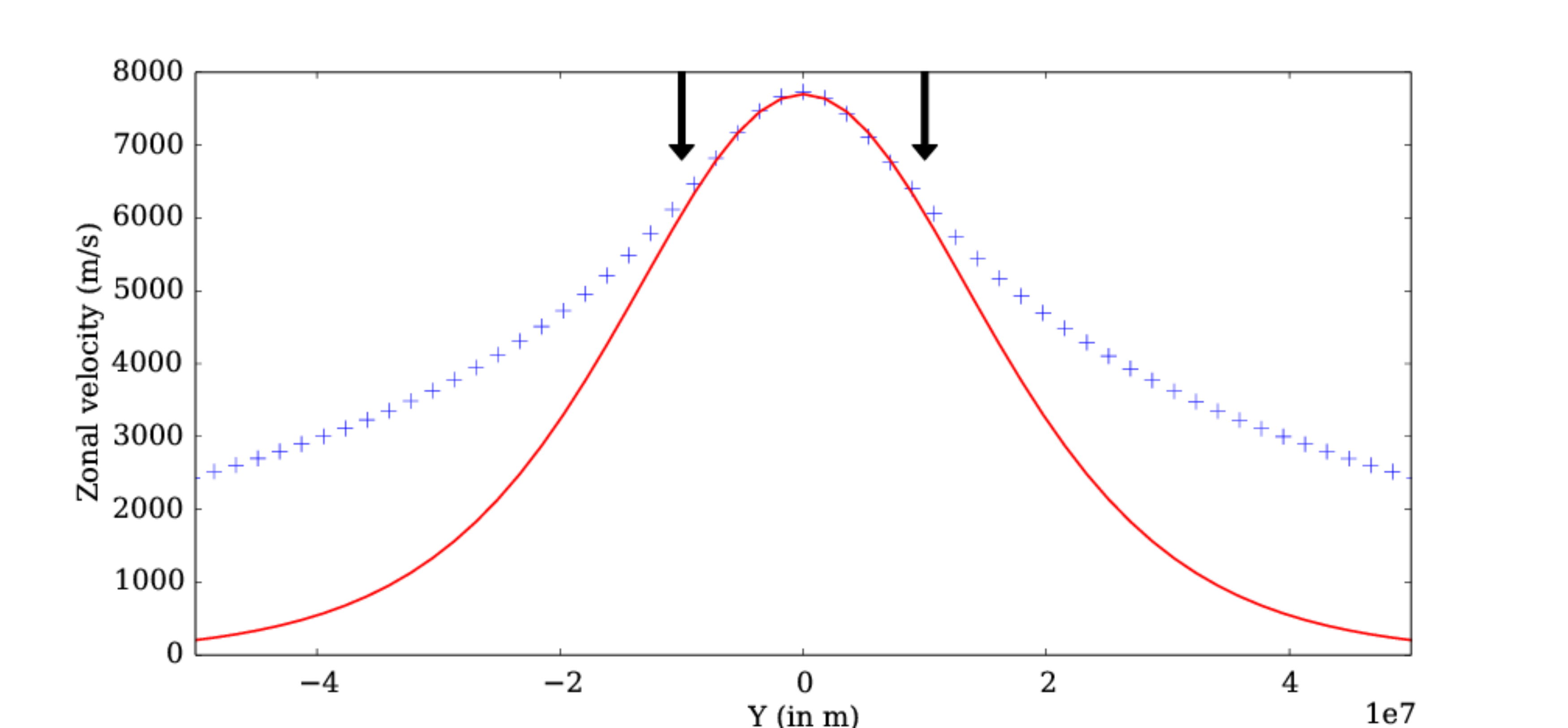}
\caption{Top: color contours show the rms of the eddy specific kinetic energy
  ($u^{\prime 2}+v^{\prime 2}$) spatial distribution in the horizontal
  plane at pressure level $P=50$ mbars for model HighRes (during phase
  II of the evolution) while arrows display the steady component of
  the wind over the same time interval. The thick solid line plots the
locus of the points where the meridional gradient of total vorticity
(planet+flow) vanishes. Bottom: y--profile of the time-averaged zonal
velocity (during Phase II) at the vicinity of the equator and at
pressure level $P=50$ mbars for model HighRes (blue crosses) compared
with a Bickney jet profile (red line -- see text for details). The two
vertical arrows mark the location of vanishing meridional gradient of
total vorticity.}
\label{fig:EKEandPV}
\end{center}
\end{figure}

\subsubsection{Equatorial jet variability} 

During ``phase II'' of model HighRes, the equatorial jet periodically 
alternates between two states, during which it is either almost
perfectly zonal in the vicinity of the equator or display meanders
(figure~\ref{fig:jet_phaseII}). The jet velocity increases when the
jet is zonal and slows down when the meanders amplitude saturates and
decreases. To make further progress, we next decomposed the horizontal
velocity as the sum of a time averaged component and a fluctuating part: 
\begin{equation}
\bb{v}(x,y,t)=\bb{\overline{v}}(x,y)+\bb{v'}(x,y,t) \, ,
\end{equation}
where the time averaged is performed using $100$ dumps evenly spaced
between $t=170$ and $t=270$. In the following, we will use the
notations $\overline{u}$ (resp. $\overline{v}$) and $u'$ (resp. $v'$)
to denote the $x$ (resp. $y$) component of these velocities. Using this
decomposition, we define the specific kinetic energy of the
fluctuations as
\begin{equation}
E_K=u^{\prime 2}+v^{\prime 2} \, , 
\end{equation}
and the meridional gradient of the total vorticity of the flow
(including the planetary scale vorticity associated with $\Omega_p$):
\begin{equation}
\xi_y=\frac{\partial \xi}{\partial y}= \beta - \frac{\partial^2
  \overline{u}}{\partial y^2}  \, .
\end{equation}
Taking advantage of the barotropic nature of the flow
(figure~\ref{fig:timeEvolHIGHRES}, top panel), we now focus on its
structure at the $50$ mbars pressure level until the end of this
section.

Figure~\ref{fig:EKEandPV} (top panel) shows that the time averaged
value of the rms of $E_K$ over ``Phase II'' reaches values of the
order of a few hundred m.s$^{-1}$ at $50$ mbars in the jet core,
interior to the region where $\xi_y$ vanishes, and drops rapidly as
one moves outward of that region. It is a well--known result of
hydrodynamics that the presence of two extrema in the flow vorticity
profile (such as shown on figure~\ref{fig:EKEandPV}) is a necessary
condition to destabilize a two--dimensional incompressible flow
\citep[see for example][]{vallis06}. Here, the flow is
compressible, stratified and three dimensional, and such a simple
criterion does not apply, strictly speaking. Nevertheless, we can still
expect that it remains a good guide for linear stability because the
deviations from both compressibility and two--dimensionality are
small. The rapid drop of $E_K$ outside of the two locations where
$\xi_y$ vanishes also argues in favor of this interpretation and
strongly suggest that the jet is subject to a Kelvin-Helmholtz
instability that originates from the velocity meridional gradient and
drives the oscillations seen during Phase II. 

\begin{figure}[!t]
\begin{center}
\includegraphics[scale=0.45]{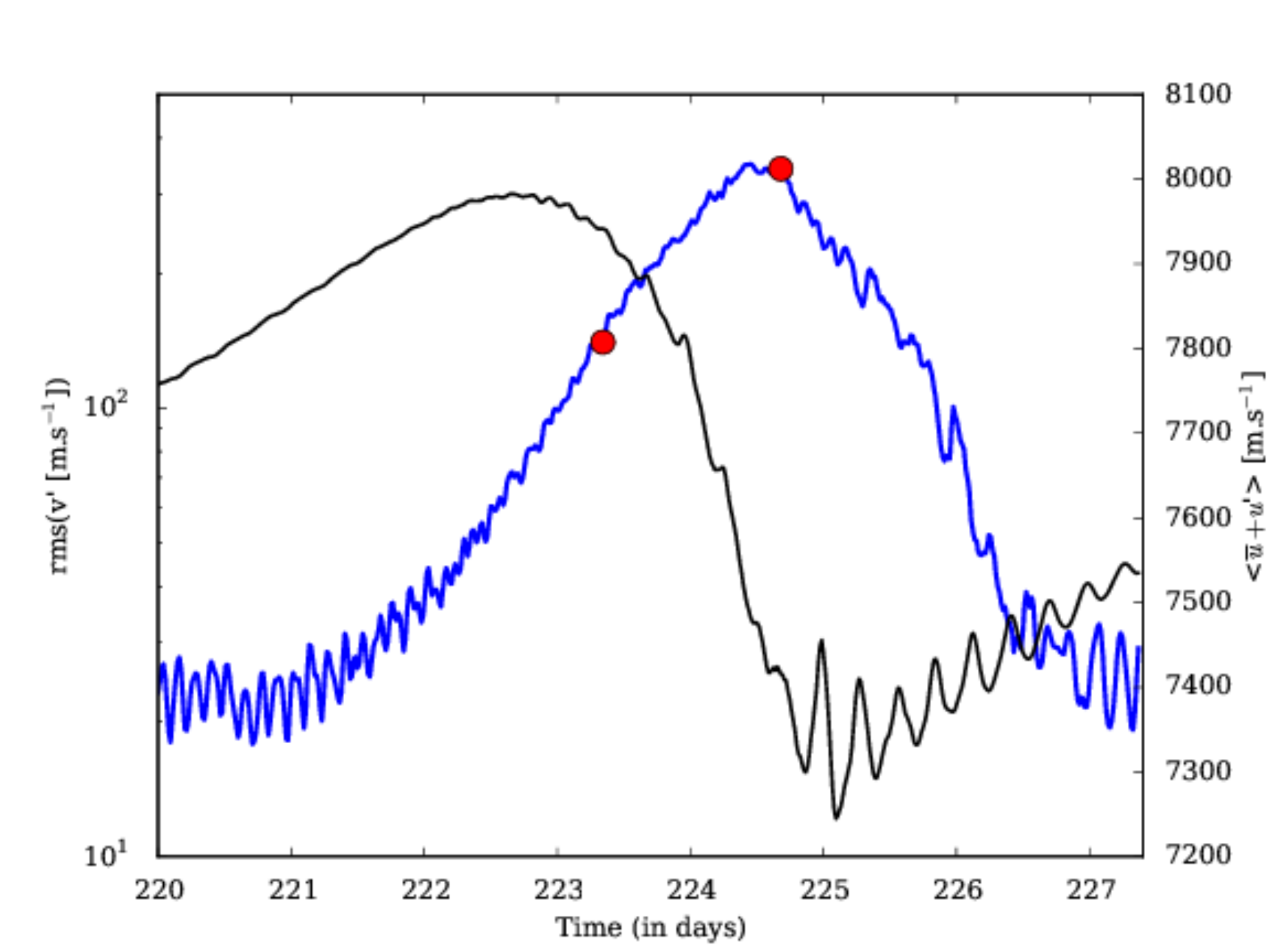}
\caption{Time evolution of the root mean square of the meridional
  velocity fluctuations $v'$ (blue curve -- see text for
  details) and mean jet velocity (black curve) at the equator. The red 
  circles indicate the different time at which the jet structure is
  plotted on figure~\ref{fig:jetandfluc}.}   
\label{fig:timehist_vprime}
\end{center}
\end{figure}

\begin{figure*}[t]
\begin{center}
\includegraphics[scale=0.33]{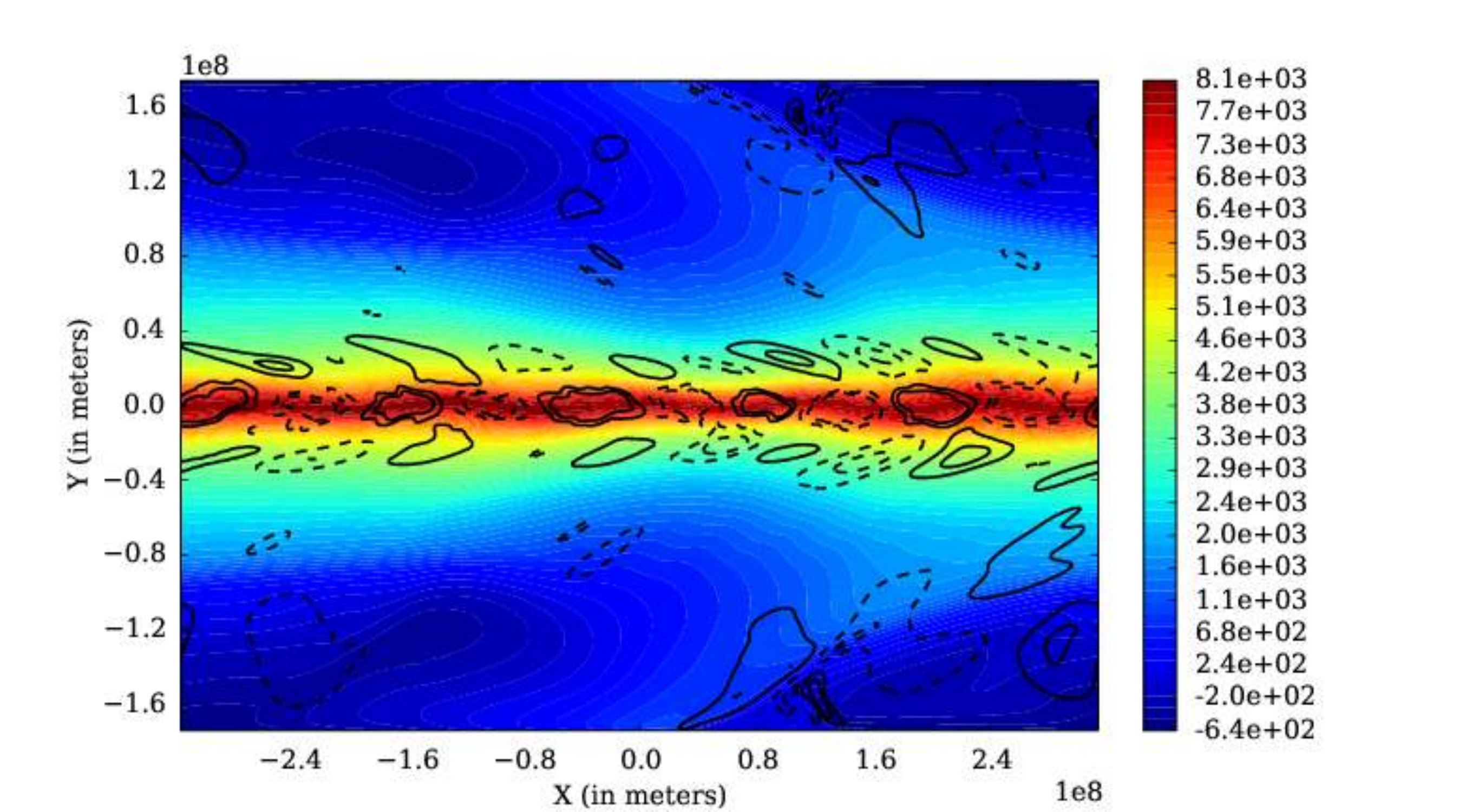}
\includegraphics[scale=0.33]{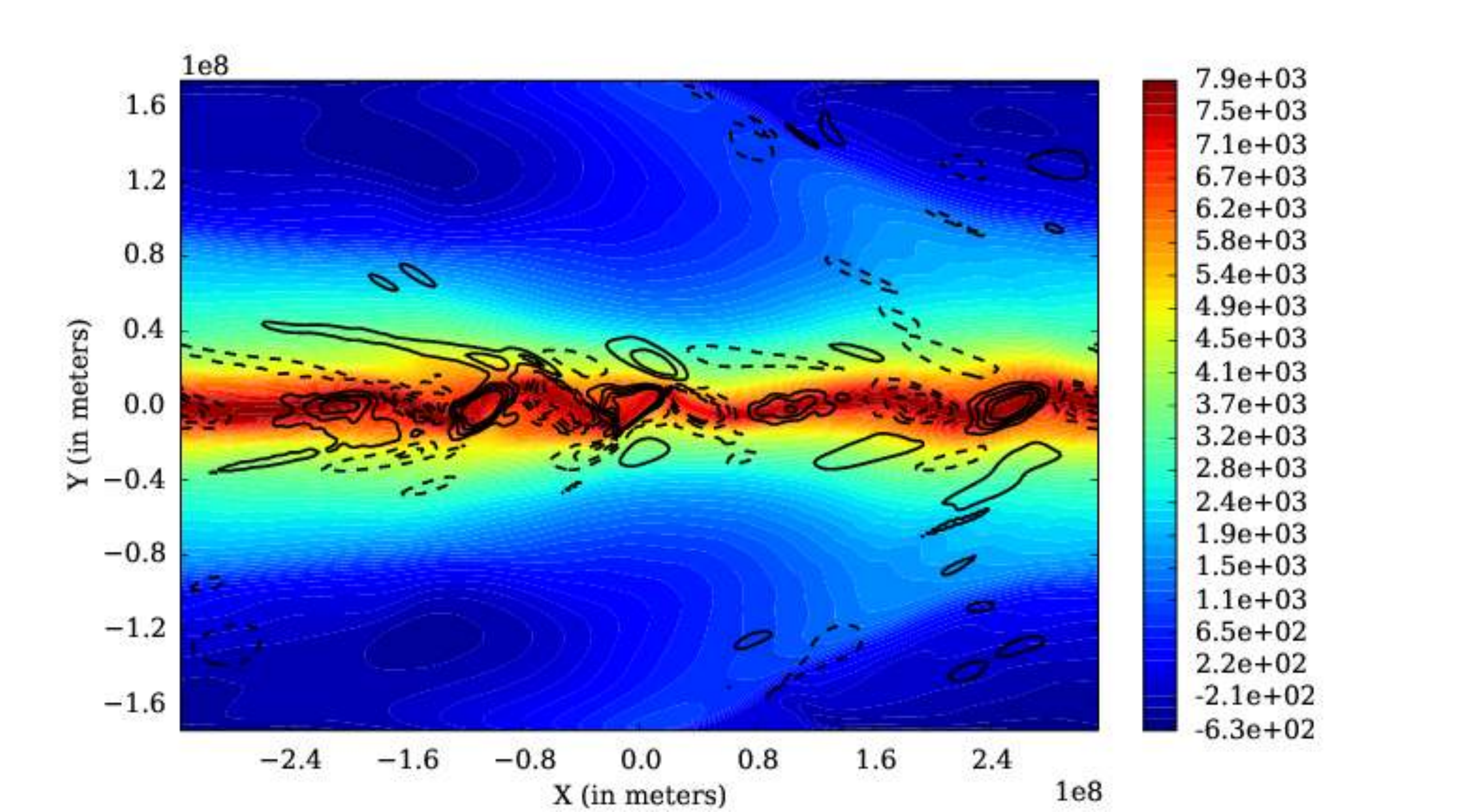}
\caption{Color contours showing the horizontal spatial distribution of
  the zonal wind at $P=50$ mbars at times indicated with red dots on
  figure~\ref{fig:timehist_vprime}, namely $t=223.3$ ({\it left
    panel}) and $t=224.7$ ({\it right panel}). Black contours plots
  the meridional velocity fluctuations $v$'. Contours are shown every $50$
  m.s$^{-1}$ from $-100$ to $100$ m.s$^{-1}$ on the left hand side panel and
  every $100$ m.s$^{-1}$ from $-400$ to $400$ m.s$^{-1}$ on the right hand
  side panel. On both panels, positive (resp. negative) contours are
  shown with solid (resp. dashed) lines and the zero contour is
  omitted.}   
\label{fig:jetandfluc}
\end{center}
\end{figure*}

\begin{figure*}[t]
\begin{center}
\includegraphics[scale=0.45]{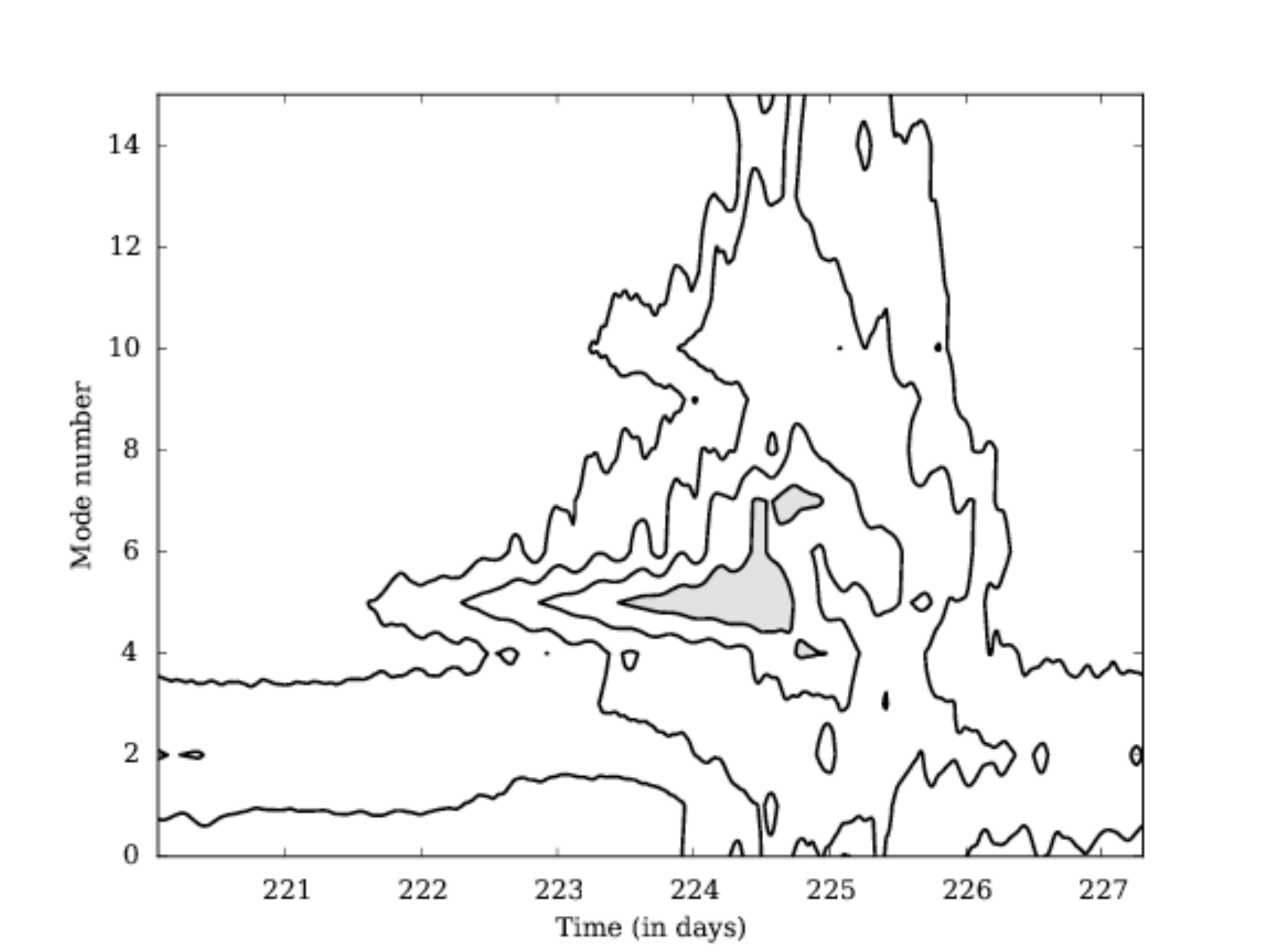}
\includegraphics[scale=0.45]{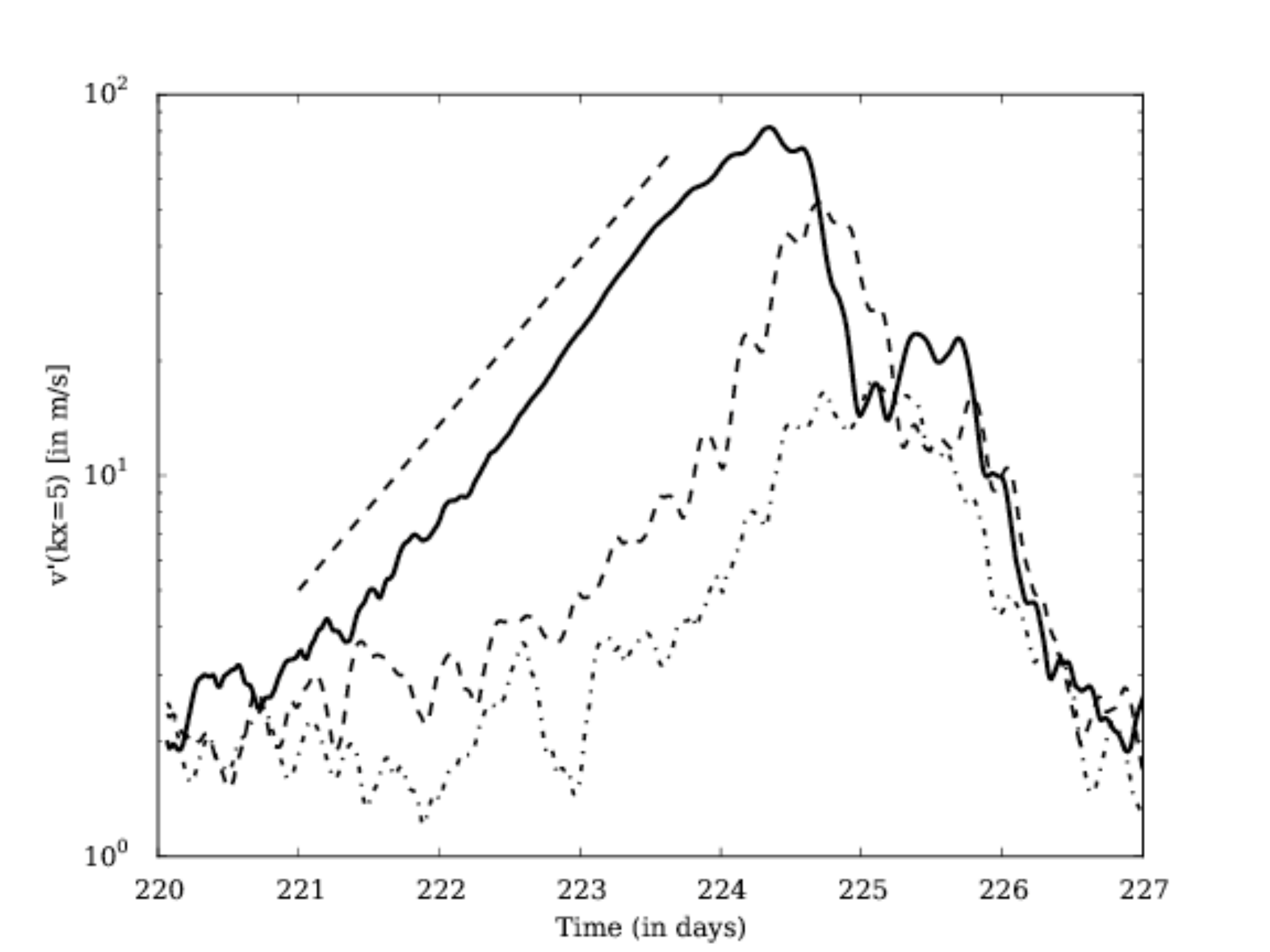}
\caption{Left: Amplitude of the meridional
  velocity fluctuations Fourier transform $\tilde{v'}(k_x,t)$ in the
  ($k_x$-time) plane for model HighRes, spatially averaged over the
  region $|y|<5 \times 10^7$ meters. Contours are for
  $log(\tilde{v'}_{max}/2)$, $log(\tilde{v'}_{max}/4)$, 
  $log(\tilde{v'}_{max}/8)$ and $log(\tilde{v'}_{max}/16)$. The region
  surrounding the maximum is filled in gray. Right: Same as
  the left panel, but showing the time evolution of particular modes,
  namely: $k_x=5$ ({\it thick solid line}), $k_x=7$ ({\it dashed line})
  and $k_x=9$ ({\it dotted--dashed line}). The straight dashed line
  represents an exponential growth with a timescale of $1$ day.}
\label{fig:fftanalysis_vprime}
\end{center}
\end{figure*}

\subsubsection{Linear instability properties} 

To investigate that possibility in more details, we next focus on a 
single oscillation of the jet. It is highlighted as the 
region shaded in light gray in figure~\ref{fig:timeEvolHIGHRES} (bottom
panel). In practice, the analysis that follows was conducted by restarting
model HighRes at $t=220$ for about $7$ days, saving the simulation
data every $4000$ seconds in order to sample the fluid evolution at
high frequency. Figure~\ref{fig:timehist_vprime} shows that the rms 
fluctuations of $v'$ at the equator (solid blue curve), initially of
the order of $20$ m.s$^{-1}$, grows during 
about $3$ days until they reach an amplitude of $\sim 350$ m.s$^{-1}$
after which they decay in roughly one day. The spatial structure of
these fluctuations is further illustrated on
figure~\ref{fig:jetandfluc} using two snapshots that illustrate the
growing (left panel) and maximum (right panel) phases of the
evolution. During the growing part (here plotted at $t=223.3$), $v'$
clearly displays a regular and oscillating pattern at the equator, with
a well--defined spatial period of $1/5$th of the domain size,
suggestive of a linear modal growth. The background jet structure is
only weakly modified 
during that phase. Note that the jet velocity reaches its maximum at
that point (black curve on figure~\ref{fig:timehist_vprime}). By time
$t=224.7$, the velocity fluctuations have grown significantly and show
clear signs of saturation, as evidenced by positive and negative contours
colliding with each other. Before looking at this saturation phase in
more details in the next paragraph, we first quantify the properties
of the fastest growing mode of the instability by performing a spectral
analysis of the flow properties. This was done by computing the
amplitude of the Fourier coefficients of $v'(x,y,t)$ in 
the zonal direction. We then spatially averaged those coefficients in the
vicinity of the planet equator and show their variations as a function
of the normalized zonal wavenumber $k_x$ and time on
figure~\ref{fig:fftanalysis_vprime} (left panel)\footnote{The zonal
wavenumber is defined such that $k_x=1$ corresponds to a wavelength
equal to the size of the domain in the x--direction.}. From $t=222$ to
$t=224$, the flow evolution is dominated by the $k_x=5$
mode. Consistent with the idea of a linear instability, that mode
grows exponentially with a typical 
timescale of about $1$ day (fig.~\ref{fig:fftanalysis_vprime}, right
panel). It reaches its maximum amplitude at $t \sim 224$, shortly
after higher $k_x$ modes amplitudes also start to grow, presumably as
a result of nonlinear interactions with the $k_x=5$ mode. This later
growth of higher $k_x$ modes is illustrated on the right panel of
figure~\ref{fig:fftanalysis_vprime} for the particular cases of the
modes $k_x=7$ and $k_x=9$. At $t=225$, a large range of spatial scales
, with wavenumber $k_x$ up to $15$ (see left panel of
figure~\ref{fig:fftanalysis_vprime}), have reached a significant 
amplitude. During the entire duration of this
additional simulation, we also find that larger scale modes (with $k_x$
ranging from $1$ to $3$) have significant amplitudes. Their evolution,
however, is quite different from the modes with $k_x \geq 5$. They
do not display any signature of an exponential growth and their
amplitude is modulated by no more than a factor of two during the
simulation. It is likely that these properties reflect the nonlinear
feedback of the instability on the large scale structure of the
equatorial jet. 

To summarize, we have determined the wavenumber of the most unstable
mode of the instability ($k_x=5$) and its growth rate ($\sigma=1$
day$^{-1}=3.5 \times 10^{-6}$ s$^{-1}$). In principle, these
properties (as well as the mode phase velocity, see below) could be
compared to the result of a numerical linear instability
analysis. However, for a 3D compressible flow, such an analysis is
tedious and beyond the scope of this paper. Here, we only provide 
a very crude estimate of these properties based on 2D incompressible 
flows. A useful and well studied example of such flows is the
so--called Bickney jet, for which the jet profile is given by 
\begin{equation}
U(y)=U_B \sech^2 \left( \frac{y}{L_{B}} \right) \, ,
\end{equation}
where the parameters $U_B$ and $L_{B}$ caraterize the jet. The
Bickney jet is known to be unstable to the Kelvin-Helmoltz instability
and the most unstable wavelength satisfies the relation $k_x L_{jet} \sim
1$ \citep{drazin&reid81}. The core of the jet in our simulations
(i.e. within the region where the vorticity is maximum) can be well
fitted by the Bickney jet meridional profile (see
figure~\ref{fig:EKEandPV}, bottom panel) for
which $L_{jet}=2 \times 10^7$ meters. Given the size of our
computational box and the above scaling, this is consistent with the
instability having a most unstable wavenumber $k_x=5$, as seen in the
simulations. However, in this case, the predicted growth rate
$\sigma_{th}$ is of the order of $\epsilon k_x U_B$, with $\epsilon
\sim 0.15$. This translates into a growth rate that is more than an
order of magnitude larger than measured in the simulation. This large
disagreement illustrates the limit of a naive 2D reasoning and
suggests that three--dimensional effects, compressibility, or the
finite resolution of our simulations likely affect the flow. More work
is needed to clarify the properties of the instability, and, more
importantly, the conditions under which it develops.

\subsubsection{Nonlinear saturation (due to shocks?)} 

\begin{figure}[!t]
\begin{center}
\includegraphics[scale=0.33]{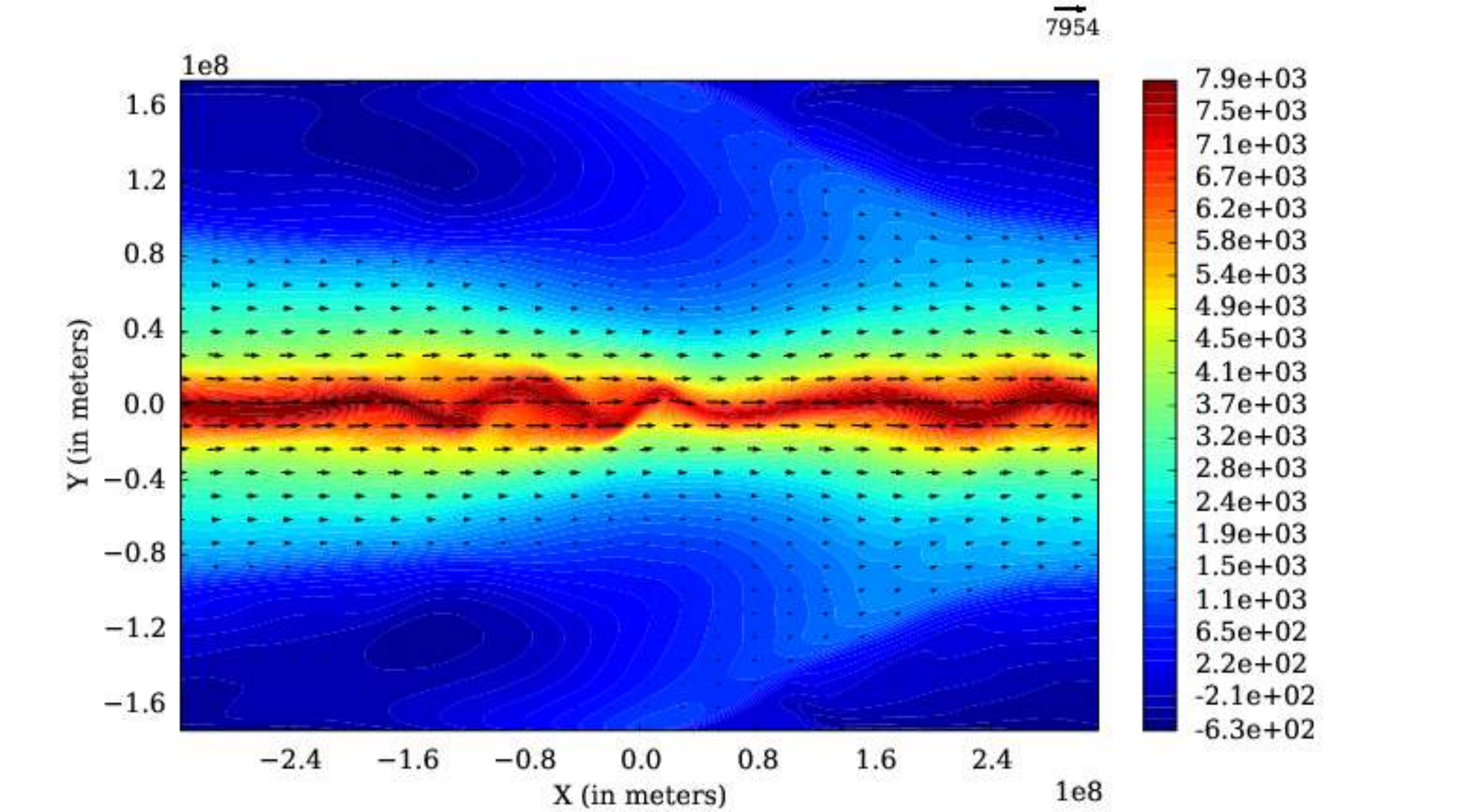}
\includegraphics[scale=0.33]{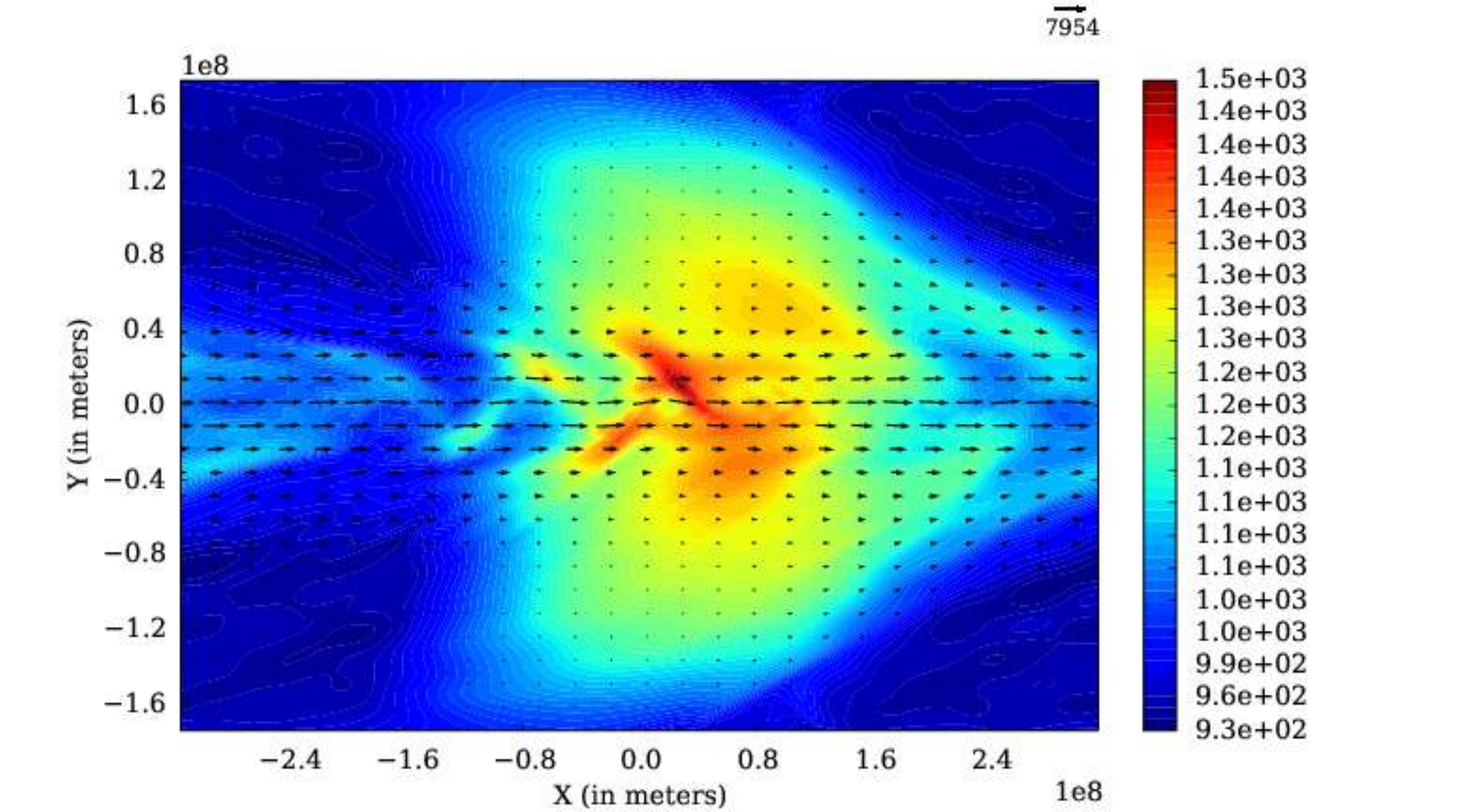}
\caption{Zonal velocity ({\it top panel}) and temperature ({\it bottom
    panel}) at time $t=224.7$ in model HighRes (i.e. when velocity
  fluctuations reach their maximum value) at pressure level
  $P=50$ mbars. Jet meanders upstream of the substellar point, located
at $(x,y)=(0,0)$, are clearly associated with temperature fluctuations.} 
\label{fig:vx_T_saturation}
\end{center}
\end{figure}

As mentioned
above, the background jet becomes significantly distorted when the
amplitude of the perturbations reaches large amplitudes (see right
panel of figure~\ref{fig:jetandfluc}). The nonlinear interaction that
results quickly damps the flow velocity fluctuations, the jet slows
down (black curve on figure~\ref{fig:timehist_vprime}) and return to a
more zonal structure such as shown on figure~\ref{fig:jet_phaseII}
(top panel). We have found temperature fluctuations of a few hundred
Kelvin during that phase (figure~\ref{fig:vx_T_saturation}, bottom 
panel). They tend to be associated with the region of the jet that
displays the largest meanders (figure~\ref{fig:vx_T_saturation}, top
panel). In addition, sharp features in the zonal velocity (such as
seen at the location $x\sim-0.2\times 10^8$ and $y\sim-0.1\times 10^8$
for example) are ubiquitous in snapshots of the flow during the
nonlinear stage of the instability.

For these reasons, and because of the supersonic nature of the jet
velocity, it is natural to ask whether saturation occurs via
shocks. Detecting shocks in 3D non--steady flows is known to be
extremely difficult to achieve in a systematic manner. Here, we only
give a couple of simple arguments that suggest that shocks play no
role in the instability saturation. First, as in
section~\ref{sec:shocks}, we computed the distribution of $\eta$ given
by Eq.~(\ref{eq:eta_shocks}). We again found that the maximum value of
$\eta$ never exceeds a few percents at $50$ mbars (and, in fact, at
all pressure levels). The sharp gradients mentioned above in the
zonal velocity are in fact compensated by regions with significant
vertical downward flow. This suggests that the nonlinear
saturation of the instability is not associated with large compressive
events.

\begin{figure}[!t]
\begin{center}
\includegraphics[scale=0.38]{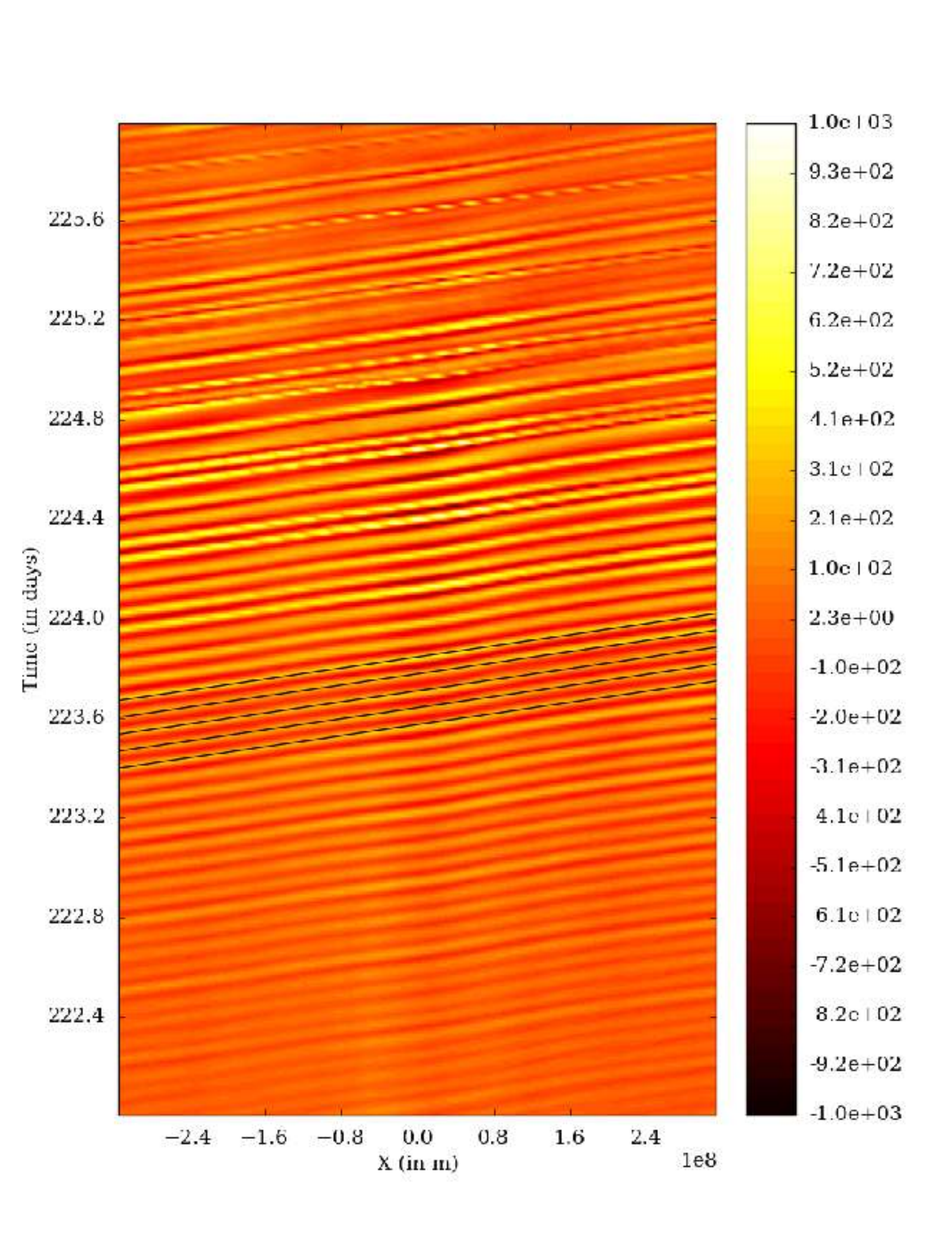}
\caption{Hovm\"oller space--time diagram of the meridional velocity
fluctuations $v'$ at the equator in model HighRes at pressure level
$P=50$ mbars. The thin black lines have a $6$ km.s$^{-1}$ slope.} 
\label{fig:hovmoller}
\end{center}
\end{figure}

A second argument is that, in the frame where the shock would be
stationary, the flow is actually just about sonic. This can be seen
by plotting the equatorial meridional velocity fluctuations in a
spacetime diagram in the $(x,t)$ plane, also known as an Hovm\"oller
plot (see figure~\ref{fig:hovmoller}). It shows that the eastward
velocity of the growing wave pattern amounts to about $6$ km.s$^{-1}$
(see solid black line)\footnote{This velocity also 
corresponds to the phase velocity of the linear instability discussed
above}. Because the flow velocity in the frame rotating with the
planet reaches at most 8 km.s$^{-1}$ (see upper panel of
figure~\ref{fig:vx_T_saturation}), this means that the zonal wind
velocity upstream of a putative shock would be smaller than $2$
km.s$^{-1}$ in the frame in which the flow structures are
stationary. This is comparable to (and in fact, slightly smaller than)
the sound speed, so that the flow Mach number is at best of order 
unity. If shocks exist, they are only weak shocks, with upstream Mach
number of order $1$ at best. More work is needed to characterize the
dynamical mechanism responsible for the saturation and to investigate
whether that result holds across the entire parameter space, but the
present simulation seems to rule out the presence of shocks, at least
for the set of parameters we considered.

\subsection{A vertical shear instability (phase III)}
\label{sec:longterm}

\begin{figure*}[!t]
\begin{center}
\includegraphics[scale=0.45]{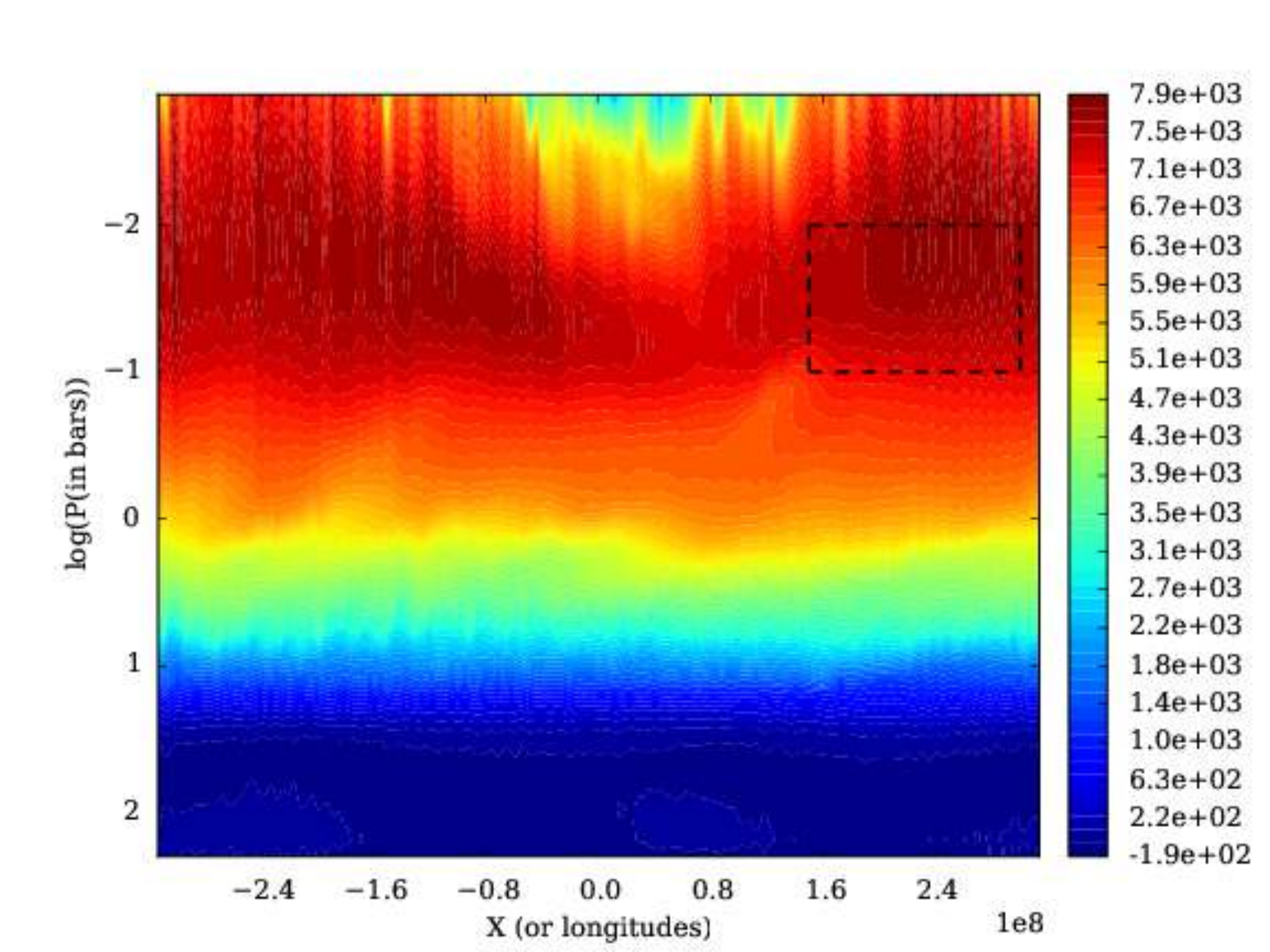}
\includegraphics[scale=0.45]{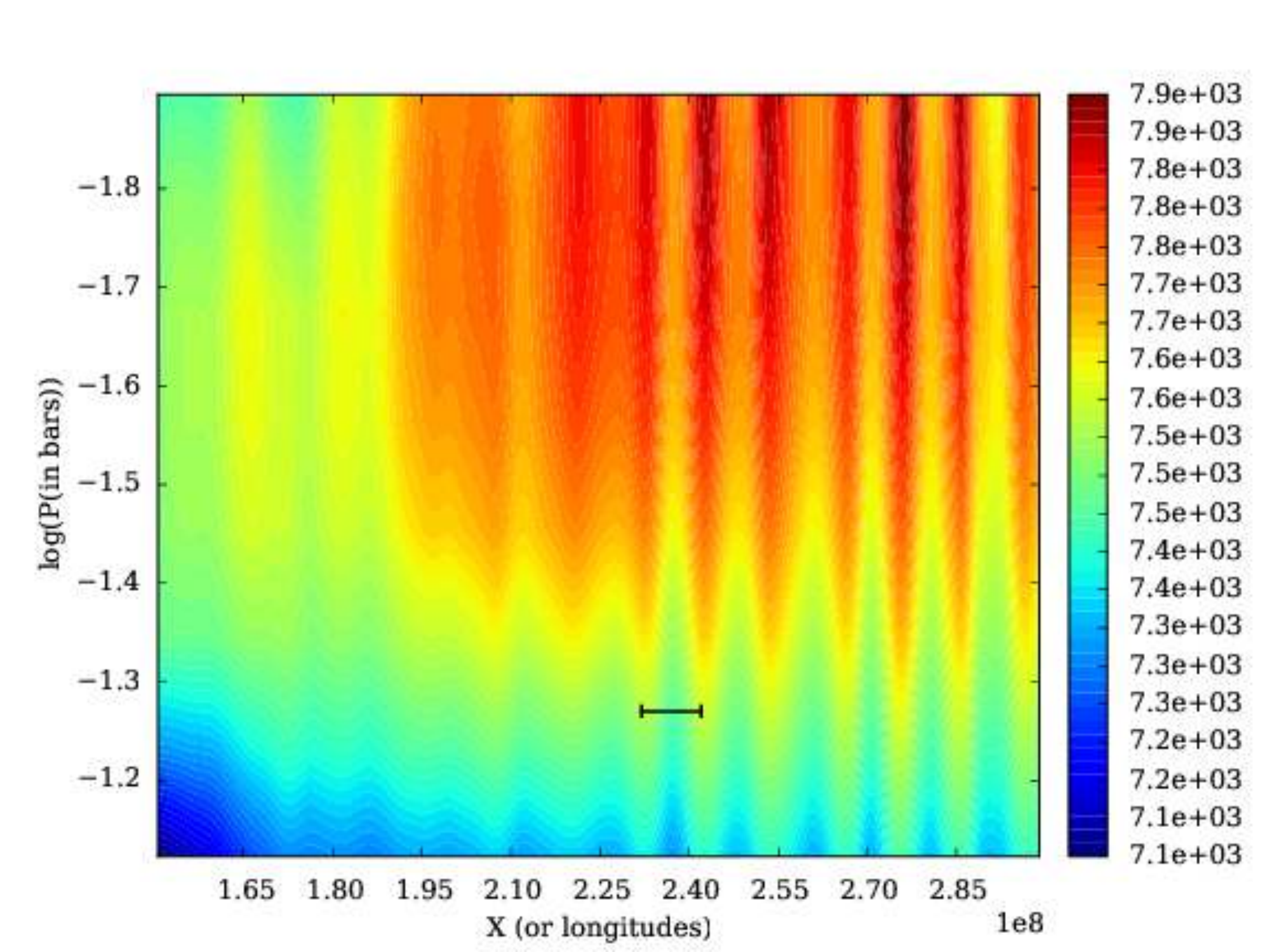}
\caption{Zonal velocity at the equator in a $(x,P)$ plane at time
  $t=319$ in model HighRes ({\it left panel}). Note the high frequency
  variations of the velocity (particularly easy to see in the
  atmosphere upper layers), superposed on the large scale zonal and
  vertical variation of the equatorial jet velocity. The right panel
  shows an enlargement of the dashed box shown on the left panel. The
  horizontal bar has a length of $10000$ km.} 
\label{fig:vxXZ_t319}
\end{center}
\end{figure*}

\begin{figure*}[!t]
\begin{center}
\includegraphics[scale=0.45]{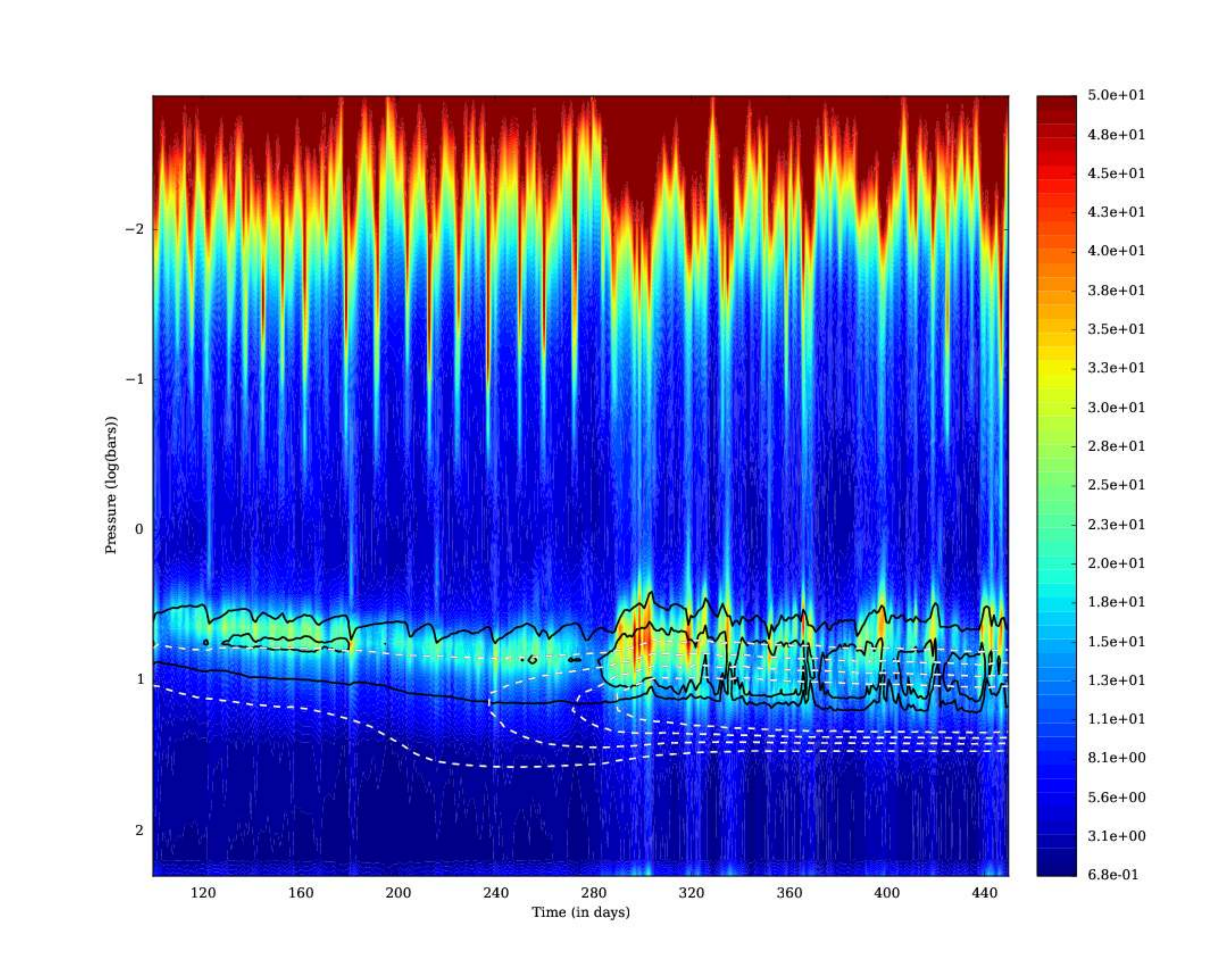}
\caption{Spacetime (time-pressure) diagram of the zonally averaged
  high frequency 
component of the zonal wind (see text for details) in color contours
(note that the color table has been saturated at $50$
m.s$^{-1}$). Contours of $Ri$ (solid lines) and $T$ (dashed lines) are
overplotted. For $Ri$, contours are for $Ri=0.1$ and $0.25$. For the
temperature, contours are for $T=1900$, $2000$, $2100$ and $2200$ K.} 
\label{fig:spacetime_vxprime}
\end{center}
\end{figure*}

The nature of the flow changes during phase III of the
simulation. A typical snapshot of the equatorial zonal wind in the
$(x,P)$ plane during that phase is plotted on
figure~\ref{fig:vxXZ_t319} (left panel, here shown at $t=319$). It
demonstrates that the zonal wind at the equator displays high
frequency variations in the $x$--direction that seem to extend over
the entire atmosphere.

\subsubsection{Identification of the instability}
By-eye measurement of the oscillations
spatial period (figure~\ref{fig:vxXZ_t319}, right panel, see the black
horizontal line) gives a typical value of about $8000$--$10000$ 
km. By comparison, the most unstable mode of the vertical
shear instability has a typical wavenumber $k_x$ that satisfies the
relation $k_xH \sim 0.5$ \citep{li&goodman10}. For temperature of
about $1800$ K such as we use here, the associated wavelength amounts
to about $11000$ km, remarkably close to our measurement. As an
additional diagnostic, we plot in figure~\ref{fig:spacetime_vxprime}
a spacetime diagram of the high frequency component of
the zonal wind (hereafter noted $\delta u$). The latter is calculated
by applying a high--pass filter to individual snapshots such as shown on
figure~\ref{fig:vxXZ_t319}. To do so, we use a Hamming function with a
half width of $30$ zonal cells as a low pass filter, and subtract the
low--pass filtered data to the raw data. The comparison with the
Richardson number (solid contours on
figure~\ref{fig:spacetime_vxprime}) shows a clear positive
correlation: at the bottom of the  atmosphere ($P \sim 1$--$10$ bars),
$\delta u$ displays a local maximum at 
pressure depth for which $Ri$ drops below $1/4$. In addition, that
maximum is larger when $Ri$ is smaller (see the difference between
phase II and phase III of the simulation, or the increase of $\delta
u$ at $t \sim 150$ that is coincident with a period of smaller $Ri$
values). Taken together, these results are consistent with the analysis of
\citet{li&goodman10} and strongly suggest that the vertical shear
instability operates at pressure levels where the Richardson number
reaches zonally averaged values smaller than
$1/4$. 

\subsubsection{Formation of shocks in the upper atmosphere}

Figure~\ref{fig:spacetime_vxprime} also shows that $\delta u$
increases strongly in the atmosphere upper layers, reaching values of
up to a few hundred m.s$^{-1}$ at pressure levels of a few mbars. It
is also clear that these velocity fluctuations are connected to the
atmospheric activity at $10$ bars. Future work is needed to
investigate their exact origin, but it is plausible, as recently
suggested by \citet{choetal15} in another context, that they are
gravity waves triggered by the vertical shear instability that
propagate upward and amplify as the density decreases. 

\begin{figure}[!t]
\begin{center}
\includegraphics[scale=0.35]{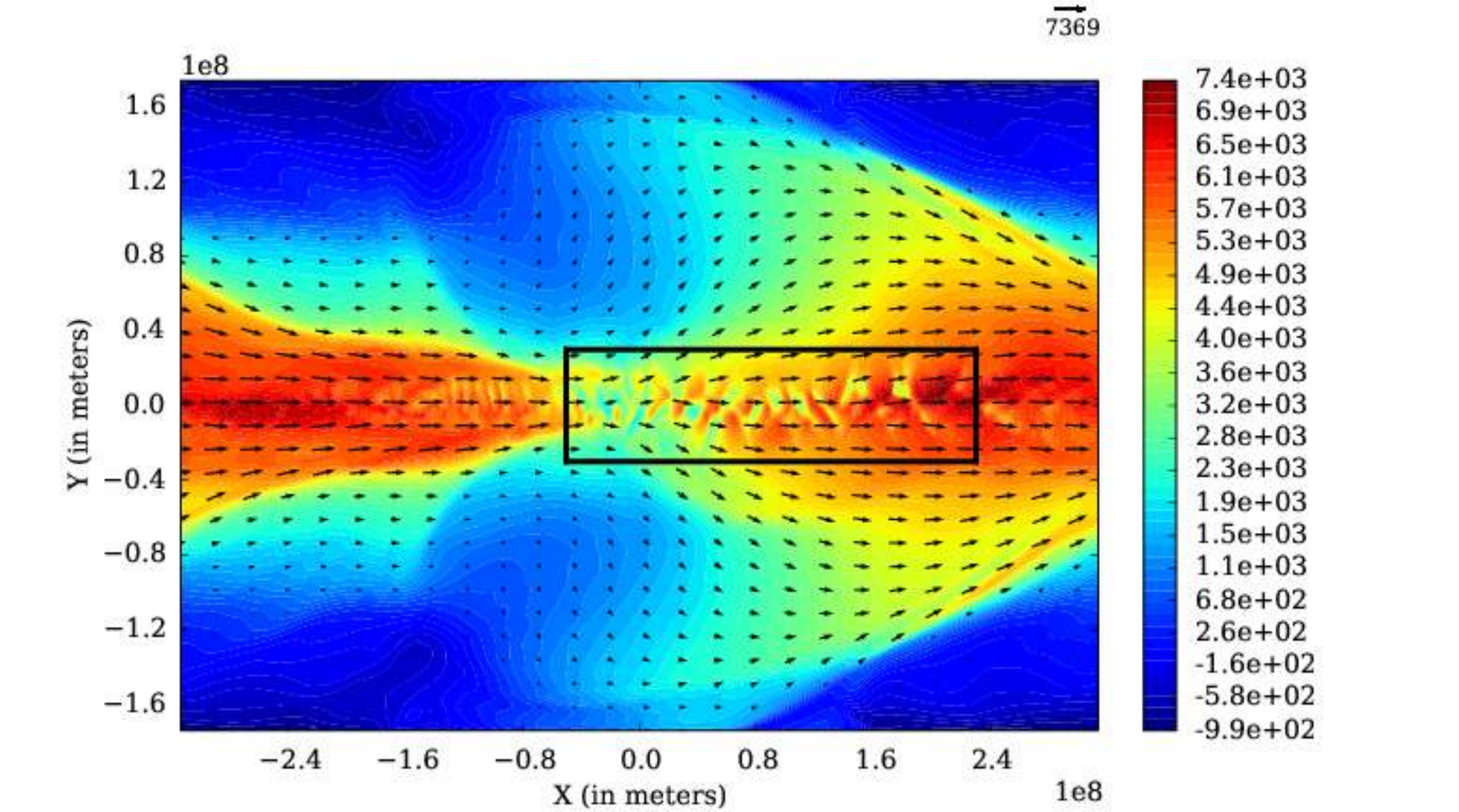}
\caption{Zonal velocity in model HighRes at $t=335$ at the pressure
  level $P=1.66$ mbars. Arrows indicate velocity vectors and the
  square marks the region studied in figure~\ref{fig:shocksII}.} 
\label{fig:shocksI}
\end{center}
\end{figure}

\begin{figure*}[!t]
\begin{subfigure}{.5\textwidth}
  \centering
  \includegraphics[scale=0.22]{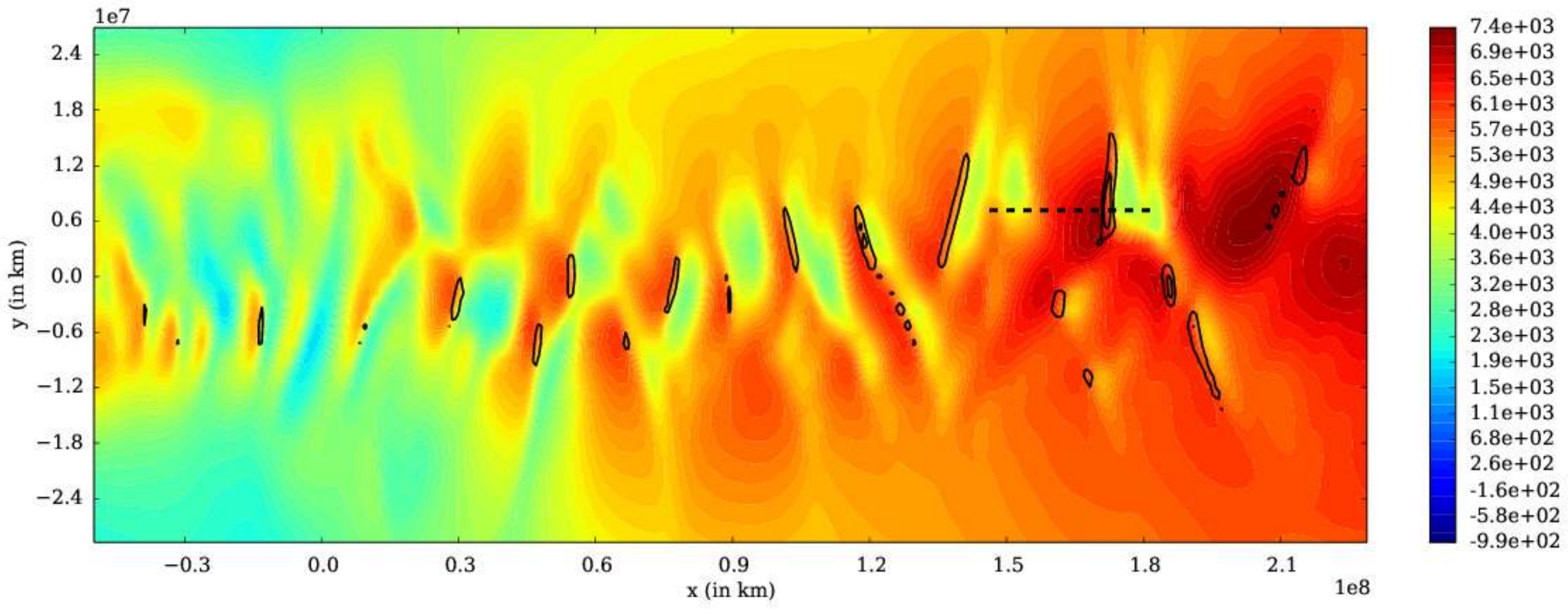}
  \includegraphics[scale=0.22]{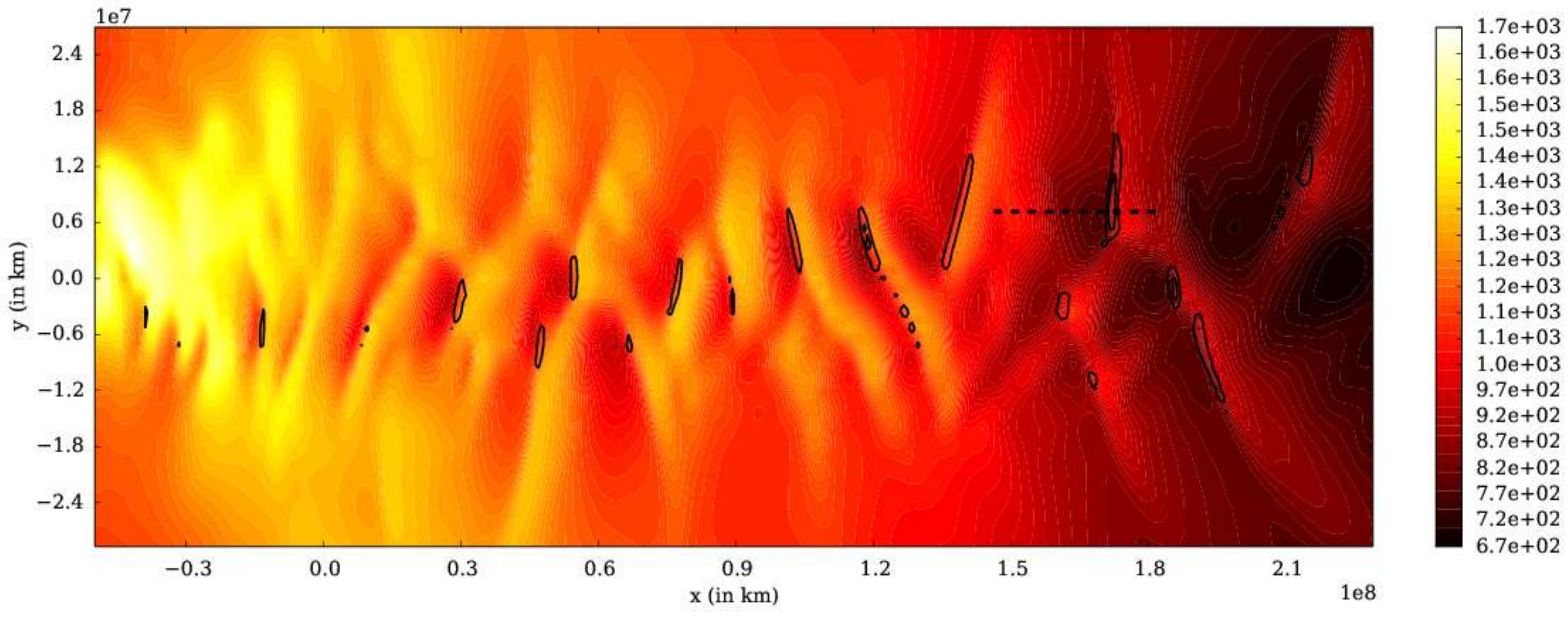}
  \label{fig:sfig1}
\end{subfigure}%
\begin{subfigure}{.5\textwidth}
  \centering
  \includegraphics[scale=0.45]{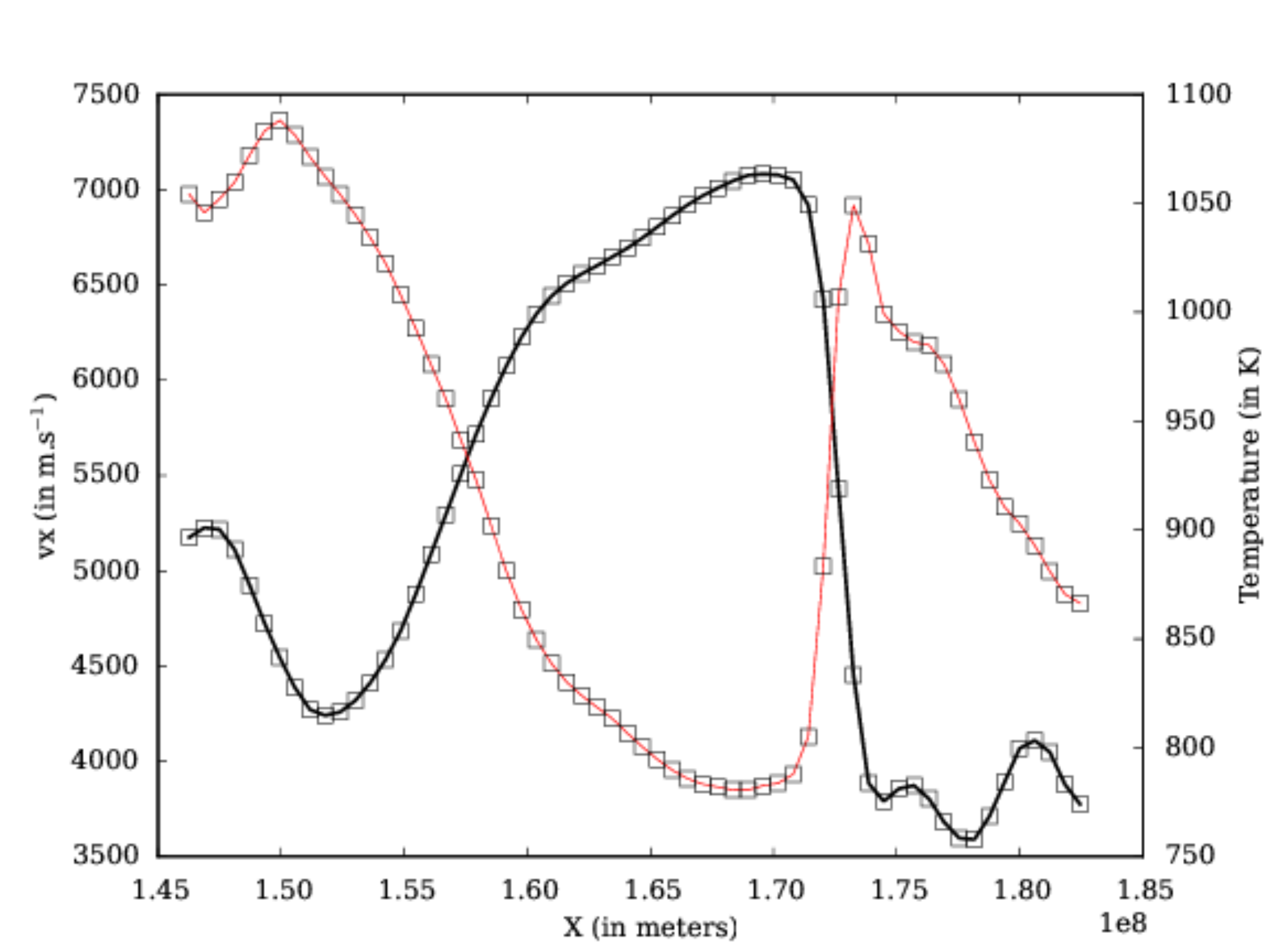}
  \label{fig:sfig2}
\end{subfigure}
\caption{Left: Color contours of the zonal wind ({\it top panel}) and
  temperature ({\it bottom panel}) at time $t=335$ in the rectangle box
  depicted in figure~\ref{fig:shocksI}. The contours shows the
  distribution of $\eta$, as given by
  Eq.~(\ref{eq:eta_shocks}). Levels are for $\eta=0.1$ and
  $0.2$. Right: Profiles of the zonal velocity ({\it black curve}) and
  temperature ({\it red curve}) along the dashed line shown on the
  left panels. For both curves, the empty squares marks the locations
  of the cells centers.} 
\label{fig:shocksII}
\end{figure*}

Because these fluctuations reach large amplitudes in the upper
atmosphere, we have closely investigated whether or not they
eventually steepen into shocks. In agreement with the findings of the
preceding sections, the flow remains devoid of such shocks at
pressures larger than $\sim 10$ mbars, with $\eta$ being of 
the order of a few percent at most. However, we have found signatures
of weak shocks in the top layers of the atmosphere. As an example, we
show on figure~\ref{fig:shocksI} that the zonal wind at $P=1.66$ mbars 
displays large and high frequency variations near the equator,
accross which the zonal velocity appears to decrease by a 
few km.s$^{-1}$. These
fluctuations are associated with values of $\eta$ up to $35 \%$
and also display significant increase of the temperature by a few
hundreds Kelvin (see figure~\ref{fig:shocksII}, left panel). Given the
short radiative timescale at this pressure ($\sim 10^4$ seconds),
such significant temperature fluctuations are indications of fast 
dynamics. We thus focused on one of these structure (marked as a
dashed line of the left panels of figure~\ref{fig:shocksII}), and plotted the
profiles of the zonal wind and of the temperature along that line
(figure~\ref{fig:shocksII}, right panel). There is a clear
discontinuity in both profiles at $x \sim 1.72 \times 10^8$m that we
identify as a shock. At that 
location, the zonal velocity rapidly decreases from $7$ km.s$^{-1}$ to
$4$ km.s$^{-1}$ within a handful of cells. Simultaneously, the
temperature increases from $\sim 800$ K to about $1050$ K. We
measured the shock front velocity $V_{sh}$ by 
restarting the simulation for a short amount of time, saving output
data at a high frequency. We found for $V_{sh}$ a value of the order of $3.9$
km.s$^{-1}$ (not shown), which, along with a typical sound speed of $2$
km.s$^{-1}$ at this location, indicate that the shock upstream Mach
number is roughly $M_1 \sim 1.5$. The associated temperature
discontinuity shown on figure~\ref{fig:shocksII} (right panel) can
then be compared with the prediction of the Rankine-Hugoniot
condition:
\begin{equation}
\frac{T_2}{T_1}=\frac{[(\gamma-1)M_1^2+2][2 \gamma
    M_1^2-(\gamma-1)]}{(\gamma+1)^2M_1^2} \, ,
\label{eq:rh}
\end{equation}
where $T_1$ and $T_2$ are the upstream and downstream gas temperatures,
respectively. For the case considered here ($M_1=1.5$ and
$\gamma=1.4$), Eq.~(\ref{eq:rh}) gives $T_2/T_1=1.32$, in very good
agreement with the simulation 
(for which the ratio is about $1.35$). This confirms that the numerous
structures identified in figure~\ref{fig:shocksII} (left panel) with
the contours of $\eta$ are indeed shocks. We have not attempted to
quantify the frequency, Mach number distribution and preferred
locations of these shocks. Besides being a difficult task on its own
to be carried in a systematic manner, there are also limitations and
artifacts of the present setup (see below) that have led us to
postpone a detailed characterization of these shocks
to future work.  

\subsubsection{Dissipation in the deep atmosphere and implications for the interior}

\begin{figure}[!t]
\begin{center}
\includegraphics[scale=0.45]{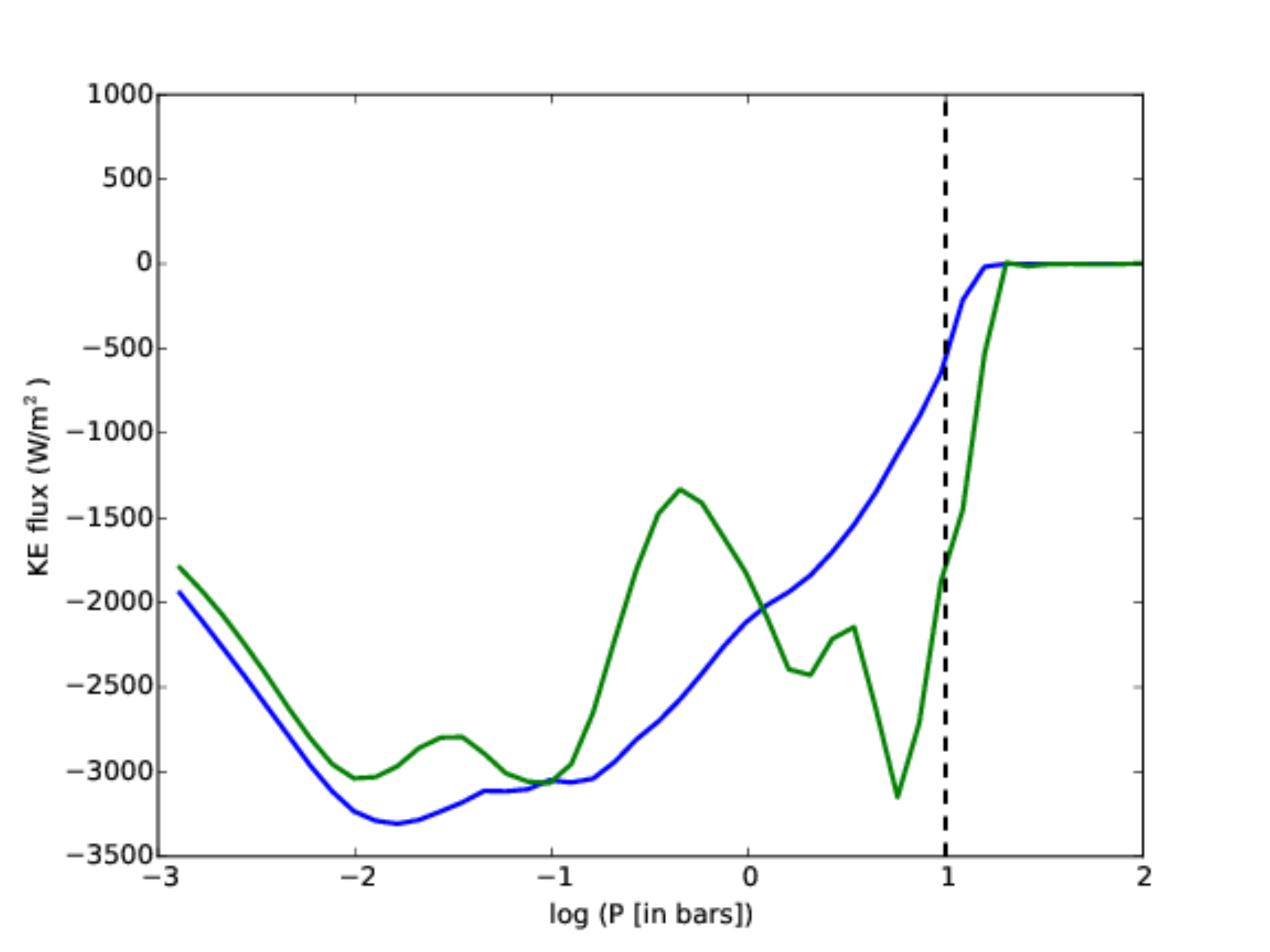}
\caption{Vertical profile of the kinetic energy flux during Phase III
  of model HighRes ({\it green curve}) and in model LowRes ({\it blue
  curve}), calculated according to Eq.~(\ref{eq:KEflux}). The dashed line
  marks the location of the top of the inert layer at $10$ bars. Both
  curves are time averaged between $t=350$ and $t=450$.} 
\label{fig:KEflux}
\end{center}
\end{figure}

Figure~\ref{fig:spacetime_vxprime} also shows a significant increase
of the temperature at $P \sim 10$ bars (see the white dashed contours) up to
a temperature of about $2400$ K (which should be compared to
$T_{eq}=1800$ K at that level). Again, there is a strong correlation
between that rise and the flow activity, which suggests that the
former is a consequence of the latter. This is not surprising and only
a result of small scale kinetic energy being dissipated and
transformed into heat. In RAMSES, such a thermalization is naturally
captured because we solve the total energy equation and does not
require using explicit dissipation coefficient (in that sense, the
dissipation is solely numerical in origin). At $10$ bars and below,
the radiative timescale $\tau_{rad}$ goes to 
infinity and heated gas cannot cool radiatively anymore : in other
words, heat deposited in the inert layer as a result of kinetic energy
dissipation accumulates and temperature rises. As shown on
figure~\ref{fig:spacetime_vxprime}, such an increase of the
temperature at $10$ bars also increases the vertical temperature
gradient (in absolute value) and helps decrease the Richardson number
through its dependence on the Brunt-Vaisala frequency. This effect can
be quantified by measuring the kinetic energy flux in the vertical
direction. To do so, we computed the kinetic energy flux per unit area
according to the relation:
\begin{equation}
F_{KE}(P)=\frac{1}{L_x L_y} \iint \frac{1}{2} \rho v^2 v_z dxdy \, ,
\label{eq:KEflux}
\end{equation}
where the integral is calculated at the pressure level
$P$. In both models LowRes and HighRes, figure~\ref{fig:KEflux} shows
that we recover a negative vertical flux of kinetic energy, in
quantitative agreement with \citet{showman&guillot02} and with a
similar value of about $-2000$ to $-3000$ W/m$^2$. Note that this 
agreement is somewhat fortuitous and should not be taken too
seriously: in the equatorial $\beta$--plane 
model such as used here, motions gradually go to zero away from the
equator, so that $F_{KE}$ depends on the arbitrary location of the
boundaries in the $y$--direction. However, this qualitative
agreement with \citet{showman&guillot02} supports the fact that the
existence of this flux is robust.  

Figure~\ref{fig:KEflux} also shows
that a larger negative flux is present in the deep atmosphere ($P \sim
5$--$10$ bars) in model HighRes than in model LowRes, precisely at the
location where we see the temperature increase during Phase III. This
additional flux is likely 
associated with the vertical shear instability and illustrates the 
idea of \citet{showman&guillot02} that downward kinetic energy
transport can be associated with extra heating in the deep layers of
the atmosphere. In addition to the increased kinetic energy flux
suggested by figure~\ref{fig:KEflux}, turbulence in stably stratified
atmosphere also transports heat downward \citep{youdin&mitchell10} and
is likely at play here as well. 

As first recognized by \citet{GS02}, such downward energy fluxes
  are comparable to, and even greater than, the internal cooling flux
  of a typical inflated hot Jupiter. This mechanism---conversion of
  stellar energy into kinetic energy to be transported and dissipated
  deep in the atmosphere---is thus still a viable explanation to the
  problem of the large radius of highly irradiated giant planets. The
  main difficulty---a difficulty faced by most proposed
  mechanisms---is that this energy must be deposited deep enough
  inside the planet to significantly affect its thermal evolution
  \citep{ginzburg&sari15}, possibly in the deep adiabat
  \citep{GS02}. Depending on the precise planet and its age,
  \citet{GS02} predict this level to be in the 100--1000\,bar range.   
Although figure~\ref{fig:KEflux} seems to imply that not much energy
is deposited below the 10--20\,bar level, it should be clear that,
in our model, this
\textit{barrier} is most probably an artifact of the numerical
setup. As discussed above, the presence of an inert layer below
10\,bars, where $\tau_{rad}$ becomes infinite, provides a strong
positive feedback on the wave activity and the heating at this
level. If we had 
a more realistic, gentler increase of the radiative timescale, the
level of maximum energy deposition would probably be pushed
deeper. Indeed, we remark that 
figure~\ref{fig:spacetime_vxprime} displays 
clear evidence of significant small scale activity at pressure levels
of a few bars, i.e. largely above the inert layers, early in the
simulation ($t<300$ days). This is suggestive evidence that an
infinite radiative timescale is \textit{not} needed to trigger the 
vertical shear instability. But we cannot yet assess the maximum depth
at which energy deposition will occur. 

Therefore, while our model illustrates the possibility that
vertical shear instability can develop in hot Jupiter atmospheres, their
long--term effect on the flow remains to be determined with a more
realistic treatment of the radiative properties of the deep
atmosphere. Only then will we be able to properly quantify whether the
mechanism first proposed by \citet{GS02} can realistically account for
the inflated radius of strongly irradiated giant planets. 

Before ending this section, and with the above caveats in mind, we
stress that the effect of the vertical shear instability on the
equatorial jet velocity is significant. While its time averaged value
over phase II of the flow evolution is about $7700$ m.s$^{-1}$ at $50$
mbars, the equatorial wind goes down to $6700$ m.s$^{-1}$ when averaged
between $t=400$ and $t=450$. We have check that the spatial
distribution of the temperature above $1$ bar is almost identical
during the two phases (not shown) and can not be incriminated for the
jet velocity decrease. Rather, it is likely that the small scale
velocity fluctuations associated with the vertical shear instability
act as a form of drag that slows down the jet, possibly mediated by
gravity waves \citep[see][]{watkins&cho10} and, as shown above, by
shocks, both excited by the vertical shear instability. Given the
limitations of the numerical setup used here, a detailed and
quantitative investigation of that 
possibility is beyond the scope of this paper but opens up the
possibility for developing a physically motivated subgrid scale model
for the dissipation in the atmosphere of hot Jupiter that could be 
incorporated in traditional GCMs.

\section{Discussion and conclusion}
\label{sec:conclusion}

\subsection{Main results}

Using a series of high resolution idealized simulations of
hot--Jupiters atmospheres that solve the Euler equations with a finite
volume shock-capturing scheme, we have found the following
results: 
\begin{enumerate}[topsep=0pt,leftmargin=10pt,itemsep=0pt]
\item[$\bullet$] Numerical simulations performed in the framework of
  the equatorial $\beta$--plane model are in good agreement with
  results published in the literature using a wide range of models and
  elaborate treatments of the radiative effects.
\item[$\bullet$] A supersonic, equatorial, eastward jet forms
  quickly in the upper layers of the atmosphere. At $P\sim 50$ mbars,
  its zonally averaged velocity reaches about $\sim 7$ km.s$^{-1}$ and
  is found to be sensitive to the meridional resolution (or,
  equivalently, to the dissipation).
\item[$\bullet$] At large enough spatial resolution (or low enough
  dissipation), the jet displays large velocity fluctuations that can
  be attributed to meanders upward of $\sim 1$ bars and smaller scale
  fluctuations with an amplitude of a few tens of m.s$^{-1}$ at
  pressure levels $P \sim 1$--$10$ bars.
\item[$\bullet$] The meanders are clearly associated with a barotropic
  Kelvin-Helmholtz instability that results in quasi--periodic
  modulations of the jet velocity with a typical period of $\sim 10$
  days. The properties of the flow are broadly consistent with the
  expectations of linear stability analysis. Temperature fluctuations
  of a few hundred Kelvins are found at the photosphere of the
  planet at the peak of the instability, most likely a result of
  adiabatic compression associated with the equatorial jet
  meanders. Future work is needed to determine whether these
  variations are compatible with the observed upper limit of $2.7 \%$
  of the dayside variability of HD189733b 
  \citep{agoletal10,knutsonetal12} and whether they could
  be observable with the JWST using brightness mapping such as
  discussed by \citet{dewitetal12}.   
\item[$\bullet$] The smaller scales fluctuations are likely associated
  with a vertical shear instability. They correlate nicely
  with the locations where the Richardson number is smaller than
  $1/4$. They create zonal variations of the jet
  velocity with a spatial scale of $\sim 10^4$ km that is consistent
  with the most unstable mode predicted by a linear analysis
  \citep{li&goodman10}. 
\item[$\bullet$] The dissipation of the 
  kinetic energy associated to the vertical shear instability results
  in a substantial increase of the temperature at $P\sim 10$
  bars. This thus confirms that the atmosphere converts stellar energy
  into kinetic energy that is transported downward to be deposited at
  deeper levels \citep{showman&guillot02}. A better treatment of the
  lower boundary is needed to know whether the deposition level can be
  deep enough to affect the interior and help explain the radius
  anomaly \citep{GS02}.
\item[$\bullet$] We find weak shocks in the upper layers of the
  atmosphere ($P<10$ mbars). They have typical upstream Mach
  numbers between $1$ and $2$ and create temperature fluctuations of
  a few hundreds Kelvins. At larger pressure ($P>10$ mbars), we find
  no shocks despite the supersonic nature of the equatorial
  jet.
\end{enumerate}

\subsection{Limitations and future work}

These results should not hide the limitations of the work presented
here that are as many avenues for progress. As noted in
section~\ref{sec:setup}, the model depends on a number of free
parameters. The motivation of the present paper was to choose a unique
set of these parameters so that the flow properties matches that of the
benchmark calculation described by \citet{hengetal11}. Even if the
dynamical mechanisms responsible for the instabilities highlighted
here are fairly general, future work is needed to investigate
systematically the sensitivity of the results presented here to each
of these parameters and to make quantitative predictions that would be
valuable for more realistic atmospheres. For example, we note that the
zonal wind velocity we obtained tends to be larger than commonly found
by other authors. Is it because we work in the framework of the
equatorial $\beta$--plane model? Or is it related to the form of the
thermal forcing we use, and in particular to the value of $L_{th}$?
Another question is that of convergence with spatial 
resolution: it is possible (and even likely!) that numerical
dissipation is still affecting the instabilities described in
section~\ref{sec:phaseII} and \ref{sec:longterm}. This may affect
their saturation properties and their quantitave effect of the mean
flow, and particularly on the velocity of the jet. The present work
shows that a proper convergence study (systematically varying both the
meridional and the vertical resolution)---even though it is very
computationally demanding---is definitely needed.

Perhaps the most stringent limitation of the present work 
comes from the presence of an ``inert'' layer below $10$ bars where
the radiative timescale $\tau_{rad}$ goes to infinity. The large
temperature increase we see at this location is probably overestimated
because of the inability of the gas to cool radiatively. This problem
may be somewhat mitigated upon noting that $\tau_{rad}$ is expected to
increase rapidly in the deep layers of hot Jupiters atmospheres ($P
\leq 10$--$100$ bars). For example, \citet{showmanetal08} use
$\tau_{rad}=10^8$ seconds at $20$ bars. This is more than $300$ planet
days and much faster than the timescale of a few days that is
associated with the vertical shear instability (see
fig.~\ref{fig:spacetime_vxprime}), so that the effect of choosing an
infinite value for $\tau_{rad}$ may not be as severe as naively
expected. Nevertheless, it is clear that a large temperature 
increase such as found here acts as a positive feedback onto the
vertical shear instability by reducing the Richardson
number $Ri$. Whether or not it affects the findings presented in this
paper, and how much, remains to be clarified. One possibility to do so
is to replace the Newtonian relaxation scheme with a simplified
radiative transfer scheme \citep{hengetal11b,rauscher&menou12}. This 
may alleviate that problem in the deep atmosphere.

The need to strengthen these results is made even more important
because standard GCM codes that use the primitive formulation of the
hydrodynamic equations are unable to account for such vertical shear
instabilities. By assuming hydrostatic equilibrium in the vertical
direction, they assume {\it de facto} that such an instability does not
exist. Instead, they rely on subgrid scale modelling to include their
effect on the flow. Future work is therefore needed to validate these
subgrid models and/or develop more appropriate approaches if
necessary. This is an important step to validate the long--term use of
such codes as the primary tool to model hot Jupiter atmospheric flows.

\appendix

\section{Equilibrium temperature and cooling timescale}
\label{app:radiative_eq}

\begin{figure}[!ht]
\begin{center}
\includegraphics[scale=0.45]{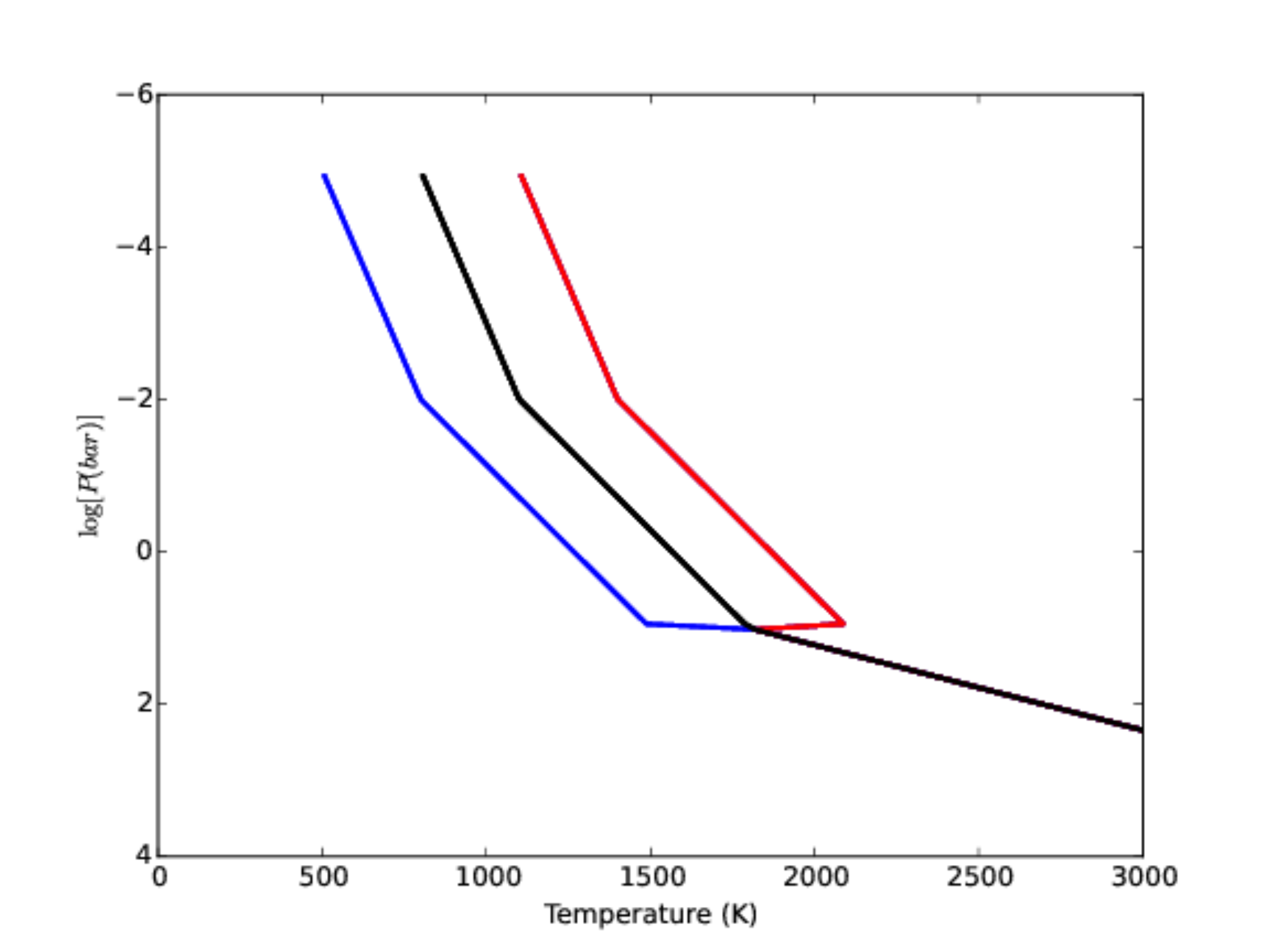}
\includegraphics[scale=0.45]{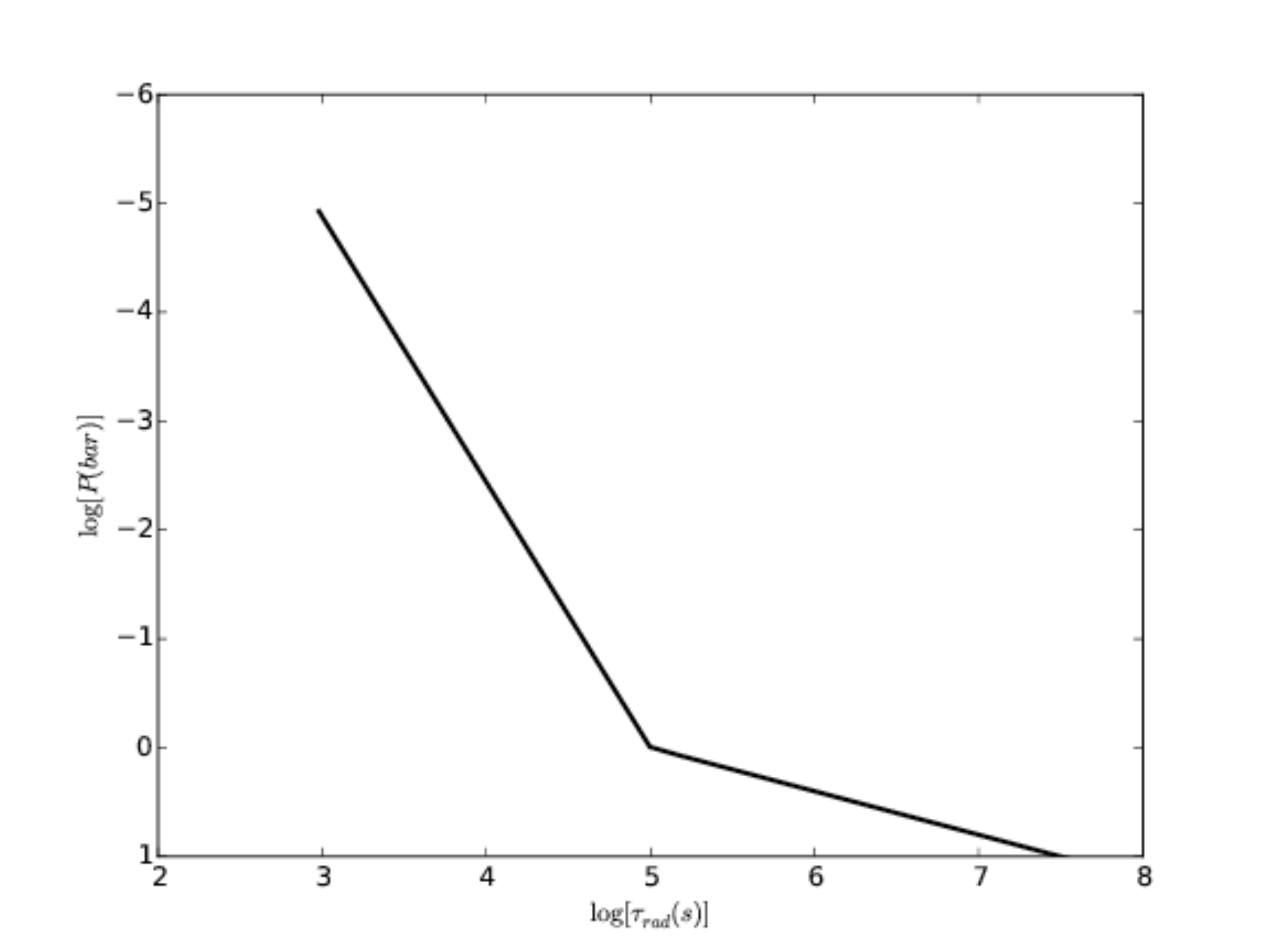}
\caption{Pressure variations of the equilibrium temperature $T_{eq}$
  ({\it top panel}) and the Newtonian relaxation time $\tau_{rad}$
  ({\it bottom panel}) used to calculate the cooling function $\cal{L}$
  in the deep hot Jupiter model presented in this paper.} 
\label{fig:Teq_tau_prof}
\end{center}
\end{figure}

We give here for reference the pressure profiles of $T_P^0$
and $\tau_{rad}$ that we have used in this paper. They are a simplified
version of the profiles presented by \citet{hengetal11}. The thermal
timescale $\tau_{rad}$ is calculated with the following relation: 
\begin{equation}
\tau_{rad}=\left\{ \begin{array}{ll}
  \tau_0\left( P_0/P \right)^{\alpha_0} & \textrm{if } P<P_0 \\
 \tau_1\left( P_1/P \right)^{\alpha_1} & \textrm{if } P_0<P<P_1 \\
 +\infty & \textrm{if } P>P1 \end{array} \right. \, ,
\end{equation}
where $P_0=1$ bar, $P_1=10$ bars, $\tau_0=10^5$ and $\tau_1=10^{7.5}$
seconds. The dimensionless exponent $\alpha_0$ and $\alpha_1$
respectively amounts to $-0.41$ and $-2.5$. Likewise,
$T_P^0$ is given as a combination of linear functions of
the logarithm of the pressure:
\begin{equation}
T_{iro}=\left\{ \begin{array}{ll}
  T_{-2} - \gamma_{-2} \log \left( P_{-2}/P \right) & \textrm{if } P<P_{-2} \\
 T_1 - \gamma_{1} \log \left( P_1/P \right) & \textrm{if } P_{-2}<P<P_1 \\
 T_2 - \gamma_2 \left( P_2/P \right) & \textrm{if } P>P1 \end{array}
\right. \, , 
\end{equation}
where $P_{-2}=10^{-2}$ bar and $T_{-2}$, $T_1$ and $T_2$ are
respectively set to $1100$, $1800$ and $3000$ K. The slopes of the
linear relations are given by $\gamma_{-2}=100$ K, $\gamma_{1}=233$ K
and $\gamma_{2}=983$ K. The relation between $\tau_{rad}$ and $P$ and
between $T_P^0$ and $P$ are shown in figure~\ref{fig:Teq_tau_prof}
where we also display the variation of $T_{day}$ and $T_{night}$ such
as defined in section~\ref{sec:cooling_func}. All curves are meant to
be compared with, for example, figure~7 of \citet{hengetal11} with
which they are in good agreement.

\section{Numerical tests}

\subsection{Baroclinic instability in an adiabatic atmosphere}
\label{app:baroclinic_wave}

\begin{figure}[!t]
\includegraphics[scale=0.45]{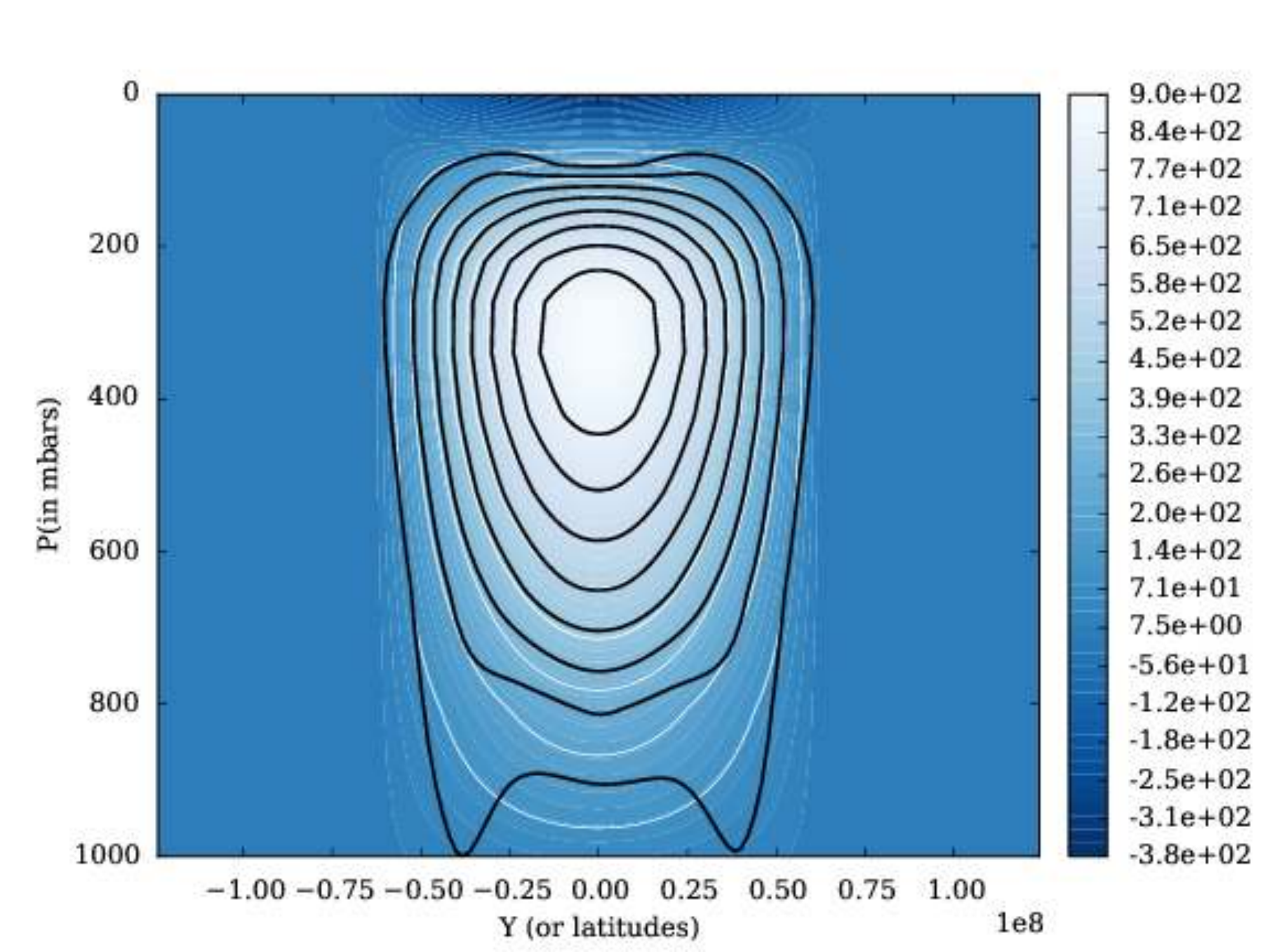}
\caption{Initial (i.e. at $t=0$) zonally averaged zonal wind used for
  the baroclinic instability simulations (colors and white
  contours). The black contour show the zonally averaged zonal wind at
the end of the simulation for the model with resolution
$(N_x,N_y,N_z)=(256,128,32)$. Contours are drawn every $100$
m.s$^{-1}$ from $100$ to $900$ m.s$^{-1}$.}
\label{fig:zonalwind_baro}
\end{figure}

\begin{figure}[!t]
\includegraphics[scale=0.45]{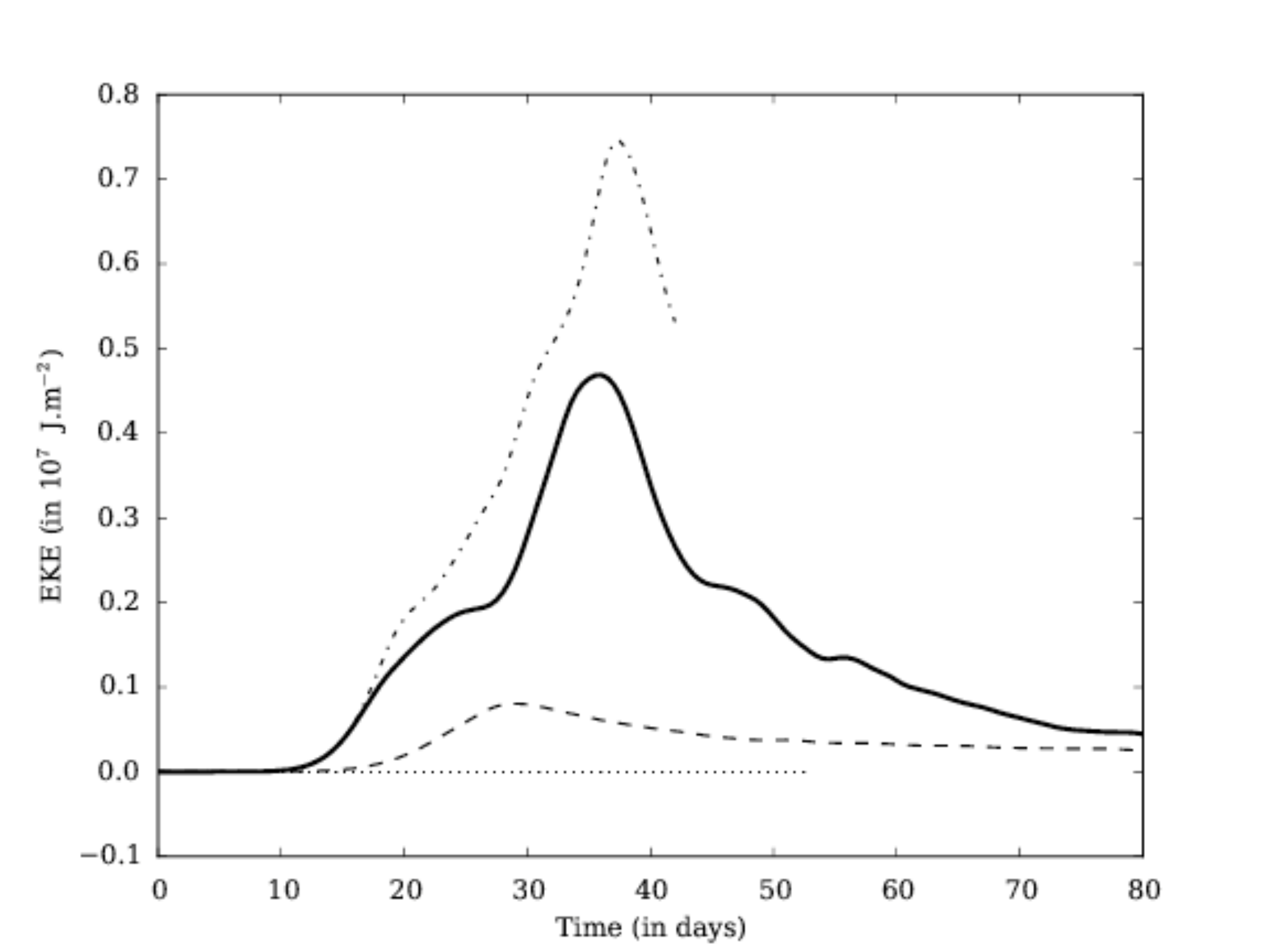}
\caption{Time evolution of the volume integrated eddy kinetic energy
  (per unit area) in the baroclinic instability test. The solid and
  dotted lines both share the resolution $(N_x,N_y,N_z)=(256,128,32)$
  and corresponds to the model with and without an initial temperature
  perturbation, respectively. The dashed and dotted--dashed lines
  shows the results of the models with resolution
  $(N_x,N_y,N_z)=(128,64,32)$ and $(N_x,N_y,N_z)=(512,256,32)$,
  respectively. }
\label{fig:EKE_baro}
\end{figure}

\begin{figure}[!t]
\includegraphics[scale=0.24]{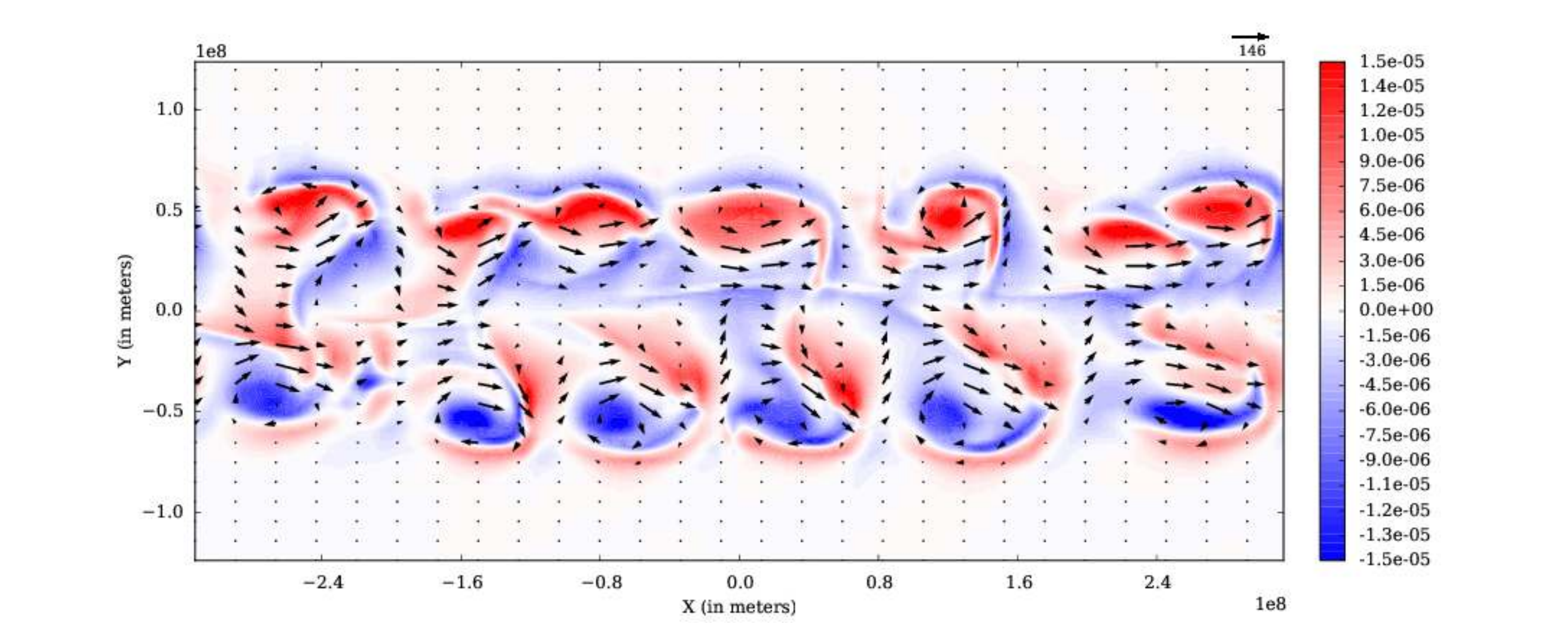}
\includegraphics[scale=0.24]{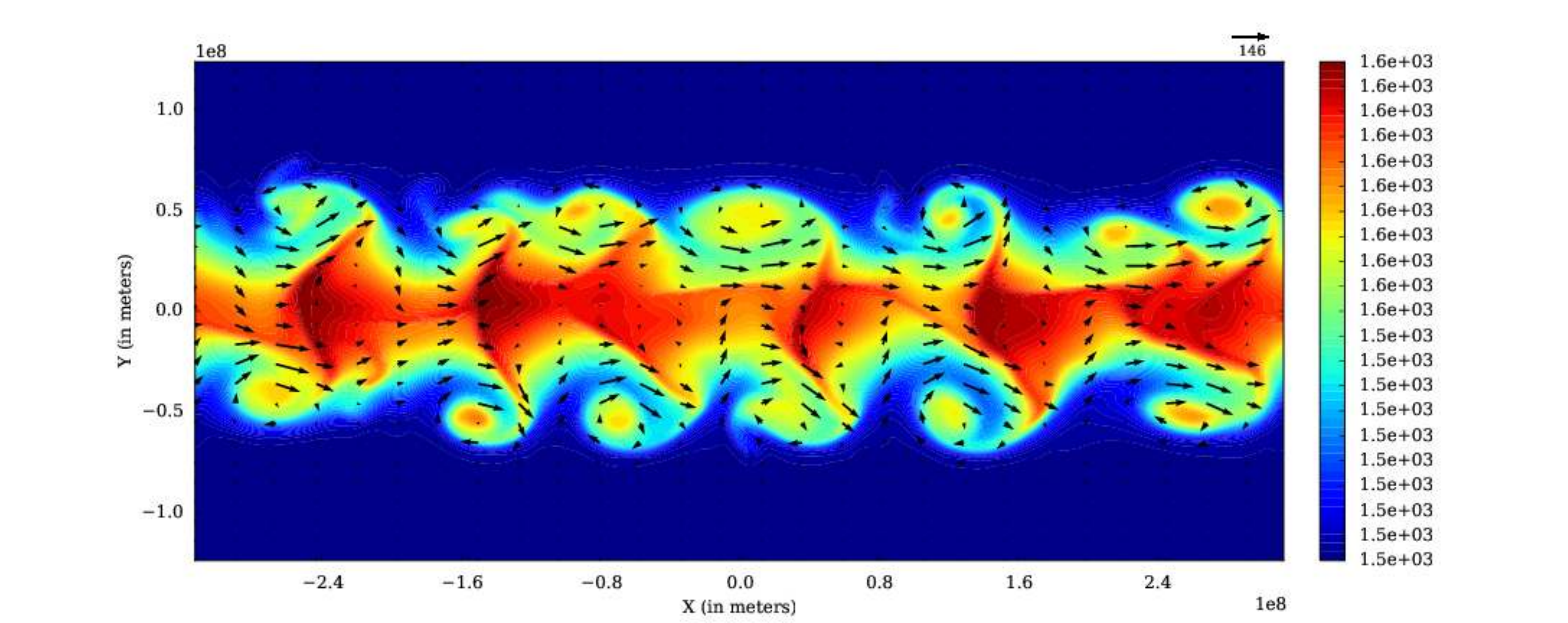}
\caption{Relative vorticity ({\it top panel}) and temperature ({\it
    bottom panel}) distribution in the horizontal plane at $930$
  mbars, at time $t=32$ showing the development of a baroclinic wave
  that grows on top of a zonal equatorial jet. The parameters are chosen
  after \citet{polichtchouketal13} and the resolution is
  $(N_x,N_y,N_z)=(256,128,32)$.}
\label{fig:baro_snapshots}
\end{figure}

\begin{figure}[!t]
\includegraphics[scale=0.24]{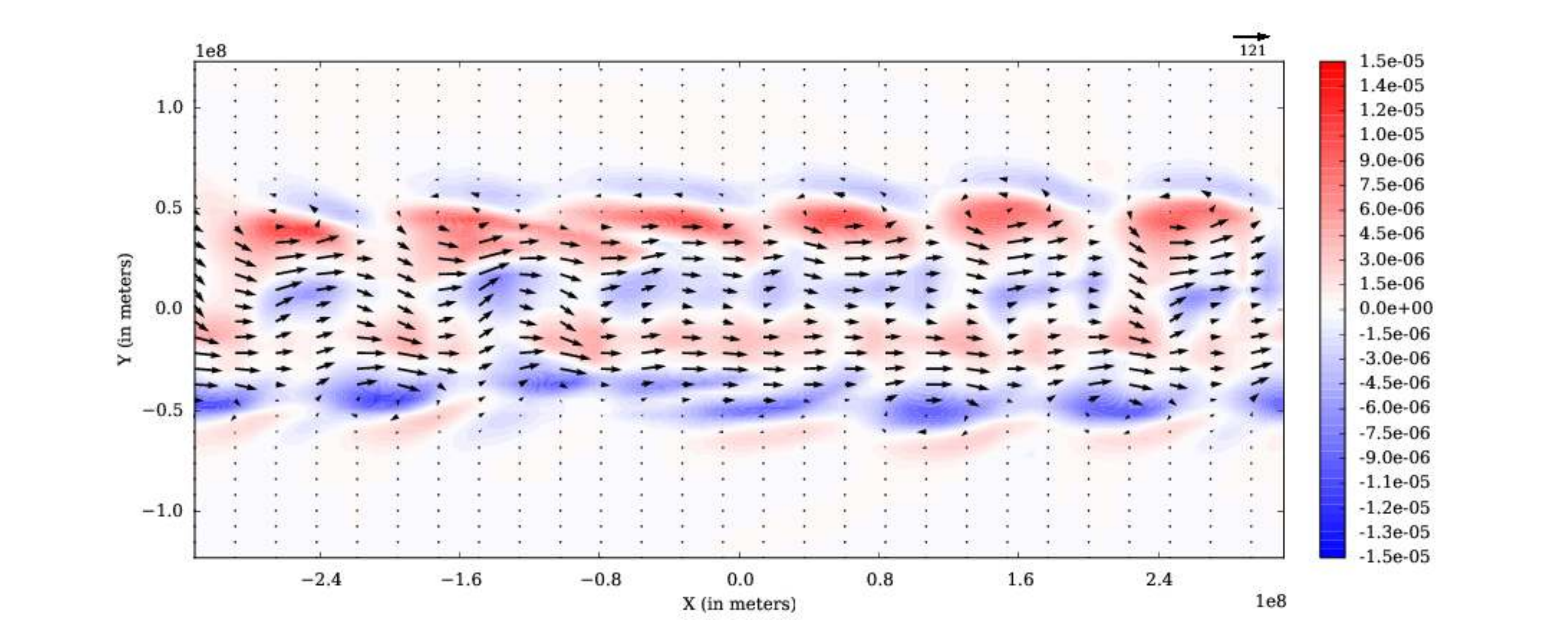}
\includegraphics[scale=0.24]{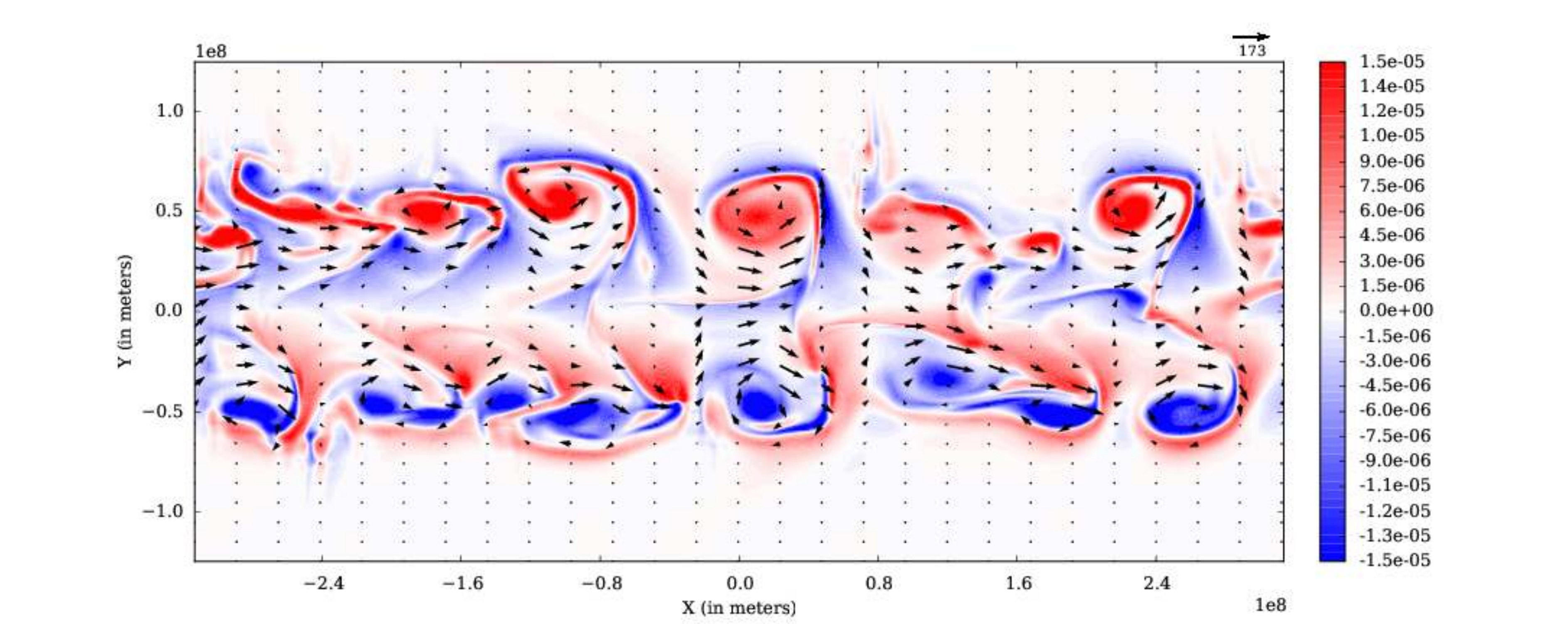}
\caption{Relative vorticity at $930$ mbars at time $t=32$ for the
  baroclinic instability growth with resolution $(N_x,N_y)=(128,64)$
  ({\it top panel}) and $(N_x,N_y)=(512,128)$ ({\it bottom panel}).}
\label{fig:baro_snaphots_res}
\end{figure}

As argued by \citet{polichtchouketal14}, the growth of a baroclinic
wave is a severe test for codes that pretend to describe accurately
atmospheric flows because it grows slowly from infinitesimal
perturbations. This is even more so for finite volume codes such as
RAMSES that have problems handling hydrostatic equilibria, and we
found that problem to be very helpful in assessing the robustness of
our setup. In this appendix, we thus qualitatively reproduce one of
the model presented by \citet{polichtchouketal13}, namely their
equatorial jet case, since the base flow is close to the jet
configuration studied in the present paper. As noted by 
\citet{polichtchouketal14}, the detailed evolution of the atmosphere
is very sensitive to the exact structure of the jet and the initial
perturbation, so that our goal here is not to reproduce quantitatively
the results of \citet{polichtchouketal13}, but rather to show that we
obtain the same qualitative evolution of the flow. Indeed, since we
neither use the same equations (Euler vs. primitive) nor the same
geometry (cartesian vs. spherical), a one to one quantitative
comparison is not possible.

The structure of the atmosphere at $t=0$ is computed assuming thermal
wind balance after having specified the jet zonal velocity. The latter
is computed using the following relation, adapted to the equatorial
$\beta$--plane geometry from \citet{polichtchouketal13}: 
\begin{equation*}
u(x,y)=\begin{cases}
 U_0  G(z) \sqrt{\sin \left[ \pi \sin^2 \left( \frac{\pi}{2}
     \frac{4y+L_y}{2L_y} \right)\right]} \, , & |y|<L_y/4  \\
 0 \, , & \textrm{otherwise}
\end{cases}
\end{equation*}
where $G$ is defined as:
\begin{equation*}
G(z)=\frac{1}{2} \left[ 1 -\tanh^3 \left( \frac{z-z_J}{\Delta z_J}
  \right) \right] \sin \left(\pi \frac{z}{L_z} \right) \, .
\end{equation*}
The previous relations depend on a number of free parameters. In this
appendix, we use $z_J=0.8 L_z$, $\Delta z_J=0.2 L_z$, $U_0=1000$
m.s$^{-1}$ and $L_z=2 \times 10^6$ m.s$^{-1}$. The initial structure
of the jet is shown on figure~\ref{fig:zonalwind_baro} and resembles that
shown on their figure 4 (right panel) by \citet{polichtchouketal13}
. Next, a localized temperature perturbation of amplitude $\delta T_0$
is added at all pressure levels in the atmosphere. It takes the form:  
\begin{equation}
\delta T(x,y)=\delta T_0 \sech^2 \left( \frac{6 \pi x}{L_x} \right) \sech^2
\left( \frac{\pi}{3}\frac{4y-L_y}{2 L_y} \right) \, ,
\end{equation}
so that it is localized initially at $x=0$ and $y=L_y/4$, i.e. on the
jet northern flank. 

We first present and contrast the results of two simulations with
resolution $(N_x,N_y,N_z)=(256,128,32)$ and for which we respectively
set $\delta T_0=0$ (control run) and $1$ K (perturbed run). 
We computed the volume integral of the specific eddy kinetic energy
(EKE) as a diagnostic of the baroclinic instability growth: 
\begin{equation}
EKE=\int_{V} \rho \left[ u^{\prime 2}+v^{\prime 2} \right] dV \, .
\end{equation}
As described by \citet{polichtchouketal13}, the specific EKE grows  
after a few days in the perturbed case (see figure~\ref{fig:EKE_baro}) as
a result of a baroclinic wave but stays very small in the control
run during the entire simulation. In fact, the specific EKE remains
at the level of $10^{-10}$ J/m$^{-2}$ (i.e. $17$ orders of magnitude
smaller than the perturbed model after the growth of the baroclinic
wave!), indicating that our numerical scheme accurately conserves the 
symmetry of the flow and the initial atmospheric hydrostatic
equilibrium (the maximum meridional velocity at the end of that run at
$975$ mbars is only of the order of $8$ cm.s$^{-1}$!). In the
perturbed run, the specific EKE peaks between $t=30$ and $t=35$. The
flow at $975$ mbars shows five well defined cyclones that 
roll on both sides of the jet (see figure~\ref{fig:baro_snapshots}),
clearly correlated with 
temperature fluctuations. This is again in very good qualitative
agreement with the results of \citet{polichtchouketal13}. We note that
the number of cyclones and the amplitude of the vorticity
perturbations they display is also in good quantitative agreement with
those results. Finally, after a few tens of days, the specific EKE
saturates and decays. At the end of the model evolution, the
equatorial jet structure is modified and displays a vertical structure
that is more barotropic (see figure~\ref{fig:zonalwind_baro}, black
contours), in agreement with the finding of \citet{polichtchouketal13}.

We have also performed a resolution study, keeping the number of
vertical levels fixed to $N_z=32$, and gradually varying the
horizontal resolution from $(N_x,N_y)=(128,64)$ to
$(N_x,N_y)=(512,256)$, although the highest resolution simulation is
only integrated to $t=45$. The vorticity distribution at $975$ mbars (see
figure~\ref{fig:baro_snaphots_res}) shows that the minimum resolution
required to capture the instability is $(N_x,N_y)=(256,128)$ and that
the structure of the flow is qualitatively captured at that
resolution, although quantitatively, figure~\ref{fig:EKE_baro} shows
that the growth rate is not converged yet. This resolution study gives
a useful comparison with the analysis of \citet{polichtchouketal13}
who used a spectral numerical scheme and somewhat mitigate their claim
that finite difference shemes require resolution larger by an order of
magnitude to capture the growth of the baroclinic wave. Here, it
appears that a difference of about a factor of two is sufficient.

Overall, the results presented here are in good qualitative agreement
with the results of \citet{polichtchouketal13} and
\citet{polichtchouketal14} and give credit to our implemantation.

\subsection{A shallow hot Jupiter model}
\label{app:shallow_hj}

\begin{table}[t]\begin{center}\begin{tabular}{@{}ccc}\hline\hline
Parameters & Symbol & Value  \\
\hline\hline
Box vertical size (m) & $L_z$ & $3 \times 10^6$ \\
Pressure at bottom boundary (Pa) & $P_0$ & $10^5$ \\
Day-night temperature difference (K) &  $\Delta T$ & $300$ \\
Cooling time scale (sec) & $\tau_{rad}$  & $1.5 \times 10^5$ \\
Cut--off length of cooling function (m) & $L_{th}$ & $3.5 \times 10^7$ \\
Temperature at bottom boundary (K) & $T_{\textrm{surf}}$ & $1600$ \\
Tropospheric lapse rate (K.m$^{-1}$) & $\Gamma_{\textrm{trop}}$ & $2
\times 10^{-4}$ \\ 
Tropopause temperature increment (K) & $\delta T_{\textrm{stra}}$ & $10$ \\
Tropopause pressure (Pa) & $\sigma_{\textrm{stra}}$ & $1.25 \times 10^4$ \\
\hline\hline
\end{tabular}
\caption{Parameters used for the shallow hot Jupiter model.} 
\label{tab:shallow_model}
\end{center}
\end{table}

\begin{figure}[!t]
\includegraphics[scale=0.45]{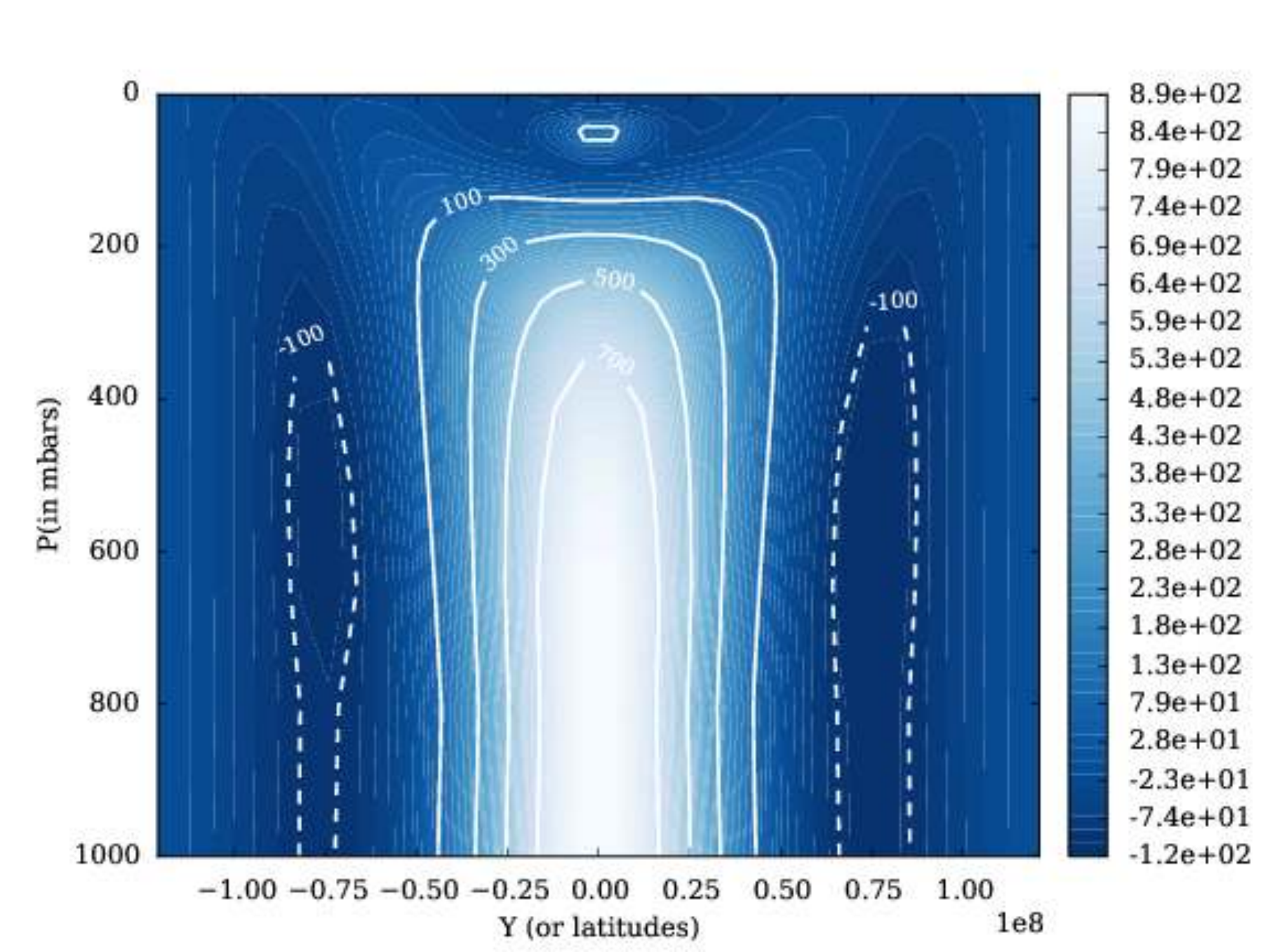}
\includegraphics[scale=0.45]{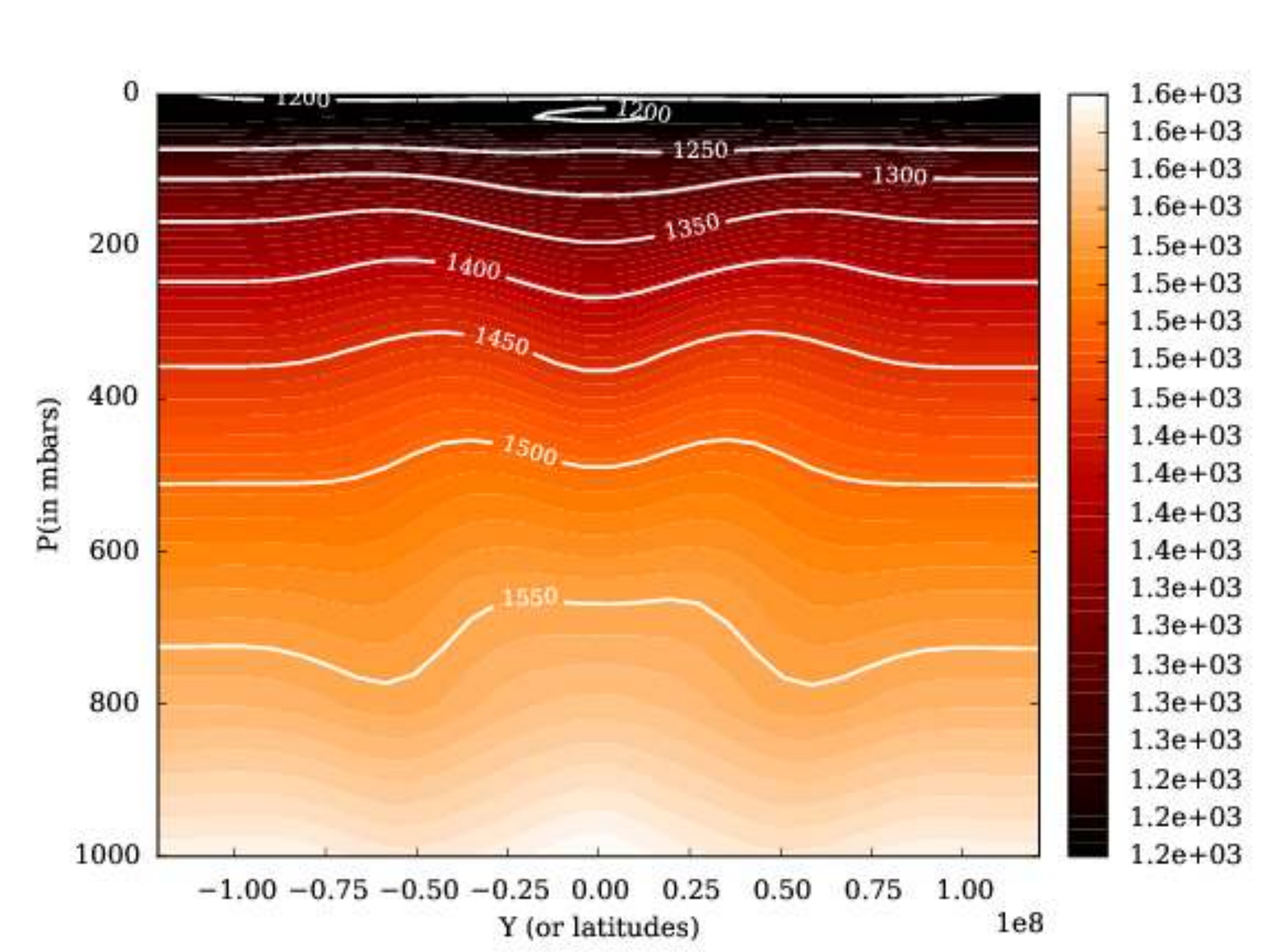}
\caption{Zonally averaged zonal wind ({\it top panel}) and temperature
  ({\it bottom panel}) in the shallow hot Jupiter model. The raw
  simulation data are averaged over $200$ planet days starting at
  $t=100$. See text and table~\ref{tab:shallow_model} for the model
  parameters.}
\label{fig:zonalplot_shallow}
\end{figure}

\begin{figure}[!t]
\includegraphics[scale=0.24]{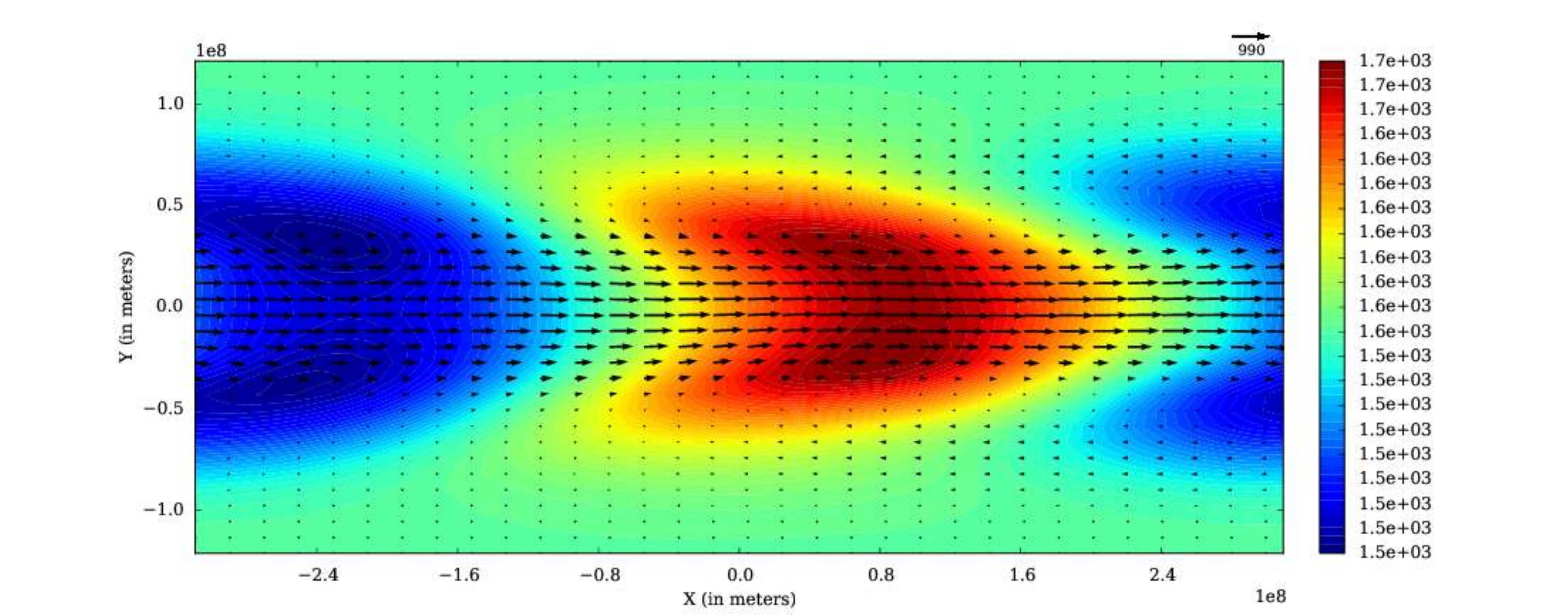}
\caption{Temperature ({\it color contours}) and wind velocities at the
$750$ mbars pressure level in the shallow hot Jupiter model, averaged
  in time over $200$ planet days starting at $t=100$.}
\label{fig:XYplot_shallow}
\end{figure}

We next reproduce the benchmark simulation of a shallow
hot Jupiter, such as presented by \citet{menou&rauscher09},
\citet{hengetal11}, \citet{bendingetal13} and \citet{mayneetal14}. As
for the case of the baroclinic instability presented above, we have
adapted the setup described by these authors to the equatorial
$\beta$--plane. It differs from the deep hot Jupiter model by the
depth of the  atmosphere, which extends downward only to $1$ bar, and
by the parameters entering in the cooling function $\cal{L}$. More
specifically, $\tau_{rad}$ is a constant equal to half a planet day
and $T_{eq}$ is given by the function:
\begin{equation}
T_{eq}=T_{perp}+\beta_{\textrm{trop}} \Delta T \cos \left(\frac{2 \pi x}{L_x} \right) \exp
  \left(-\frac{y^2}{2 L_{th}^2} \right) \, ,
\end{equation}
where $T_{perp}$ and $\beta_{\textrm{trop}}$ depend on $z$ according to
\begin{eqnarray}
T_{perp} = T_{\textrm{surf}} &-& \Gamma_{\textrm{trop}}\left(
z_{\textrm{stra}}+\frac{z-z_{\textrm{stra}}}{2}\right) \nonumber \\
 &+& \sqrt{\delta T_{\textrm{stra}}^2+\left(\frac{\Gamma_{\textrm{trop}}}{2}[z-z_{\textrm{stra}}] \right)^2}
\end{eqnarray}
and
\begin{equation}
\beta_{\textrm{trop}}=\begin{cases}
  \sin\left( \frac{\pi}{2}
     \frac{\sigma-\sigma_{\textrm{stra}}}{1-\sigma_{\textrm{stra}}} \right) \, ,&
     \textrm{if } z<z_{\textrm{stra}} \\
 0 \, ,& \textrm{ otherwise.}
\end{cases}
\end{equation}
The different parameters of the problem are taken identical
to the original papers mentioned above and are recalled for
completness in table~\ref{tab:shallow_model}. In our case, we also
need to specify the 
location of the upper boundary of the domain, which we take to be
$L_z=3 \times 10^6$ meters, while we retain for $L_x$ and $L_y$ the
same values as for the deep model described in
section~\ref{sec:num_params}. $L_{th}$ and $\beta$ also take values
identical to those of the main body of the paper. We start our
simulation with an atmosphere initially at rest and in hydrostatic
equilibrium with $T=T_{\textrm{perp}}$. We integrate the hydrodynamics
equations for $300$ planet days, of which the last $200$ days are used
to produce time averaged fields. The grid resolution is chosen to be
$(N_x,N_y,Nz)=(64,32,24)$.

As for the deep hot Jupiter model described in section~\ref{sec:LR},
we find that a fast equatorial wind develops with typical amplitude of $1000$
m.s$^{-1}$. Here, however, the flow display significant
fluctuations with typical wind velocity fluctuations of a few tens of
m.s$^{-1}$ (not shown). These fluctuations have also been reported by
previous authors and likely result from the interaction between the
atmosphere and the bottom domain boundary. The time averaged spatial
distributions of the zonally averaged zonal wind and temperature are
both in good agreement with the papers mentionned above (see
figure~\ref{fig:zonalplot_shallow}). Likewise, the temperature 
distribution at $750$ mbars display the familiar chevron--like shape
that has been identified as the consequence of planetary scale waves by
\citet{showman&polvani10,showman&polvani11}.

The good agreement of our results with previously published
simulations of shallow hot Jupiter atmospheres demonstrate that the
equatorial $\beta$ plane model is appropriate for the study of slowly
rotating tidally locked gas giant planet.

\section*{ACKNOWLEDGMENTS}
SF research is funded by the European Research Council under the
European Union's Seventh Framework Programme (FP7/2007-2013) / ERC
Grant agreement n 258729. SF acknowledges the hospitality of the Kavli
Institute of for Theoretical Physics, Santa Barbara, where some of
this work was completed during the program "Wave--flow interaction in
Geophysics, climate, astrophysics and plasmas'' as well as enlightening
discussions with Gwendal Rivi\`ere.

\bibliographystyle{aa}
\bibliography{author_atm}

\end{document}